\documentclass[
  aps,prc,twocolumn,superscriptaddress,floatfix,
  amsmath,amssymb,nofootinbib,longbibliography
]{revtex4-2}

\usepackage{url}
\usepackage{mathtools}
\usepackage{braket}
\usepackage{physics}
\usepackage{soul}
\usepackage[english]{babel}
\usepackage{bm}
\usepackage{comment}
\usepackage{color}
\usepackage{graphicx}
\usepackage{orcidlink}

\usepackage{hyperref}
\hypersetup{
  colorlinks = true,
  citecolor  = blue,
  linkcolor  = blue,
  urlcolor   = blue
}

\makeatletter
\let\label\ltx@label
\makeatother

\newcommand{\IdentityMat}{\mathbf{1}}

\newcommand{\oprsPH}{$k_F$}
\newcommand{\oprsMaxuv}{$\mu$}

\newcommand{\acroExc}{ \ell }
\newcommand{\acroOmega}{ \omega_{n_q} }

\newcommand{\CB}[1]{{\color{black} #1}}

\begin{document}

\author{F.~Marino~\orcidlink{0000-0001-7743-1982}}
\email{frmarino@uni-mainz.de}
\affiliation{Institut f\"{u}r Kernphysik and PRISMA+ Cluster of Excellence, Johannes Gutenberg-Universität Mainz, 55128 Mainz, Germany}

\author{C. Barbieri~\orcidlink{0000-0001-8658-6927}}  
\email{carlo.barbieri@unimi.it}
\affiliation{Dipartimento di Fisica ``Aldo Pontremoli'', Universit\`a degli Studi di Milano, 20133 Milano, Italy}
\affiliation{INFN,  Sezione di Milano, 20133 Milano, Italy}

\author{G.~Col\`{o}~\orcidlink{0000-0003-0819-1633} } 
\email{colo@mi.infn.it}
\affiliation{Dipartimento di Fisica ``Aldo Pontremoli'', Universit\`a degli Studi di Milano, 20133 Milano, Italy}
\affiliation{INFN,  Sezione di Milano, 20133 Milano, Italy}

\title{Gorkov algebraic diagrammatic construction for infinite nuclear matter 
}

\begin{abstract}
We propose \CB{a novel} many-body truncation for Gorkov self-consistent Green's function (SCGF) theory where pairing correlations are handled at first order, while \CB{dynamical correlations} are described using the particle-number-conserving Dyson-SCGF scheme up to third order in the algebraic diagrammatic construction. 
The new method is enabled by the introduction of a scheme that allows to approximate the Gorkov propagator in terms of a particle-number-conserving optimized reference state.
The approach provides state-of-the-art predictions of the equation of state and spectral properties of infinite nuclear matter at zero temperature and in the presence of pairing. We find satisfactory results using modern saturating Hamiltonians at next-to-next-to-leading order in chiral effective field theory.
\end{abstract}


\maketitle

\section{Introduction}
\label{sec: intro}
Achieving accurate \emph{ab initio} descriptions of nuclear phenomena is a long-term endeavor of theoretical nuclear physics.  
At present, advanced quantum-mechanical methods to solve the many-nucleon problem~\cite{computational_nuclear,Coraggio2020FrontBook,Hergert2020} incorporate the use of chiral effective field theory ($\chi$EFT), which provides the most widely used framework for understanding nuclear interactions in terms of the emergent degrees of freedom, that is, nucleons and pions~\cite{Machleidt2011,MACHLEIDT2024104117,Epelbaum2009,Epelbaum2024}. The combination of accurate many-body techniques and interactions enables to provide first-principle predictions with controlled estimates of the theoretical errors~\cite{EkstromAbInitio,papenbrock2024}.

Applying \textit{ab initio} techniques to the study of homogenous nuclear matter is of interest because the nuclear matter equation of state (EOS)~\cite{Gandolfi2015,rocamaza2018,Burgio2020,chatziioannou2024neutronstarsdensematter,margueron2025nucleardatapytoolkitsimpleaccess,ji2025equationstateneutronstars,Compose} enters as a key input in the description of neutron star structure~\cite{Piekarewicz2018,chatziioannou2024neutronstarsdensematter} and complex astrophysical scenarios, such as supernova explosions~\cite{Oertel2017,Burrows:2020qrp} or neutron-star mergers~\cite{Baiotti2019,Burgio2021,Burns:2019byj}.
Symmetric nuclear matter (SNM), made of protons and neutrons in equal number, is relatively well constrained around saturation density $\rho_0=0.16\,\rm{fm}^{-3}$ by measurements of finite nuclei~\cite{rocamaza2018,Piekarewicz2019NeutronSkin,Sorensen:2023zkk,Piekarewicz:2024tun}. On the contrary, isospin-asymmetric matter, and pure neutron matter (PNM) in particular, are less well-known.
PNM can be described as an interacting Fermi liquid~\cite{Gandolfi2015,Schwenk2012} at densities of the order of $\rho_0$.
However, its high-density behavior is still largely unexplored~\cite{Gandolfi2015,chatziioannou2024neutronstarsdensematter}, while, at very low densities, a superfluid phase transition is expected to take place~\cite{Haskell2018,Sedrakian2019,Ramanan2020}, which can leave observable signatures in e.g. the cooling of neutron stars or in the phenomenon of glitches~\cite{Glitches,Burgio2020}.
Information on the symmetry energy can be linked to isovector properties of nuclei, such as the neutron skin thickness~\cite{Piekarewicz2019NeutronSkin,Piekarewicz:2024tun} or the electric dipole polarizability~\cite{rocamaza2018,PbAbInitio,NeumannCosel2025}.
Still, uncertainties remain large since neutron-rich nuclei are hard to study experimentally. Phenomenological energy density functional calculations~\cite{colo2020,Schunck2019,PASTORE20151}, being fitted to nuclei close to stability, may loose predictive power when the neutron fraction increases and they have a large spread in their predictions for PNM~\cite{Burgio2020,DftTables}.

The \textit{ab initio} approach can help answer the above questions by providing controlled predictions anchored in microscopic nuclear interactions.
Earlier simulations, including Br\"{u}ckner-Hartree-Fock~\cite{Burgio2021} and self-consistent Green's function (SCGF) theory at finite temperature~\cite{Frick2003scgfT,Soma2009scgfT,Rios2020}, have been capable of handling hard interactions to pursue predictions at large densities. More recently, $\chi$EFT potentials have been embedded in several methods, such as coupled-cluster (CC) theory~\cite{Hagen2014,LietzCompNucl}, many-body perturbation theory (MBPT) ~\cite{Drischler2019,Drischler2021Review,Keller2023,Marino2024}, coordinate-space Quantum Monte Carlo~\cite{Gandolfi2015,Tews2020,Lonardoni2020LocalChiral}, 
configuration-interaction Monte Carlo~\cite{Roggero2013CIMC,Roggero2014CIMC,Arthuis2023} and recently full configuration-interaction Quantum Monte Carlo~\cite{Hu2026Fciqmc}, lattice effective field theory~\cite{Lu2020,Lee2025}, and in-medium similarity renormalization group~\cite{ZHEN2025139350}, as well as finite-temperature SCGF~\cite{Carbone2013Sym,Carbone2014,Carbone2020,Rios2020}.
Besides the EOS, these studies have addressed single-particle (s.p.) properties of nuclear matter, such as occupation numbers~\cite{Arthuis2023,Rios2020}, nucleon mean free paths~\cite{Rios2012}, effective masses~\cite{Gezerlis2019}, PNM pairing gaps~\cite{Rios2017Pairing,Gezerlis2021Twist,Drissi2022}, and the static~\cite{Gezerlis2016,Gezerlis2017} and dynamic response~\cite{Shen2014,PASTORE20151,sobczyk2024spinresponseneutronmatter} of PNM.

The SCGF approach is a powerful method to investigate strongly-correlated systems, with the added advantage that the language of propagators makes it natural to investigate the s.p.~dynamics in the nuclear medium~\cite{DickhofVanNeck,Barbieri2004,Barbieri2017,Soma2020,Rios2020}. 
Current applications (see e.g.~\cite{Carbone2020,Rios2020}) for infinite matter focus on a T-matrix resummation of two-particle (pp) and two-hole (hh) ladder diagrams. 
The finite-temperature formalism is exploited to circumvent pairing instabilities at low densities, while extrapolations to $T=0$ are performed when required. Refs.~\cite{Muther2005pairing,Ding2016pairing,Drissi2022} have investigated pairing gaps by including fragmentation effects in a generalised gap equation, but still relied on computing the spectral functions at finite temperatures. 
A Nambu covariant formalism that, instead, embeds both pairing and the thermal ensemble in a systematic fashion was proposed in Refs.~\cite{Drissi2024NCPT1,Drissi2024NCPT2}.

An alternative approach is based on the algebraic diagrammatic construction (ADC) scheme, which provides an improvable hierarchy of many-body approximations~\cite{Schirmer2018,Barbieri2017}. State-of-the-art simulations of closed-shell finite nuclei apply ADC at third-order [ADC(3)] in the so-called Dyson formulation---where reference states conserve particle number symmetry.
A superfluid extension of ADC, rooted in Gorkov Green's functions theory, has been developed in Refs.~\cite{Soma2011,Soma2014Numerical,Barbieri2022Gorkov} and applied successfully to study open-shell nuclei.
Numerous Gorkov calculations have been performed to date for open-shell isotopes with the second-order truncation, ADC(2), see for example Refs.~\cite{Soma2014Chains,Soma2020Chiral,Soma2020}, while the Gorkov-ADC(3) working equations have been derived only recently~\cite{Barbieri2022Gorkov}.

In this work, we present a novel approach that is grounded in Gorkov theory and achieves ADC(3) for nuclear matter.
Exploratory ADC(3) computations for nuclear matter were reported in Refs.~\cite{Barbieri2017,McilroyChristopher2020Sgfs} but based on Dyson theory, which is unstable to pairing effects.
The issue could be resolved by embedding the self-energies computed in the Dyson formalism within the Gorkov framework to include a static self-consistent pairing field~\cite{MarinoPhdThesis,Marino2024,Marino2025Qnp}.
Then, the bulk of dynamical contributions remains described with the Dyson-SCGF scheme up to ADC(3), allowing to recover correlation energies accurately.
A further improvement of our method exploits a combination of the ADC method with amplitudes determined from CC computations~\cite{Barbieri2017,Marino2024} to effectively include even higher-order many-body correlations.

Our method has been recently applied in Refs.~\cite{Marino2024,Marino2025Qnp,Colo:2025ejt} to predicting the EOS, and its accuracy has been demonstrated in Ref.~\cite{Marino2024} by a comparison of nuclear matter energies between ADC, CC, and MBPT at third-order.
However, a full account of our ADC-SCGF approach for nuclear matter was not yet available.
The purpose of the present paper is twofold.
First, we give an in-depth description of the method (Sec.~\ref{sec:formalism}), introducing the hybrid Gorkov-Dyson scheme and discussing its novel features, in particular, the modification of the ADC self-energy by the inclusion of CC corrections. 
Second, after carefully validating the numerical convergence of our approach (Sec.~\ref{sec:validation}), we present extensive new results extending the scope of ADC-SCGF to the calculation of both EOS and s.p.~properties, notably, momentum distributions and spectral functions as obtained with different $\chi$EFT interactions.
Finally, conclusions and perspectives are outlined in Sec.~\ref{sec: Conclusions}.
Further details regarding the working equations specific to homogeneous infinite matter and relations to other conventions used in the Gorkov SCGF literature are collected in the Appendices.

\section{Formalism}
\label{sec:formalism}

The starting point of our calculations is a Hamiltonian $H$ comprising two-body (2B) and three-body (3B) interactions, respectively $V$ and $W$,
\begin{align}
    \label{eq:H}
    H = H_0 + H_1,
\end{align}
where
\begin{subequations}\label{eq:H0_H1}
\begin{align}
    \label{eq:H0vsTU}
    & H_0 = \sum_{\alpha} \left(\frac{\hbar^2 }{2\, m} \mathbf{k}_{\alpha}^{2}  + u_\alpha \right)  c^\dagger_\alpha c_\alpha, \\
    \label{eq:H1vsVW}
    & H_1 = - \sum_\alpha u_\alpha c^\dagger_\alpha c_\alpha + \frac 1 4 \sum_{\alpha\, \beta\, \gamma\, \delta} \
    {v}_{\alpha \beta, \gamma \delta} \, c^\dagger_\alpha c^\dagger_\beta c_\delta c_\gamma \nonumber \\
    &\qquad + \frac 1 {36} \sum_{\alpha\, \beta\, \gamma\, \mu\, \nu\, \rho} {w}_{\alpha \beta \gamma, \mu \nu \rho} \, c^\dagger_\alpha c^\dagger_\beta c^\dagger_\gamma c_\rho c_\nu c_\mu .
\end{align}
\end{subequations}
Greek indices denote a complete set of s.p.~states.
Creation (annihilitation) operators are denoted as $c^{\dagger}_{\alpha}$ ($c_{\alpha}$), respectively.
$\mathbf{k}_{\alpha}$ denotes the momentum carried by the s.p.~state, $m$ is the nucleon mass, and 2B and 3B matrix elements are antisymmetric and Hermitian.
In Eqs.~\eqref{eq:H0_H1}, we have added and subtracted an external one-body (1B) field, $U$, that we will use to define a self-consistent reference state.
Both the kinetic energy and the external field are assumed to be diagonal in our s.p.~basis because of the translational invariance of homogeneous matter.

In this work, the 3B interaction is incorporated as an effective 2B interaction via the normal-ordered 2B (NO2B) approximation~\cite{Carbone2013,Hebeler3nf,Marino2024,Hagen2014}. The NO2B scheme consists in averaging the 3B interaction over the states occupied in the mean-field reference (see App.~\ref{sec: adc3 diagrams} for details).

\subsection{Model space}
\label{sec: model space}
We simulate infinite nuclear matter as a finite system of $A=N+Z$ nucleons, with $N$ ($Z$) being the number of neutrons (protons), enclosed in a cubic box of size $L$ and open boundaries~\cite{Barbieri2017,LietzCompNucl,Marino2023,MarinoPhdThesis}. Given the nucleon density $\rho$, the size of the box is $L = (A/\rho)^{1/3}$ and its volume $V=L^3$.
Translational invariance imposes a discretized spectrum of momenta as
\begin{align}
    \label{eq:momgrid}
    \mathbf{k} = \frac{1}{L} \bigg( 2\pi \mathbf{n} + \boldsymbol{\theta} \bigg),
\end{align}
where $\mathbf{n} = (n_x,n_y,n_z)$ is a triplet of integer numbers, $n_i =0, \pm 1, \pm 2 ...$, and $\boldsymbol{\theta}$ denotes a set of three real numbers, called twist angles (see below). Assuming time-reversal invariance, $0 \le \theta_i < \pi$~\cite{Hagen2014}.
In this work, we truncate the momentum basis according to
\begin{equation}
    \label{eq:MStrunc}
    \abs{ \mathbf{k} }^2 = k_x^2 + k_y^2 + k_z^2 \le \frac{4\pi^2}{L^2} N_{max}.
\end{equation}
The standard case of periodic boundary conditions (PBCs) is obtained by neglecting all offset angles ($\theta_i=0$), which involves degeneracies between several different basis states $\bf{n}$. 
When using twisted-angle boundary conditions (TABC)~\cite{CeperleyTABC,Hagen2014}, a non-vanishing offset $\boldsymbol{\theta}$ allows the degeneracy to be lifted and generates a much finer mesh of momenta%
\footnote{In the limit of PBC, Eq.~\eqref{eq:MStrunc} reduces to the simpler form $\abs{\mathbf{n}}^2 = n_x^2 + n_y^2 + n_z^2 \le N_{max}$. When using TABCs, the condition~\eqref{eq:MStrunc} retains the same number of basis states---imposed through $ N_{max}$---but ensures that the model space remains nearly spherical and centered around $\mathbf{k}=0$.}.

The s.p.~basis is made of momentum eigenstates,
\begin{align}
    \label{eq: sp states}
    \ket{\alpha} = \ket{ \mathbf{k}_\alpha, \sigma_\alpha, \tau_\alpha},
\end{align}
where momenta $\mathbf{k}_{\alpha}$ have the form of Eq.~\eqref{eq:momgrid}, $\sigma_\alpha = \pm 1/2$ is the spin projection, and $\tau_\alpha = p,n$ is the isospin projection.

As a consequence of the isotropy of homogeneous matter, the energy of s.p.~states depends only on $\abs{ \mathbf{k}_{\alpha} }$ and the isospin, but not on the direction of the momentum.
PBCs give rise to a shell structure in momentum space with the filling of spherical momentum shells at particle numbers $A/g=$1, 7, 19, 27, 33 etc.~\cite{LietzCompNucl,Marino2023}, where $g$ is the spin-isospin degeneracy (2 for spin-saturated PNM, 4 for spin-saturated SNM).
These degeneracies can result in large gaps at the Fermi surface, hence mimicking a stable closed-shell system but possibly introducing sizeable finite-size (FS) effects. 
Among the possible choices, simulations with $33g$ fermions are known to minimize the FS effects for the EOS at the mean field level~\cite{Hagen2014,Gezerlis2017,Marino2023}. Hence, PBC simulations routinely employ $N=66$ neutrons for PNM and $A=132$ nucleons for SNM~\cite{Marino2024,Hagen2014}.

The twisted angles in Eq.~\eqref{eq:momgrid} remove the degeneracies in the values of $\abs{ \mathbf{k}_{\alpha} }$ and generate a finer momentum mesh. This improves the description of s.p.~quantities, including the momentum distributions, as well as the EOS, and it generally results in a far better approximation of the thermodynamic limit (TL).
On the other hand, the momentum and energy gaps at the Fermi surface are reduced, potentially leading to numerical instability. 
This issue is solved by the mixed Dyson-Gorkov formulation of ADC-SCGF introduced in the following subsections.
In this work, we will use the special-point TABC (sp-TABC) prescription~\cite{CeperleyTABC,Hagen2014}, which consists in exploiting only one specific choice of twist angles. The optimal angles are determined by minimizing the deviation of the Hartree-Fock (HF) energy in the finite-$A$ system with respect to its value at the TL~\cite{Hagen2014,McilroyChristopher2020Sgfs}, a procedure that has proven successful and substantially less demanding than taking the full TABC averages among many sets of angles~\cite{Hagen2014}.

\subsection{Gorkov equations}
\label{sec:GkvDys_formalism}

In this work, we follow the Gorkov formulation of SCGF~\cite{Soma2011,Soma2014Numerical,Soma2020,Barbieri2022Gorkov} and consider the zero-temperature grand canonical Hamiltonian
\begin{align}
 \Omega = H - \mu_p \, Z - \mu_n \, N
\label{eq:Odef}
\end{align}
where $\mu_p$ and $\mu_n$ are the chemical potentials and $Z$ and $N$ are the particle number operators for protons and neutrons, respectively. 
The ground state (g.s.)~eigenvector and eigenenergy of Eq.~\eqref{eq:Odef} are labeled $|\Psi_0\rangle$ and $\Omega_0$, respectively, and correspond to a fixed even number of particles only on average, while they have conserved particle number parity. 
Eigenstates of $\Omega$ over the Fock space are the solutions to
\begin{align}
    \label{eq:Om_eig_sts}
    \Omega \ket{\Psi_q} = \Omega_q \ket{\Psi_q}.
\end{align}
The Gorkov propagators associated to $|\Psi_0\rangle$ are\,%
\footnote{In this work we adopt different definitions of the Gorkov propagators than in Refs.~\cite{Soma2011,Soma2014Numerical,Barbieri2022Gorkov}, which relied on the dual model space to $\{\alpha\}$ and an anti-unitary phase $\eta_\alpha$. The relations among the same quantities in the two formalisms are summarized in App.~\ref{app:g_vs_G}.} 
\begin{subequations}
\label{eq:gkv_def}
\begin{align}
g^{11}_{\alpha \beta}(t,t') =& - \frac{i}{\hbar}
\mel{\Psi_0} {T [ c_\alpha(t) c^\dagger_\beta(t') ]} {\Psi_0} ,\\
g^{12}_{\alpha \beta}(t,t') =& - \frac{i}{\hbar} 
\mel{\Psi_0}{T [ c_\alpha(t) c_\beta(t') ] } {\Psi_0} ,\\
g^{21}_{\alpha \beta}(t,t') =& - \frac{i}{\hbar}
\mel{\Psi_0}{T [ c^\dagger_\alpha(t) c^\dagger_\beta(t') ] } {\Psi_0} ,\\
g^{22}_{\alpha \beta}(t,t') =& - \frac{i}{\hbar} 
\mel{\Psi_0}{T [ c_\alpha^{\dagger}(t) c^\dagger_\beta(t') ] } {\Psi_0},
\end{align}
\end{subequations}
where the superscript `1' and `2' represent the so-called \emph{normal} and \emph{anomalous} Nambu indices, respectively, and $T[...]$ is the time ordering operator.
Following Ref.~\cite{Soma2011}, the propagators~\eqref{eq:gkv_def} can be interpreted as components of two-dimensional matrices in Nambu space,
\begin{align}
    \mathbf{g}_{\alpha\beta}(\omega) =
    \begin{pmatrix}
        g^{11}_{\alpha\beta}(\omega) & g^{12}_{\alpha\beta}(\omega) \\
        g^{21}_{\alpha\beta}(\omega) & g^{22}_{\alpha\beta}(\omega)
    \end{pmatrix}.
    \label{eq:gkv_Nambu}
\end{align}
The complete Gorkov propagator $\mathbf{g}_{\alpha\beta}(\omega)$ is a solution to the Gorkov equation
\begin{align}
    \mathbf{g}_{\alpha\beta}(\omega) = \mathbf{g}_{\alpha\beta}^{(0)}(\omega) +
    \sum_{\gamma\delta} \mathbf{g}_{\alpha\gamma}^{(0)}(\omega) \mathbf{\Sigma}^{\star}_{\gamma\delta}(\omega) \mathbf{g}_{\delta\beta}(\omega),
    \label{eq:GorkovEq}
\end{align}
which generalizes the Dyson-Schwinger equation.
In Eq.~\eqref{eq:GorkovEq}, $\mathbf{g}^{(0)}_{\alpha\beta}(\omega)$ denotes a reference propagator associated to $H_0$ in Eq.~\eqref{eq:H}.  
Many-body correlations are encoded in the irreducible self-energy operator, $\mathbf{\Sigma}^{\star}_{\gamma\delta}(\omega)$, which includes all possible one-particle irreducible diagrams stripped of their external legs. This can be expresses as the sum of a static contribution, $\mathbf{\Sigma}^{(\infty)}$, that accounts for the interaction of a particle with the correlated mean-field, and an energy-dependent component $\mathbf{ \widetilde{\Sigma} }(\omega)$,
\begin{align}
\mathbf{\Sigma}^{\star}(\omega) = - \mathbf{U} + \mathbf{\Sigma}^{(\infty)} + \mathbf{ \widetilde{\Sigma} }(\omega) \,.
\end{align}
The auxiliary 1B potential $\mathbf{U}$ cancels exactly its contributions from Eq.~\eqref{eq:H0vsTU} so that a fully self-consistent solution of Eqs.~\eqref{eq:GorkovEq} will be completely independent of the choice of $\mathbf{U}$. In this work, we use a partially self-consistent scheme as discussed in Sec~\ref{sec: oprs}, where $\mathbf{ \widetilde{\Sigma} }(\omega)$ is still implicitly affected by the choice of the auxiliary potential.

The Lehmann representation of the propagator is obtained by inserting a complete set of excited states~\eqref{eq:Om_eig_sts} into Eqs.~\eqref{eq:gkv_def} and~\eqref{eq:gkv_Nambu}. It reads
\begin{align}
    \label{eq:g_SpectRep}
    \mathbf{g}_{\alpha\beta}(\omega) \!=\!
    \sum_q 
    \frac{  \begin{pmatrix}
        \mathcal{U}_{\alpha}^{q} \\
        \mathcal{V}_{\alpha}^{q}
    \end{pmatrix} 
     \begin{pmatrix}
        \mathcal{U}_{\beta}^{q *} &
        \mathcal{V}_{\beta}^{q *}
    \end{pmatrix} }{ \hbar\omega - \hbar\omega_q + i\eta} \!+\!
    \sum_q 
    \frac{ \begin{pmatrix}
        \mathcal{V}_{\alpha}^{q *} \\
        \mathcal{U}_{\alpha}^{q *}
    \end{pmatrix} 
     \begin{pmatrix}
        \mathcal{V}_{\beta}^{q} &
        \mathcal{U}_{\beta}^{q}
    \end{pmatrix} }{ \hbar\omega + \hbar\omega_q - i\eta},
\end{align}
where we have introduced the spectroscopic amplitudes 
\begin{subequations} \label{eq:UV_def}
\begin{align}
    \mathcal{U}_{\alpha}^{q} ={}& \mel{\Psi_0}{c_\alpha}{\Psi_q}, \\
    \mathcal{V}_{\alpha}^{q} ={}& \mel{\Psi_0}{c_\alpha^\dagger}{\Psi_q},
\end{align}
\end{subequations}
and the index $q$ labels all possible s.p.~excitations with respect to $\ket{\Psi_0}$:
\begin{align}
    \hbar\omega_q = \Omega_{q} - \Omega_{0}.
\end{align}
Note that the Gorkov propagators are defined with respect to a g.s.~$\ket{\Psi_0}$ with even particle number. Hence, the poles $\omega_q$ in Eq.~\eqref{eq:g_SpectRep} run only over states with either odd $N$ or odd $Z$.

Normal ($\rho$) and anomalous ($\widetilde{\rho}$) densities are related to the spectroscopic amplitudes by~\cite{Soma2011}
\begin{align}
    \label{eq: norm density def}
    & \rho_{\alpha\beta} = \mel{\Psi_0}{c_\beta^{\dagger} c_{\alpha}}{\Psi_0} = \sum_{q} \mathcal{V}_{\alpha}^{q *} \mathcal{V}_{\beta}^{q}, \\
    \label{eq: anom density def}
    & \widetilde{\rho}_{\alpha\beta} = \mel{\Psi_0}{c_\beta c_{\alpha}}{\Psi_0} = \sum_{q} \mathcal{V}_{\alpha}^{q *} \mathcal{U}_{\beta}^{q}.
\end{align}

Whenever particle number is conserved -- as is the case for nuclear matter -- the Hamiltonian $H$ and~$\Omega$ share the same spectrum and all their eigenstates have exact particle number.
Since we aim at introducing a mixed approximation between the Dyson and the Gorkov formulations of SCGF, it is useful to recall the relations among these two cases. 
We use an index $n$ to label those s.p.~excitations from Eq.~\eqref{eq:Om_eig_sts} corresponding to the nucleon addition. Thus, for $q:=n$,
\begin{subequations} \label{eq:Dys_qp_ampl}
\begin{align}
    & H \, |\Psi_n^{A+1} \rangle= E_n^{A+1} \, |\Psi_n^{A+1} \rangle \,, \\
  & \mathcal{U}_{\alpha}^{n} = \mel{\Psi_0}{c_\alpha}{\Psi_n^{A+1}} , \\
  & \mathcal{V}_{\alpha}^{n} = \mel{\Psi_0}{c_\alpha^{\dagger}}{\Psi_n^{A+1}} \rightarrow 0 \, .
\end{align}
\end{subequations}
 Analogously, we use an index $q:=k$ for particle removal,
\begin{subequations} \label{eq:Dys_qh_ampl}
\begin{align}
 &   H \, |\Psi_k^{A-1} \rangle= E_k^{A-1} \, |\Psi_k^{A-1} \rangle \,, \\
 & \mathcal{U}_{\alpha}^{k} = \mel{\Psi_0}{c_\alpha}{\Psi_k^{A-1}} \rightarrow 0
 \,, \\&   \mathcal{V}_{\alpha}^{k} = \mel{\Psi_0}{c_\alpha^{\dagger}}{\Psi_k^{A-1}}
\,.
\end{align}
\end{subequations}
Hence, the amplitudes $\mathcal{U}_{\alpha}^{n}$ ($\mathcal{V}_{\alpha}^{k}$) for particle addition (removal) decouple exactly in the limit of restoration of particle number symmetry and the anomalous propagators in Eqs.~\eqref{eq:gkv_Nambu} and~\eqref{eq:g_SpectRep} vanish. Substituting Eqs.~\eqref{eq:Dys_qp_ampl} and ~\eqref{eq:Dys_qh_ampl} into~\eqref{eq:g_SpectRep} one finds:
\begin{subequations}
 \label{eq:g_to_Dys_limit}
\begin{align}
 g^{12}_{\alpha\beta}(\omega) ={}& g^{21}_{\alpha\beta}(\omega) = 0\,,\\
 g^{11}_{\alpha\beta}(\omega) ={}& - g^{22}_{\beta\alpha}(-\omega) = g_{\alpha\beta}(E=\hbar\omega+\mu_\alpha) \,,
\end{align}
\end{subequations}
where $\mu_{\alpha}$ denotes the chemical potential appropriate for the state $\alpha$,
\begin{align}
   \label{eq:Dys_SpectRep}
   g_{\alpha\beta}(E) =
    \sum_n \frac{
    \mathcal{U}_{\alpha}^{n} \,
    \mathcal{U}_{\beta}^{n *} }
    { E - \varepsilon_n + i\eta} +
    \sum_k \frac{ 
    \mathcal{V}_{\alpha}^{k *} \,
        \mathcal{V}_{\beta}^{k} }
        { E - \varepsilon_k - i\eta} 
\end{align}
is the Dyson propagator~\cite{Barbieri2017},
and
\begin{subequations}
\begin{align}
 \varepsilon_n ={}& E_{n}^{A+1} -  E_0^{A} =
    \quad \mu_n + \hbar \omega_{q=n} \,, \\
 \varepsilon_k ={}& E_0^{A} -  E_k^{A-1} =
    \quad \mu_k - \hbar  \omega_{q=k} \,.
\end{align}
\end{subequations}

The dynamic part of the Gorkov self-energy has an analogous spectral representation~\cite{Barbieri2022Gorkov,Soma2011},
\begin{align}
  \widetilde{\mathbf{\Sigma}}_{\alpha\beta}&(\omega) = 
  \begin{pmatrix}
        \widetilde{\Sigma}^{11}_{\alpha\beta}(\omega) &
        \widetilde{\Sigma}^{12}_{\alpha\beta}(\omega) \\
        \widetilde{\Sigma}^{21}_{\alpha\beta}(\omega) &
        \widetilde{\Sigma}^{22}_{\alpha\beta}(\omega)
    \end{pmatrix} 
  \nonumber \\
  ={}&\sum_{\nu \,\nu'}
  \begin{pmatrix}
        \mathcal{M}_{\alpha\,\nu}^\dagger \\
        \mathcal{N}_{\alpha\,\nu}^*
    \end{pmatrix} 
    \left[
  \frac{1}{\hbar\omega\IdentityMat - \mathcal{E} + i\eta}
  \right]_{\nu \nu'}
     \begin{pmatrix}
        \mathcal{M}_{\nu'\beta} &
        \mathcal{N}_{\nu'\beta}^T
    \end{pmatrix} \nonumber \\
    +{}& \sum_{\nu \,\nu'}
  \begin{pmatrix}
        \mathcal{N}_{\alpha\,\nu}  \\
        \mathcal{M}_{\alpha\,\nu}^T
    \end{pmatrix} 
    \left[
  \frac{1}{\hbar\omega\IdentityMat + \mathcal{E}^T - i\eta}
  \right]_{\nu \nu'}
     \begin{pmatrix}
        \mathcal{N}_{\nu'\beta}^\dagger &
        \mathcal{M}_{\nu'\beta}^*
    \end{pmatrix} \,,
  \label{eq:DynSE_Gk} 
\end{align}
where the index $\nu$ runs over a set of intermediate state configurations (ISCs) of (2$n$+1)-quasiparticles $q$, with $n\geq$~1. The matrices $\mathcal M$ and $\mathcal N$ are coupling vertices among the ISC and s.p.~excitations, while the matrix $\mathcal E$ describes the interactions among ISCs.
In Eq.~\eqref{eq:DynSE_Gk} we have put in evidence the four Nambu components of the self-energy, which are related to each other by~\cite{Soma2011}
\begin{subequations} \label{eq:SigGk_rels}
\begin{align}
  \widetilde{\Sigma}^{11}_{\alpha\beta}(\omega) ={}& - \widetilde{\Sigma}^{22}_{\beta\alpha}(-\omega) \,,\\
  \widetilde{\Sigma}^{12}_{\alpha\beta}(\omega) ={}& - \widetilde{\Sigma}^{12}_{\beta\alpha}(-\omega) \,, \\
  \widetilde{\Sigma}^{21}_{\alpha\beta}(\omega) ={}& - \widetilde{\Sigma}^{21}_{\beta\alpha}(-\omega) \,.
\end{align}
\end{subequations}
The corresponding spectral representation in the Dyson formalism is analogous and reads
\begin{align}
    \label{eq:DynSE_Dys}
    \widetilde{\Sigma}^{Dys}_{\alpha\beta}(\omega) & =
    \sum_{rr^\prime} M_{\alpha r}^\dagger \left[ \frac{1}{\hbar\omega\IdentityMat -(E^{>}+C)  + i \eta} \right]_{rr^\prime} M_{r^\prime \beta} \nonumber \\ & + \sum_{ss^\prime} N_{\alpha s} \left[ \frac{1}{\hbar\omega\IdentityMat -(E^{<}+D) - i \eta} \right]_{ss^\prime} N_{s^\prime \beta}^{\dagger} \,,
\end{align}
where the index $r$ runs over ISCs corresponding to particle addition [two-particle-one-hole (2p1h), 3p2h etc. configurations] and the index $s$ is for particle removal ones (1p2h, 2p3h...)~\cite{Raimondi2017,Barbieri2017}. Similarly to Eqs.~(\ref{eq:Dys_qp_ampl}-\ref{eq:g_to_Dys_limit}), in the limit of conserved particle-number symmetry,
we find that $\mathcal N\rightarrow$~0 ($\mathcal M\rightarrow$~0) for particle attached (particle removed) and the index $\nu$ in Eq.~\eqref{eq:DynSE_Gk} decouples in either $r$ or $s$ type of ISCs. In this situation, analogously to Eq.~\eqref{eq:g_to_Dys_limit}, 
\begin{subequations} \label{eq:SE_to_Dys_limit}
\begin{align}
 \widetilde{\Sigma}^{12}_{\alpha\beta}(\omega) ={}& 
 \widetilde{\Sigma}^{21}_{\alpha\beta}(\omega) =0\,,\\
 \widetilde{\Sigma}^{11}_{\alpha\beta}(\omega) ={}&
-\widetilde{\Sigma}^{22}_{\beta\alpha}(-\omega) =
 \widetilde{\Sigma}^{Dys}_{\alpha\beta}(E=\hbar\omega+\mu_\alpha) \,.
\end{align}
\end{subequations}
Relations~\eqref{eq:SE_to_Dys_limit} follow straightforwardly from inserting the limit~\eqref{eq:g_to_Dys_limit} for the propagator into the full Gorkov equation~$\eqref{eq:GorkovEq}$ and comparing to the standard Dyson equation of the particle-number-conserving framework.
Moreover, Eqs.~\eqref{eq:SE_to_Dys_limit} also hold separately at each order of the ADC($n$). This property can be exploited to devise many-body truncations in a hybrid Gorkov-Dyson framework.

\CB{
It should be stressed that both the Dyson and Gorkov formulations of SCGF are implemented as expansions in the full Fock space. As a consequence, the particle number can be broken by any given many-body truncation and both approaches restore particle number only  in the limit of an \emph{exact resummation} of the Feynman series. 
This is signaled by the fact that the particle-number expectation value, $\expval{A} = \sum_{\alpha} \rho_{\alpha\alpha}$, computed on the dressed propagator might differ from the exact number of particles $A$, even if this is a slight deviation in the Dyson case.
While there exist classes of approximations that guarantee that conservation laws are rigorously satisfied by the solution of the Dyson equations (see e.g.~\cite{Baym1,Baym2,Reining_Propagators,Soma2011}), they require achieving perfect self-consistency~\cite{VanNeck2001Dys2ndSC,Peirs2002Dys2ndSC}. This is impractical beyond the trivial first order. Thus, all commonly used Green's function approaches imply a violation of particle number to some extent.

The Dyson and Gorkov approaches differ in the choice of the starting point for the perturbative expansion. Dyson-SCGF is restricted to a particle-number-conserving reference state and is characterized by a many-body expansion only in terms of conserving propagators. As a consequence, it neglects any anomalous component of the self-consistent propagator and of its self-energy (anomalous components would vanish anyway for an exact solution). Conversely, Gorkov-SCGF allows for particle number breaking already in the reference state, $\mathbf{g}_{\alpha\beta}^{(0)}(\omega)$, which provides greater freedom to choose the starting point of the perturbative expansion. For our purposes, the Gorkov framework allows encoding pairing correlations already at zeroth order of the Feynman expansion, while these could not be handled by a conserving reference state.
On the other hand, the Gorkov approach entails a more complex and more computationally costly diagrammatic expansion due to the presence of anomalous terms~\cite{Barbieri2022Gorkov}. Moreover,  imposing symmetry breaking from the start implies more extensive violations of particle number conservation, compared to the corresponding Dyson formulation. Hence, the Gorkov approach requires additional iterations to adjust the average particle number by varying the chemical potentials in the grand-canonical Hamiltonian~\eqref{eq:Odef}~\cite{Soma2014Numerical}.
}

\subsection{Self-energy approximations}
\label{sec: Self-energy approx}
The Gorkov equations~\eqref{eq:GorkovEq} are an exact formulation of the many-body problem, but they require the knowledge of the self-energy, which can only be computed approximately. The SCGF approach is based on expressing the self-energy as a functional of the dressed propagator, $\mathbf{\Sigma}(\omega)=\mathbf{\Sigma}[\mathbf{g}]$, which then simultaneously determines and it is determined by $\mathbf{\Sigma}$, making the problem inherently self-consistent~\cite{Barbieri2004,Soma2020}.
Within this framework, the static self-energy $\mathbf{\Sigma}^{(\infty)}$ can be evaluated exactly as a function of the propagator and the bare 2B and 3B interactions (see App.~\ref{App:SEinfty_diags}). One is then left with the problem of devising working approximations for the dynamical self-energies, Eqs.~\eqref{eq:DynSE_Gk} and~\eqref{eq:DynSE_Dys}.

In this work, we adopt a novel hybrid scheme in which the static self-energy, at lowest order, is computed in full Gorkov formalism, but we neglect the higher-order dynamical contributions to the anomalous self-energies. Hence, only normal dynamical terms evaluated from a Dyson-ADC(3) expansion are retained.
The prescription of including pairing only at the lowest order is inspired by Refs.~\cite{Muther2005pairing,Ding2016pairing} and it is advantageous from both a physical and technical perspective. 
Exploiting a grand-canonical formulation allows to constrain the average particle number by properly adjusting the chemical potentials~\cite{Barbieri2017,Soma2011}.
Limiting $\mathbf{\widetilde\Sigma}(\omega)$ to a Dyson formulation allows us to describe strong many-body dynamical correlations while avoiding the complications that would emerge in a full particle number breaking formalism.
In addition, the Gorkov approach avoids the instabilities associated with the vanishing of the particle-hole excitation gap in low-density matter~\cite{Burgio2021,Sedrakian2019,Gandolfi2015}.
In this regime, which marks the onset of superfluidity, many-body methods that do not treat pairing explicitly tend to struggle. 
Examples include CC computations, which fail to converge in low-density neutron matter, and finite-temperature SCGF, where calculations are performed at temperatures $\ge 5\,\rm{MeV}$ to avoid the pairing instability~\cite{Carbone2013Sym,Rios2020,Ding2016pairing}.
In contrast, our Gorkov approach remains remarkably stable over a wide range of densities in both SNM and PNM.

\CB{
The effect of pairing contributions on total energies is typically modest~\cite{Dean2003}, being superfluidity predominantly a Fermi-surface phenomenon.
Thus, as far as energies are concerned, neglecting the dynamical dependence of the anomalous self-energies is a well-justified approximation.
It is harder to judge a priori what the consequences of this approximation might be for other quantities, such as the pairing gaps. To our knowledge, only a few computations have addressed a frequency-dependent anomalous self-energy (see e.g.~\cite{Sedrakian:2003wk,Urban2026}).
Instead, most calculations of the neutron matter pairing gaps (see e.g.~Refs.~\cite{Sedrakian2006,Ding2016,Sedrakian2019,Ramanan2020}) are based on the standard (static) gap equation, which is modified by using a dressed single-particle spectrum (or, equivalently, an effective mass), as well as possibly including the effect of the depletion at the Fermi surface and/or screening effects.
Notice that these in-medium corrections are naturally embedded also in our formalism.
}

Using our hybrid scheme, the nuclear self-energy is written as 
\begin{widetext}
\begin{align}
    \mathbf{\Sigma}^{\star}_{\alpha\beta}(\omega) =
    \begin{pmatrix}
        -u_{\alpha\beta} + \Sigma^{(\infty)\,11}_{\alpha\beta} + \widetilde\Sigma^{ADC}_{\alpha\beta}(\mu_\alpha + \hbar\omega) &
        \Sigma^{(\infty)\,12}_{\alpha\beta} \\
        \Sigma^{(\infty)\,21}_{\alpha\beta} & + u_{\alpha\beta}^* + \Sigma^{(\infty)\,22}_{\alpha\beta}
        - \widetilde\Sigma^{ADC}_{\beta\alpha}(\mu_\alpha - \hbar\omega)
    \end{pmatrix} \,,
    \label{eq:HybridSig}
\end{align}
\end{widetext}
where we have exploited Eqs.~\eqref{eq:SigGk_rels} and~\eqref{eq:SE_to_Dys_limit}, while $\widetilde\Sigma^{ADC}(\omega)$ is computed in Dyson formalism and takes the analytic form of Eq.~\eqref{eq:DynSE_Dys}.

The bulk of the contributions to correlation energy stems from the dynamic part of the normal self-energy, $\widetilde\Sigma^{ADC}(\omega)$, which we treat using the state-of-the-art Dyson-ADC method~\cite{Barbieri2017,Schirmer1989,Raimondi2017,Barbieri2022Gorkov,Schirmer2018}.
ADC($n$) defines a systematically improvable hierarchy of approximations, where each self-energy is determined by enforcing the spectral representation of Eq.~\eqref{eq:DynSE_Dys} (or Eq.~\eqref{eq:DynSE_Gk} for Gorkov-SCGF), while at the same time retaining all terms up to order $n$ in the perturbative expansion.
Starting from the third order, ADC resums automatically infinite classes of diagrams, including the ladder series, which is expected to give the dominant contributions in homogeneous matter~\cite{Mattuck,Rios2020}, and the (Tamm-Dancoff) ring diagrams, plus specific fourth-order diagrams~\cite{Barbieri2017}.
For $n \to \infty$, the exact $\Sigma^{\star}(\omega)$ would be recovered. 
Gorkov-ADC and Dyson-ADC have been applied successfully to finite nuclei using the second-order (see e.g.~\cite{Soma2014Chains,Soma2020Chiral,Soma2020}) and third-order approximations~\cite{Cipollone:2013zma,Arthuis:2020toz,Soma2020}, respectively. The Gorkov-ADC(3) working equations have been derived in Ref.~\cite{Barbieri2022Gorkov}.

\CB{Note that the Dyson-ADC(3) simulation accounts for the fragmentation of single particle strength at a few tens of MeVs above and below the Fermi energy. These effects are associated with collective long-range correlations. Instead, the static pairing field $\Sigma^{(\infty)\,12}$ accounts for superfluidity effects at the Fermi surface, which are mostly decoupled from collective ADC(3) dynamics. As discussed above, we expect the dynamical pairing contributions neglected by Eq.~\eqref{eq:HybridSig} to have a low impact on ground state energies. Moreover, our experience from simulations on finite nuclei is that Gorkov-ADC(2) would not alter appreciably the breaking of particle number~\cite{Soma2014Numerical}.}

We solve Eqs.~\eqref{eq:GorkovEq} with self-energy~\eqref{eq:HybridSig} to obtain the Gorkov propagators. However, $\widetilde\Sigma^{ADC}_{\alpha\beta}(\omega)$ is computed in the Dyson formalism and requires an appropriate representation of the self-consistent propagator $\mathbf{g}(\omega)$ in Dyson format. We discuss this point in Sec.~\ref{sec: oprs}. 
By inserting Eqs.~\eqref{eq:HybridSig} and~\eqref{eq:DynSE_Dys} into~\eqref{eq:GorkovEq} and using the spectral representation of $\mathbf{g}(\omega)$, the Gorkov equation is cast into a single matrix diagonalization~\cite{Raimondi2017,Barbieri2017,BlockLanczosGF}:
\begin{widetext}
\begin{align}
    \label{eq:GkvMtxEq}
    \left(
    \begin{array}{ccc|ccc}
        T \!+ \!\Lambda \!-\! \IdentityMat \mu_q & M^\dagger & N & \Delta \\
        M & E^{>}\!+\!C \!-\! \IdentityMat \mu_q& \\
        N^\dagger & & E^{<}\!+\!D \!-\! \IdentityMat \mu_q \\ \noalign{\vskip 2pt} \hline \noalign{\vskip 3pt}
        \Delta^\dagger & & & -(T\! +\! \Lambda)^* \!+\! \IdentityMat \mu_q & -M^T & -N^* \\
        & & & -M^* & -(E^{>}\!+\!C)^* \!+\! \IdentityMat \mu_q \\
        & & & -N^T & & -(E^{<}\!+\!D)^* \!+\! \IdentityMat \mu_q
    \end{array}
    \right) \!\!\!
    \left(
    \begin{array}{c}
        \mathcal{U}^{q} \\
        \mathcal{W}^{q} \\
        \mathcal{Z}^{q} \\
        \noalign{\vskip 2pt} \hline \noalign{\vskip 3pt}
        \mathcal{V}^{q} \\
        \mathcal{R}^{q} \\
        \mathcal{S}^{q}
    \end{array}
    \right)
    \!\! = \hbar \omega_q \!\!
    \left(
    \begin{array}{c}
        \mathcal{U}^{q} \\
        \mathcal{W}^{q} \\
        \mathcal{Z}^{q} \\
        \noalign{\vskip 2pt} \hline \noalign{\vskip 3pt}
        \mathcal{V}^{q} \\
        \mathcal{R}^{q} \\
        \mathcal{S}^{q}
    \end{array}
    \right),
\end{align}
\end{widetext}
where $T$ is the kinetic energy operator and $\mu_q$ is the chemical potential associated to a quasiparticle $q$. The static mean-field matrix  $\Lambda=\Sigma^{(\infty)\,11}$ and the pairing field $\Delta=\Sigma^{(\infty)\,12}$ are spelled out in App.~\ref{App:SEinfty_diags}.

Eq.~\eqref{eq:GkvMtxEq} is an Hermitian eigenvalue problem that admits pairs of opposite eigenvalues ($\omega_q$,$-\omega_q$), as can be seen from the structure of the matrix.
Only one solution for each pair, say $\omega_q>0$, is needed to reconstruct the propagator~\eqref{eq:g_SpectRep}. Note that for a vanishing dynamic self-energy (that is, when matrices $M, N \to 0$) it reduces to the usual Hartree-Fock-Bogoliubov (HFB) equations~\cite{Soma2011}.
Similarly, if the non diagonal pairing fields are neglected, $\Delta\rightarrow0$, Eq.~\eqref{eq:GkvMtxEq} decouples into two equivalent Dyson eigenvalues problems with opposite poles $\omega_q$ and~$-\omega_q$ as solutions.
The $\mathcal{W}$, $\mathcal{Z}$, $\mathcal{R}$, $\mathcal{S}$ represent auxiliary vectors in the space of 2p1h ($\mathcal{W}$, $\mathcal{R}$) or 2h1p ($\mathcal{Z}$, $\mathcal{S}$) intermediate state configurations.

We exploit the Lanczos algorithm to efficiently solve Eq.~\eqref{eq:GkvMtxEq}.
We first project matrices $E^{>}+C$ and $E^{<}+D$ separately into smaller Krylov spaces to reduce the dimensionality of the problem, as discussed in detail in Refs.~\cite{Soma2014Numerical,Barbieri2017,BlockLanczosGF}. A few tens or hundreds of Lanczos vectors for each $\mathbf{k}_\alpha$ in the momentum mesh are typically sufficient to achieve accurate g.s.~observables and quasiparticle spectroscopy near the Fermi surface. 
Once the Krylov representation is found, Eq.~\eqref{eq:GkvMtxEq} yields all Gorkov eigenstates in a single diagonalization. Despite the increased dimensionality of the matrix~\eqref{eq:GkvMtxEq}, this method has proved faster and more stable than searching for the poles of Eq.~\eqref{eq:GorkovEq} individually~\cite{Barbieri2017,Soma2014Numerical}. 

\subsection{Optimized reference state}
\label{sec: oprs}

All elements of the Gorkov matrix in Eq.~\eqref{eq:GkvMtxEq} can be expressed as a function of the complete dressed propagator~\eqref{eq:g_SpectRep}, hence requiring an iterative solution.
State-of-the-art SCGF simulations employ the so-called \emph{sc0} approximation~\cite{Soma2014Numerical,Barbieri2017}, in which the static self-energy is always computed exactly in terms of the dressed propagators, while the contributions to the dynamic self-energy are evaluated in terms of an effective propagator, which we dub \emph{optimized reference state} (OpRS)~\cite{Barbieri2009,Barbieri2022Gorkov}. The OpRS approximates $\mathbf{g}(\omega)$ with a propagator of the same dimension and number of poles as a mean-field propagator $\mathbf{g}^{(0)}(\omega)$\footnote{In general, we speak of an approximate ``\emph{self-consistency at rank n}'' (or \emph{scn}) when the reference propagator is constrained to the first moments of the spectral function up to order \hbox{$2n+1$}. The simplest case is $n$=0 that yields an effective propagator equivalent to a mean-field reference state. The limit $n\to\infty$ (\emph{sc$\infty$}) recovers the fully dressed propagator and leads to perfect self-consistency~\cite{Barbieri2022Gorkov}. }.
The \emph{sc0} scheme is particularly useful since it keeps the pole proliferation problem~\cite{Soma2014Numerical} to a minimum while allowing to recover most of the correlation energy~\cite{Barbieri2017}. 
At the same time, the OpRS propagator is associated with a 1B potential, so that it can also be interpreted as a mean-field reference state~\cite{Barbieri2022Gorkov}. 
Because of this property, it is possible to introduce systematic and controlled corrections to compensate for the lack of full self-consistency in \emph{sc0}. These are implemented in terms of the non-skeleton diagrams (see e.g.~\cite{Raimondi2017,Barbieri2022Gorkov}) that we discuss in Sec.~\ref{sec:nsk_adc3d} and App.~\ref{app: expre non skeleton}.
Note that in each sc$n$ scheme (including \emph{sc0}) the $\mathbf{\Sigma}^{(\infty)}$ is always computed fully self-consistently through Eqs.~\eqref{eq:SigGk_static}, which is required for accuracy~\cite{Barbieri2014JPhysG}.

Our implementation of the mixed Dyson-Gorkov formalism relies on the Dyson computation for the ADC(3) dynamic self-energy, Eqs.~\eqref{eq:DynSE_Dys}. Thus, it requires reducing the dressed Gorgov propagator~\eqref{eq:g_SpectRep} to an OpRS in the Dyson (particle conserving) form of Eq.~\eqref{eq:Dys_SpectRep}.
\CB{
The standard HF propagator reads
\begin{align}
    \label{eq:g HF}
    g^{HF}_{\alpha\beta}(E) =
    \delta_{\alpha\beta} \left\{
    \frac{\delta_{\alpha\notin F}}{ E - \varepsilon^{HF}_\alpha + i \eta  }
    + 
    \frac{\delta_{\alpha\in F}}{ E - \varepsilon^{HF}_\alpha - i \eta  }
    \right\} \,, 
\end{align}
where the Kronecker deltas distinguish whether $\alpha$ is a particle ($\notin F$) or a hole ($\in F$) state.
The HF energies are given by 
\begin{align}
    \label{eq: epsilon HF}
    \varepsilon^{HF}_\alpha = \frac{\hbar^2 \mathbf{k}_{\alpha}^{2} }{2\, m} + \sum_{j \in F } v_{\alpha j, \alpha j} + \frac{1}{2} \sum_{j,k \in F } w_{\alpha jk, \alpha jk }.
\end{align}
The OpRS Green's function reads
}
\begin{align}
    \label{eq:gOpRS}
    g^{OpRS}_{\alpha\beta}(E) =
    \delta_{\alpha\beta} \left\{
    \frac{|U_\alpha|^2\delta_{\alpha\notin F}}{ E - \varepsilon^{OpRS}_\alpha + i \eta  }
    + 
    \frac{|V_\alpha|^2\delta_{\alpha\in F}}{ E - \varepsilon^{OpRS}_\alpha - i \eta  }
    \right\} \,  .
\end{align}
In Eqs.~\eqref{eq:g HF} and~\eqref{eq:gOpRS}, we have specialized to homogeneous matter and taken into account translational symmetry that makes the propagator diagonal in the momentum basis.
Therefore, each s.p.~momentum state is associated with a unique mean-field pole, both in the HF and OpRS case.
As a consequence, the corresponding overlap amplitudes are either one or vanishing:
\begin{align}
    \label{eq:OpRS_UV}
    U_\alpha ={}& \delta_{\alpha\notin F} \,, \nonumber \\
    V_\alpha ={}& \delta_{\alpha\in F} \,.
\end{align}

To constrain the OpRS poles $\varepsilon^{OpRS}_\alpha$ in Eq.~\eqref{eq:gOpRS}, we follow Ref.~\cite{Barbieri2022Gorkov} and define direct ($p\ge0$) and inverse ($p\le0$) moments for the normal component of the Gorkov propagator of Eq.~\eqref{eq:g_SpectRep} as
\begin{align}
    \label{eq:Qmoments}
    Q_{\alpha}^{(p)11}
    = \sum_q \left[
    (\hbar\omega_q)^{p} 
    \abs{ \mathcal{U}_\alpha^{q} }^2 
    ~+~ (-\hbar\omega_q)^{p} \abs{ \mathcal{V}_\alpha^{q} }^2
    \right] \,,
\end{align}
which applies to a propagator diagonal in the s.p.~indices.
A similar expression holds for $g^{OpRS}(E)$ once cast in its Gorkov form using Eq.~\eqref{eq:g_to_Dys_limit}.
Imposing the equivalence of $p=0$ moments is trivial and leads to the normalization conditions~\eqref{eq:OpRS_UV}. For $p=\pm1$ we obtain
\begin{align}
    \label{eq:e_oprs}
    (\varepsilon^{OpRS}_\alpha - \mu_\alpha)^{p} &= Q_{\alpha}^{(p)11} \\
    &= \sum_q (\hbar\omega_q)^{p} 
    \left(
    \abs{ \mathcal{U}_\alpha^{q} }^2 - \abs{ \mathcal{V}_\alpha^{q} }^2
    \right) \nonumber
\end{align}
which yields two different prescriptions for generating the OpRS frequencies: we denote the case $p=+1$ as Centroid (Cen) and $p=-1$ as Inverse (Inv) moments of the s.p. strength.

Eq.~\eqref{eq:e_oprs} also provides a natural way to determine whether a given s.p.~state $\alpha$ is a particle or hole with respect to the optimized reference. Specifically,
\begin{align}
 \label{eq:maxUV}
 \alpha \;\hbox{is a}\;
\begin{cases}
\hbox{particle state} & \hbox{if\, } \varepsilon^{OpRS}_\alpha > \mu_\alpha \,, \\
\hbox{holes state} & \hbox{if\, } \varepsilon^{OpRS}_\alpha < \mu_\alpha \,.
 \end{cases}
\end{align}
According to this prescription, $\alpha$ is a particle if \hbox{$Q_{\alpha}^{(p)11} > 0$}, i.e., the contributions from $\mathcal{U}$ amplitudes dominate over that from $\mathcal{V}$ amplitudes in the moments of the target propagator, Eq.~\eqref{eq:e_oprs}.
This provides a  Dyson mean-field reference propagator that preserves the chemical potential obtained by solving the Gorkov equations.
Accordingly, we refer to prescription~\eqref{eq:maxUV} as ``$\mu$''.

However, this recipe does not guarantee that the number of occupied states is conserved, i.e., the number of states in the OpRS with energy below the chemical potential might differ from $A$.
We therefore consider a second prescription, which we dub ``$k_F$'', where we assign OpRS energies and occupations as follows:
\begin{align}
 \label{eq:ph prescription}
\begin{cases}
    \epsilon_{\alpha}^{OpRS} = 
    \mu_{\alpha} + \bar{\omega}_{\alpha}  
    \quad \hbox{and} \quad \alpha \notin F, 
    &  \hbox{if\, } |\mathbf{k}_\alpha|> k_F   
    \\
    \epsilon_{\alpha}^{OpRS} = \mu_{\alpha} - \bar{\omega}_{\alpha} 
    \quad \hbox{and} \quad \alpha \in F, 
    & \hbox{if\, } |\mathbf{k}_\alpha|<k_F
 \, ,
\end{cases}
\end{align}
where $\bar{\omega}_{\alpha} = \abs{ \left( Q_{\alpha}^{(p)11} \right)^{p} }$. 
Hence, states that are occupied (empty) in the standard HF reference retain their hole (particle) character in the OpRS.
The prescription of Eq.~\eqref{eq:ph prescription} ensures that the Fermi momentum is kept in the OpRS propagators at each iteration. 
The four OpRS schemes that result from combining the Cen/Inv and \oprsPH/\oprsMaxuv~ prescriptions are studied numerically in Sec.~\ref{sec: oprs prescriptions}.

\subsection{Non-skeleton and coupled-cluster corrections to self-consistent ADC(3)}
\label{sec:nsk_adc3d}

Exact self-consistency implies expanding the self-energy $\mathbf{\Sigma}^\star(\omega)$ uniquely in terms of the dressed propagator $\mathbf{g}(\omega)$ and skeleton diagrams~\cite{Raimondi2017,Barbieri2017}\,%
\footnote{Skeletons are defined as those diagrams that do not contain any portion that can be disconnected by cutting two different fermion lines, i.e. they do not contain any self-energy insertion~\cite{Carbone2013,Raimondi2017}. When the self-energy is expanded in terms of fully dressed propagators, only skeletons must be explicitly accounted for.}. 
In practice, it allows to maintain coherence between the many-particle--many-hole energies of the ISCs and the correlated one-nucleon addition and removal spectrum. These can be substantially shifted with respect to the HF case for strongly correlated systems and cause instabilities if they end up being too much displaced with respect to each other.
While being a partially self-consistent approximation, the \emph{sc0} prescription maintains this important feature through the use of the optimized effective s.p. energies, the poles of Eq.~\eqref{eq:gOpRS}, in the evaluation of matrices $M$, $N$ and $E^\gtrless$ in Eq.~\eqref{eq:DynSE_Dys}. On the other hand,  $\mathbf{g}^{OpRS}(\omega)$ is also associated to an external effective mean-field $\CB{u_\alpha^{OpRS}}$ that can be found identifying $\varepsilon^{OpRS}_\alpha = \hbar^2  \mathbf{k}_{\alpha}^{2} /(2 m) + \CB{u_\alpha^{OpRS}}$ in Eqs.~\eqref{eq:H0_H1}.
This allows to introduce systematic corrections to the partially self-consistent \emph{sc0} scheme by reinserting the relevant non-skeleton diagrams~\cite{Raimondi2017,Barbieri2022Gorkov}.

\begin{figure}
    \centering
    \includegraphics[width=0.35\columnwidth]{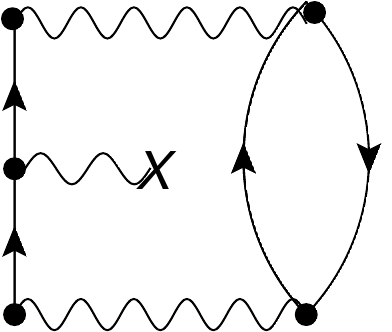}
    \includegraphics[width=0.5\columnwidth]{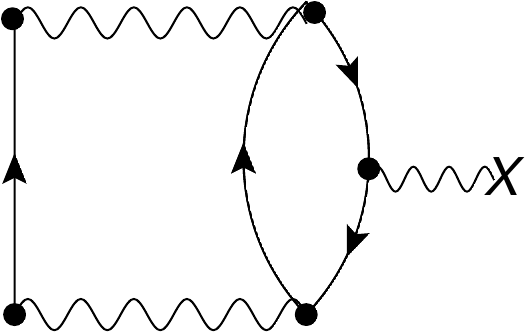}
    \caption{Third-order non-skeleton diagrams that contribute to the dynamical self-energy.
    These complement the ADC(3) diagrams of Fig.~\ref{fig: adc3 diagrams}, and should be included whenever calculations are performed with non-self-consistent propagators.
    Wiggly lines denote the effective 2B interaction \eqref{eq: eff interaction 2B} and the effective 1B interaction $\widetilde{U}^{(1)}$ \eqref{eq:Utilde1}.
    Fermion lines refer to the \CB{OpRS propagator $g^{OpRS}$.}
    }
    \label{fig:non-skeleton}
\end{figure}

The non-skeleton contributions to $\widetilde\Sigma(\omega)$ in ADC(3) are depicted in Fig.~\ref{fig:non-skeleton}. Their working equations have been derived in Ref.~\cite{Raimondi2017} and are summarized in App.~\ref{app: expre non skeleton} for translationally invariant nucleonic matter. 
Following Ref.~\cite{Raimondi2017}, we compute the diagrams in Fig.~\ref{fig:non-skeleton} in terms of the first-order contribution to the effective one-body interaction,
\begin{align}
    \widetilde{U}^{(1)} = - U^{OpRS} +  \Sigma^{(\infty)} [g^{OpRS}],
\end{align}
where $\Sigma^{(\infty)} [g^{OpRS}]$ is the static Dyson self-energy (evaluated from $g^{OpRS}(\omega)$ only) and $U^{OpRS}$ is the auxiliary potential. Its matrix elements read:
\begin{align}
    \label{eq:Utilde1}
    \widetilde{u}^{(1)}_{\alpha}
    & = \frac{\hbar^2 \mathbf{k}_{\alpha}^{2} }{2\, m} - \varepsilon^{OpRS}_{\alpha}  \\
    & + \sum_{\gamma\delta} {v}_{\alpha\gamma, \alpha\delta} \rho^{OpRS}_{\delta\gamma}
    + \frac{1}{2} \sum_{\gamma\delta\epsilon\zeta} {w}_{\alpha\gamma\delta, \alpha\epsilon\zeta} \rho^{OpRS}_{\epsilon\gamma} \rho^{OpRS}_{\zeta\delta},
    \nonumber
\end{align}
where $\rho^{OpRS}$ is the density matrix~\eqref{eq: norm density def} associated with the propagator~\eqref{eq:gOpRS}. 
We stress that whenever the HF mean-field is used for the reference potential $U$, Eq.~\eqref{eq:Utilde1} vanishes and these non-skeleton corrections cancel out~\cite{Barbieri2017,Raimondi2017}. This is not the case for a OpRS reference state and one finds
\begin{align}
  \label{eq:u1tilde_practical}
  \widetilde{u}^{(1)}_{\alpha} =
  \epsilon^{HF}_{\alpha} - \epsilon^{OpRS}_{\alpha},
\end{align}
\CB{with $\epsilon^{HF}_{\alpha}$ defined in Eq.~\eqref{eq: epsilon HF}. }
It can be seen that Eq.~\eqref{eq:u1tilde_practical} corresponds, in first approximation, to reverting the $\epsilon^{OpRS}_{\alpha}$ back to the corresponding HF s.p. energies when calculating the diagonal interaction matrices $E^{>}$, $E^{<}$ of ADC(3). However, the OpRS spectrum is still used for the interaction vertices $M$ and $N$ (see Ref.~\cite{Raimondi2017} and App.~\ref{app: expre non skeleton}).

\begin{figure*}
    \centering
    \includegraphics[width=2.\columnwidth]{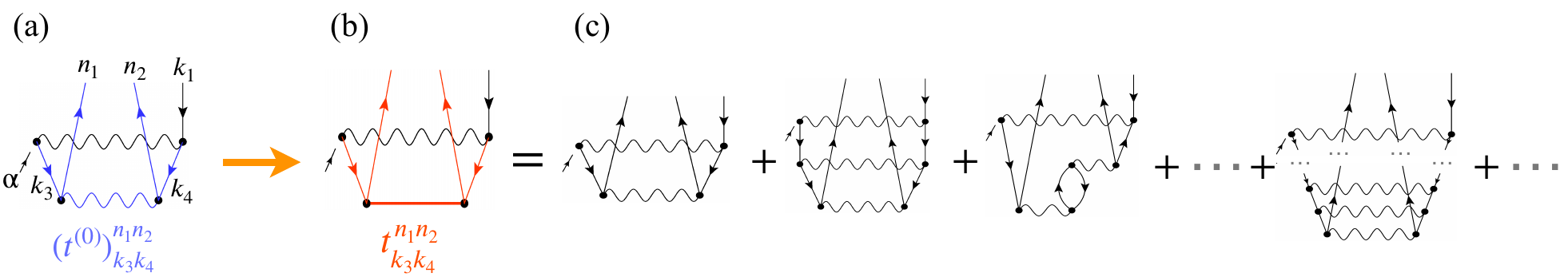}
    \caption{Diagrammatic representation of the ADC(3)-D truncation scheme.
    Diagram (a) shows a representative contribution to the $M_{r \alpha}$ 2p1h vertex in standard ADC(3) that incorporates perturbative 2p2h excitations through the amplitude $t^{(0)}$. See e.g.~Ref.~\cite{Raimondi2017}, Fig. 7(a).
    All fermion lines are explicitly labeled.
    The substitution in diagram (b) of $(t^{(0)})^{n_1\,n_2}_{k_3\,k_4}$ with the converged 2p2h amplitude $t^{n_1\,n_2}_{k_3\,k_4}$ from the solution of the CCD equations, represented as a thick horizontal line, amounts to incorporating additional infinite classes of diagrams within the ADC self-energy, some of which are shown in (c).
    Diagrams contributing to the CCD amplitudes can be found e.g.~in Refs.~\cite{ShavittBartlett,Marino2024}.
    }
    \label{fig:adc3d}
\end{figure*}

In the following, we also implement the so-called ADC(3)-D truncation for the nuclear self-energy~\cite{Barbieri2017,Soma2013,Marino2024}.
ADC(3)-D is a refinement of the standard ADC(3) based on combining elements of the Goldstone (that is, time-ordered) diagrammatic expansion that are common to both Green's functions and CC~\cite{ShavittBartlett,Hagen2014Review} theories. 
Its content is depicted diagrammatically in Fig.~\ref{fig:adc3d} for a particular contribution to the 2p1h $M$ matrix that, for standard ADC(3), is at second order in the interaction $V$. The tensor
\begin{align}
    \label{eq: amplitude mbpt}
    (t^{(0)})^{n_1\,n_2}_{k_3\,k_4} = 
    \sum_{\alpha\beta\gamma\delta}
    \frac{
     (\mathcal{U}^{n_1}_\alpha \mathcal{U}^{n_2}_\beta)^*{v}_{\alpha\beta,\gamma\delta} (\mathcal{V}^{k_3}_\gamma \mathcal{V}^{k_4}_\delta)^*}
    { \varepsilon^-_{k_3}+\varepsilon^-_{k_4}-\varepsilon^+_{n_1}-\varepsilon^+_{n_2}},
\end{align}
is responsible for 2p2h excitations in perturbative expansions, and it also contributes to CC theory as a lowest-order approximation. 
\CB{By directly substituting $t^{(0)}$ with the $t$ amplitudes from a CC calculation, one can incorporate additional infinite classes of diagrams in the ADC self-energy~\cite{Barbieri2017,Hodecker2019}, while preserving the Lehmann representation of the ADC self-energy by construction.
}

\CB{
In our approach, the CC amplitudes are computed once by solving the CC equations based on the HF reference at the doubles truncation level (CCD)---hence the naming ``ADC(3)-D''--- by using a custom implementation of the method described in Refs.~\cite{Hagen2014,LietzCompNucl}}
The CCD amplitudes introduce corrections to the ADC coupling vertices (namely, the $M$ and $N$ matrices) that pertain to higher ADC($n\geq4$) truncation schemes. 
However, the scaling with the model-space dimension remains the same as for ADC(3)~\cite{Marino2024}. In contrast, full ADC(4) would require including 3p2h configurations~\cite{SchirmerAdc4,Schirmer2018}, leading to a large increase in the computational load. 
In practice, ADC(3)-D improves the accuracy of ADC when correlations are strong~\cite{Marino2024}, while also making the method more stable when the system is close to degeneracy, mitigating the issues associated with potentially vanishing denominators~\cite{Barbieri2017}.
To the best of our knowledge, Ref.~\cite{Soma2013} reports the earliest application of the ADC(3)-D truncation.
Further connections between Green's functions and CC theories are discussed in Refs.~\cite{Coveney2025GFvsCC,coveney2025nonhermitiangreensfunctiontheory}.

\CB{
As a final comment, notice that, in principle, the coupled-cluster and non-skeleton corrections can be combined seamlessly.
Also, one may compute the CC amplitudes at each OpRS iteration using the updated energy spectrum, albeit at the cost of a much higher computational load.
We leave these extensions of ADC for future developments.
}

\section{Validation}
\label{sec:validation}
We implemented our hybrid Gorkov-Dyson scheme for PNM and SNM using two- plus three-body interactions derived from $\chi$EFT~\cite{Epelbaum2009,MACHLEIDT2024104117,Epelbaum2024}. Although no systematic study of uncertainties is attempted, we employ three different Hamiltonians at next-to-next-to-leading order (NNLO) in the chiral expansion as a way to gauge the performance of our method as the cutoff of the chiral potential is varied, as well as the spread of observables due to the choice of the $\chi$EFT model.
In particular, we employ the NNLO$_{ \rm{sat} }\,(450)$ potential from Ref.~\cite{NNLOsat}, and the $\Delta$NNLO$_{\rm{go}}\,(394)$ and $\Delta$NNLO$_{\rm{go}}\,(450)$ interactions from Ref.~\cite{DeltaGo2020}, where numbers in parentheses indicate the momentum cutoff in $\rm{MeV/c}$.

When not stated otherwise, PBCs are used with \hbox{$N=66$} neutrons in PNM and $A=132$ nucleons in SNM, as it is standard in most works~\cite{Hagen2014,DeltaGo2020,PbAbInitio,Marino2024}.
Typically, at each iterative step of the SCGF \emph{sc0} scheme, we
perform about 10-20 diagonalizations of Eq.~\eqref{eq:GkvMtxEq} to tune the chemical potential $\mu$ and the static self-energy $\mathbf{\Sigma}^{(\infty)}$. After this, the OpRS propagator~\eqref{eq:gOpRS} is updated and the new ADC(3) dynamical self-energy computed for each different $\mathbf{k}_\alpha$ value. We demand convergence of the ground state energy to within 10 keV/A between successive macro iterations, which is typically reached within 10 OpRS cycles. We verified that 100 Lanczos iterations for each Krylov projection of matrices $E^{>}+C$ and $E^{<}+D$ are sufficient to guarantee highly accurate results. 
For SNM with PBCs, 100-200 CPUh often suffice for a complete simulation~\cite{Marino2024}, since different combinations of momentum states $\mathbf{n}$ (see Eq.~\eqref{eq:momgrid}) are equivalent and need to be computed only once~\cite{Barbieri2017,MarinoPhdThesis} (see App.~\ref{app:implementation}). For TABCs, the lack of degeneracy among basis states imposes simulating each different $\mathbf{k}_\alpha$ value separately. This may result in an increase of the computing time by an order of magnitude or more.

The convergence with respect to model-space dimension, the choice of the OpRS prescription and the many-body truncation are discussed in the following subsections. Our ADC-SCGF method has also been further validated by comparison against the CC and MBPT(3) methods in Ref.~\cite{Marino2024}.

\subsection{Model-space convergence}
\label{sec: model space convergence}

\begin{figure}
    \centering
    \includegraphics[width=\columnwidth]{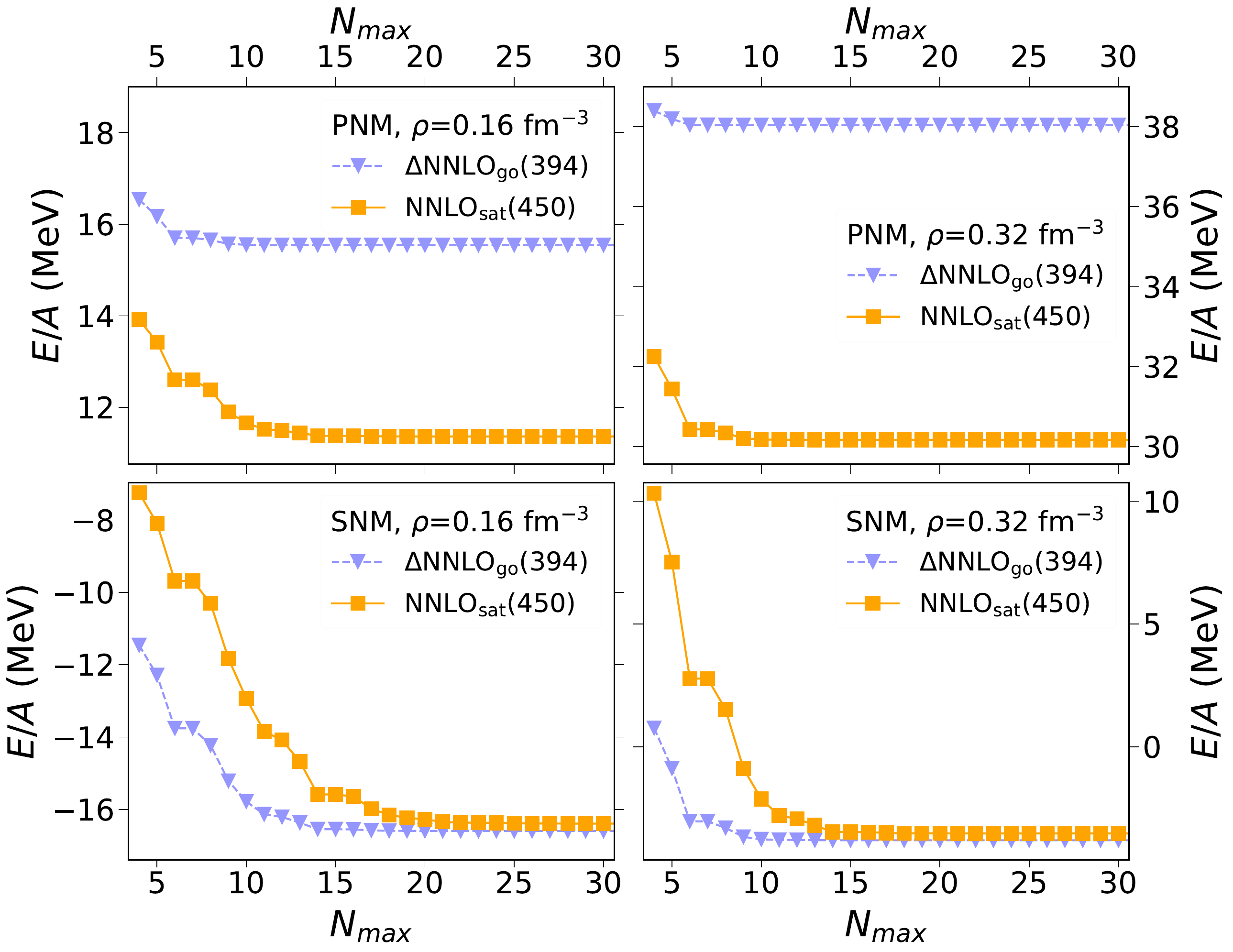}
    \caption{
    MBPT(2) energies per nucleon as a function of the cutoff $N_{max}$ of the s.p. basis. 
    Calculations are performed with the NNLO$_{ \rm{sat} }\,(450)$ (squares) and $\Delta$NNLO$_{\rm{go}}\,(394)$ (circles) interactions in both PNM (upper panels) and SNM (lower panels), at densities $\rho=$ 0.16 fm$^{-3}$ (left) and 0.32 fm$^{-3}$ (right).
    Lines are a guide to the eye.
    }
    \label{fig: mbpt convergence}
\end{figure}
Basis expansion methods are sensitive to the dimension of the employed model space, which in our approach is controlled by the momentum space cutoff $N_{max}$ on the s.p.~basis states.
As a preliminary step, we investigate the convergence of the energies per particle as a function of $N_{max}$ for MBPT(2).
For simplicity, we include only the 2B MBPT(2) diagram 
\begin{align}
    \label{eq: mbpt2 NN}
    \Delta E^{(2)} =
    \frac{1}{4}
    \sum_{n_1 n_2 k_1 k_2}(t^{(0)})^{n_1 n_2}_{k_1 k_2} \,v_{k_1 k_2, n_1 n_2},
\end{align}
and neglect the explicit 3B diagram~\cite{Marino2024}, as it does not influence the convergence pattern.
Binding energies are shown in Fig.~\ref{fig: mbpt convergence} for both PNM and SNM at densities of $\rho=0.16\,\rm{fm}^{-3}$ and $\rho=0.32\,\rm{fm}^{-3}$.
We focus on the $\Delta$NNLO$_{\rm{go}}\,(394)$ and NNLO$_{ \rm{sat} }$ potentials that are respectively the softest and hardest of the three interactions used in this work.
As expected, binding energies decrease monotonically with $N_{max}$ although some small plateaus are sometimes observed. 
The convergence is more rapid in PNM, where binding energies at $N_{max} = 20$ are already indistinguishable from those obtained for $N_{max} = 30$, with discrepancies $<1$ eV. Correlations are stronger in SNM so that larger model spaces are required and the convergence pattern is also influenced significantly by the choice of the potential. In particular, with the softer $\Delta$NNLO$_{\rm{go}}\,(394)$ force the model space truncation error becomes almost negligible (within a few eV) already at $N_{max} = 20$. In contrast, the same quality is reached only for $N_{max}= 25$ in the case of the harder NNLO$_{ \rm{sat} }\,(450)$ model.

Convergence with respect to $N_{max}$ is faster for dense matter at $\rho=0.32\,\rm{fm}^{-3}$ than for $\rho=0.16\,\rm{fm}^{-3}$, for both PNM and SMN. 
This behavior can be explained as a consequence of the procedure we use, namely, employing a fixed particle number in a finite box.
The number of nucleons $A$ is kept fixed, and we always use $A/g=33$ to minimise finite-size effects~\cite{LietzCompNucl,Marino2023}.
Therefore, the box size $L$ scales proportionally to $\rho^{-\frac{1}{3}}$, decreasing as the density increases.
On the contrary, the lattice spacing $\Delta k = 2\pi/L$ and the maximum momentum in the s.p.~basis,
\begin{align}
    \label{eq: kmax of rho}
    k_{max} = \frac{2\pi}{L} \sqrt{ N_{max} } = 2\pi \left(  \frac{\rho}{A} \right)^{1/3} \sqrt{ N_{max} },
\end{align}
both scale proportionally to $\rho^{\frac{1}{3}}$ for a given nucleon number. 
Therefore, at a given $N_{max}$, the highest momentum reached in the s.p.~basis is actually larger for larger densities.

The $N_{max}$ convergence pattern can now be understood. 
First, s.p.~levels are progressively farther from each other (in both momentum and s.p.~energy) as $\rho$ increases.
Thus, the energy denominators entering the MBPT diagram, as seen in Eq.~\eqref{eq: mbpt2 NN}, are larger in magnitude at higher densities, and the contribution due to particle states carrying high momentum tends to be suppressed more rapidly.
In addition, at a given $N_{max}$, for larger density the highest momentum is actually larger.

\begin{figure}
    \centering
    \includegraphics[width=0.9\columnwidth]{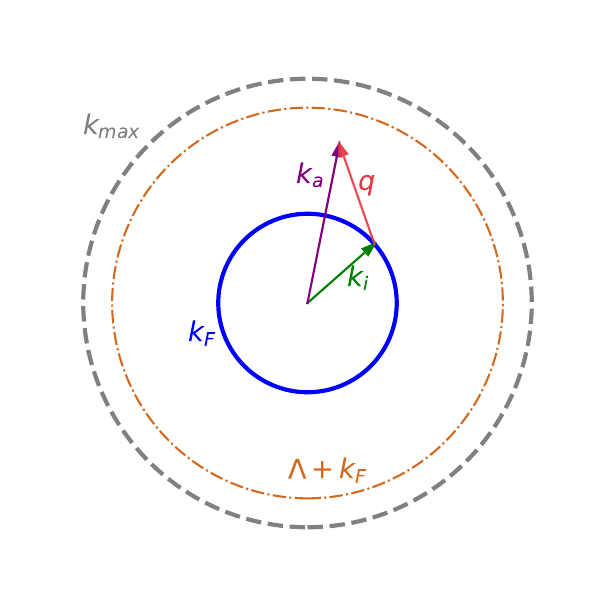}
    \caption{
    Schematic representation of the momentum space.
    The Fermi sphere is represented as a circle of radius $k_F$.
    The model space is defined by the region within the dashed circle of radius $k_{max}$.
    A representative `hole' state lying on the Fermi surface with momentum $\mathbf{k}_{i}$ is shown.
    The red arrow represents an excitation process in which $\bf{k}_{i}$ is scattered into a particle state $\bf{k}_{a}$ by the 2B interaction.
    We indicate the momentum transfer $\bf{q} = \bf{k}_{a} - \bf{k}_{i}$ with a red arrow.
    }
    \label{fig:fermi_sphere}
\end{figure}

It should also be noted that model-space convergence can be achieved if the basis is extended enough to resolve the momentum cutoff $\Lambda$ of the interaction, as noticed in Ref.~\cite{Hagen2010}.
This is depicted schematically in Fig.~\ref{fig:fermi_sphere} (see e.g.~Refs.~\cite{Masios2024,Shepherd2012} for related discussions).
Interaction matrix elements relevant for MBPT(2) involve the scattering (via the 2B interaction) of a hole state (with momentum $\mathbf{k}_{i}$) into a particle state ($\mathbf{k}_{a}$) outside the Fermi sphere, shown as a blue circle of radius $k_F$.
The corresponding transferred momentum~\cite{MACHLEIDT2024104117} (red arrow) is labelled as $\mathbf{q} = \mathbf{k}_{a} - \mathbf{k}_{i}$.
Chiral potentials are suppressed for momentum transfers $\abs{\mathbf{q}} > \Lambda$.
Thus, particle states contributing significantly to the perturbative correction should satisfy $\abs{ \mathbf{k}_a } < \Lambda + k_F$ and lie within the dashed-dotted circle.
Therefore, fully converged calculations can be obtained provided that the model space can take into account all the scattering processes described above, implying that the maximum momentum in the model space, $k_{max}$, should satisfy the condition~\cite{McilroyChristopher2020Sgfs}
\begin{align}
    \label{eq:inequality kmax}
    k_{max} > k_F + \Lambda.
\end{align}
But, for a fixed value of $A$, Eq.~\eqref{eq: kmax of rho} implies that the model space extends to larger $k_{max}$ when the density is increased~\cite{McilroyChristopher2020Sgfs}.
Thus, model-space convergence can be achieved for a smaller $N_{max}$ at larger densities.

Notice that at the same time the lattice spacing is higher at larger densities, thus potentially implying larger FS effects. This should be compensated in principle by increasing the number of particles~\cite{Lu2020}.
However, the convergence to the TL as a function of the nucleon number has rarely been discussed in \textit{ab initio} nuclear theory (see e.g.~\cite{Ismail,Lu2020,Gezerlis2017}), given the high computational cost of these simulations. TABCs are an effective alternative way of controlling FS errors (Sec.~\ref{sec: fs effects}).

\begin{figure}[t]
    \centering
    \includegraphics[width=\columnwidth]{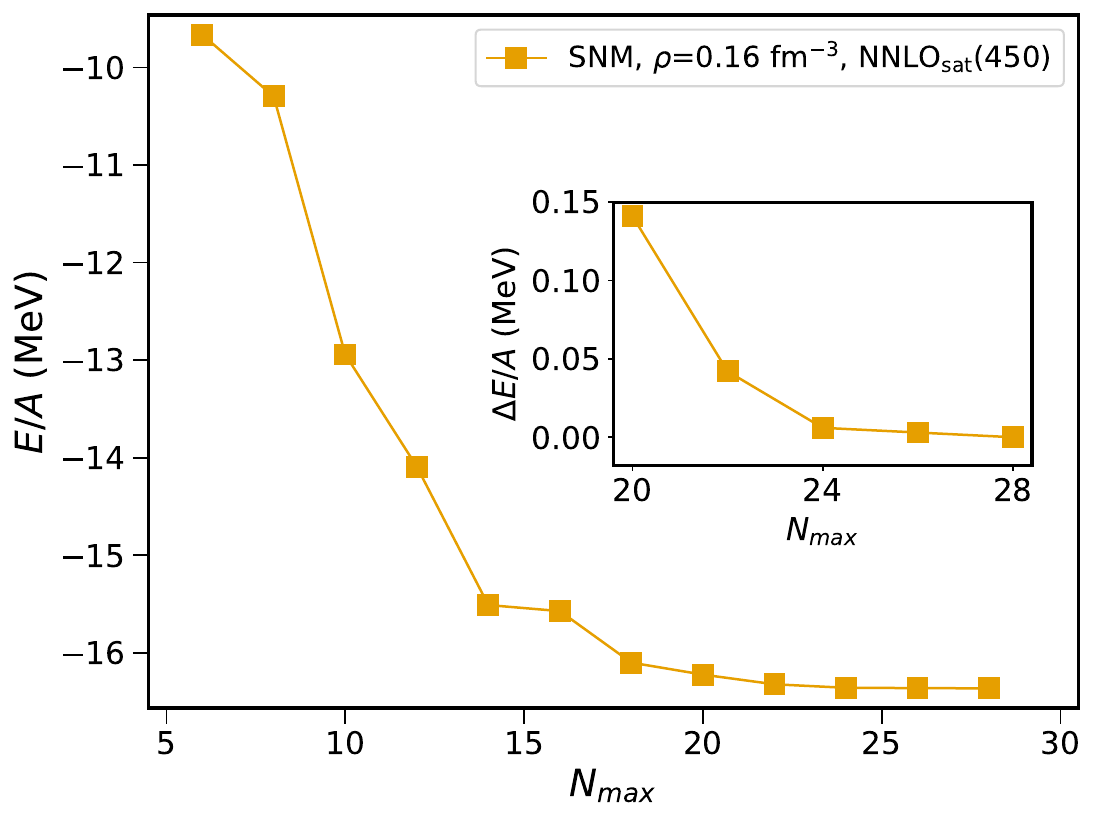}
    \caption{Energy per nucleon as a function of $N_{max}$ for SNM, $A=132$ and density $\rho=$ 0.16 fm$^{-3}$. Calculations are performed at the level of Gorkov-ADC(3) with the Cen-\oprsPH~prescription (see text). Inset: difference between the energy per particle at a given cutoff and the energy for $N_{max} =28$.
    The NNLO$_{ \rm{sat} }\,(450)$ interaction is employed. 
    }
    \label{fig:Convergence_NNLOsat_SNM_rho0.16_Gorkov_Centroid}
\end{figure}

We further test the model space convergence for an actual ADC(3) simulation, using the Cen-\oprsPH~prescription for generating the OpRS. Fig.~\ref{fig:Convergence_NNLOsat_SNM_rho0.16_Gorkov_Centroid} demonstrates this trend for SNM at $\rho=0.16$ fm$^{-3}$ and with NNLO$_{ \rm{sat} }$, which is the case that manifests the slowest convergence pattern in MBPT(2). Up to cutoffs of $N_{max}=28$ the pattern is completely analogous to the one observed in Fig.~\ref{fig: mbpt convergence} and energies for $N_{max} =24$ are already within $\approx 20\,\rm{keV}/A$ from $N_{max} =28$ (see the inset).
We note that the computing time for the present ADC(3) simulations is always manageable. Rather, the main limiting factor in accessing very larger model spaces is the memory required to store the matrix elements of the interaction and the Dyson matrix, which can increase quickly with $N_{max}$. 

Based on the above finding, all subsequent calculations reported in this work were performed with $N_{max} =25$, as it guarantees that uncertainties related to the model space truncation are substantially negligible for our purposes.
Note that, at very low densities or with hard interactions, larger model spaces might become necessary.

\subsection{OpRS prescriptions}
\label{sec: oprs prescriptions}

\begin{figure}[t]
    \centering
    \includegraphics[width=\columnwidth]{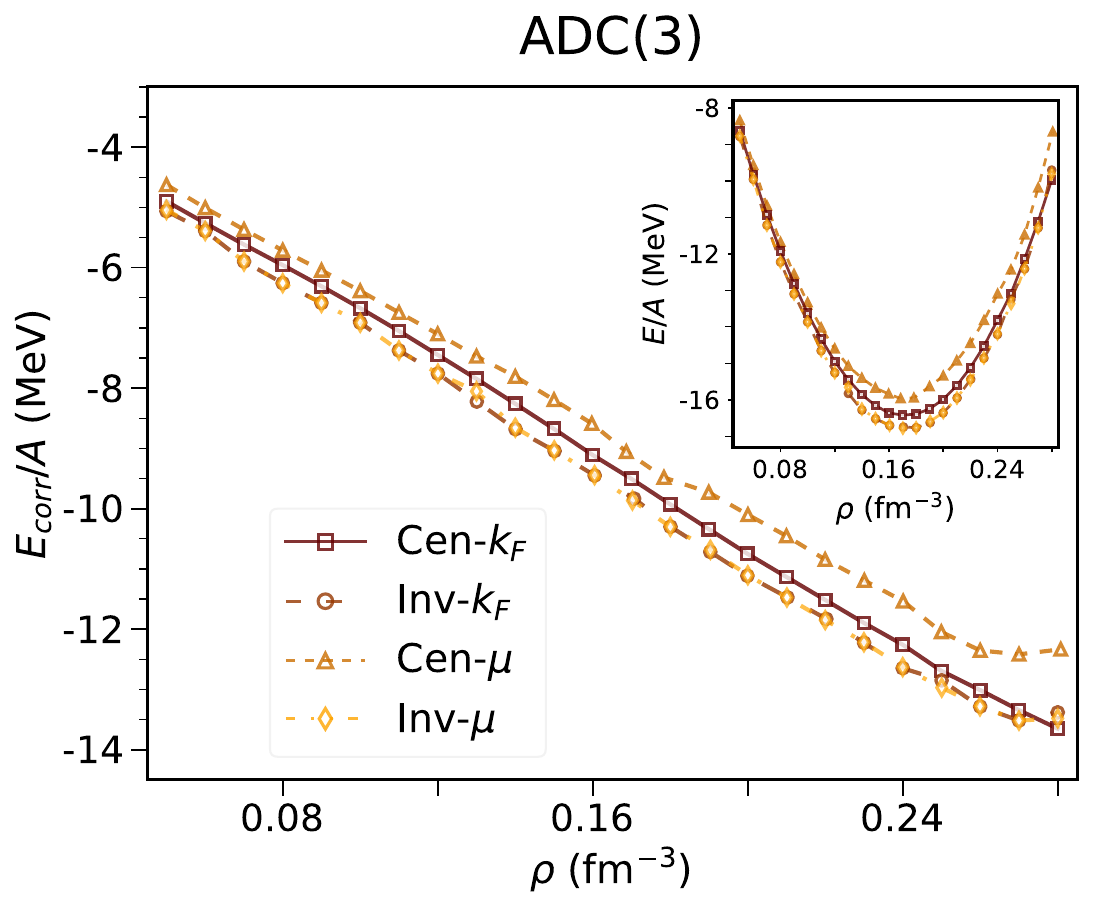}
    \includegraphics[width=\columnwidth]{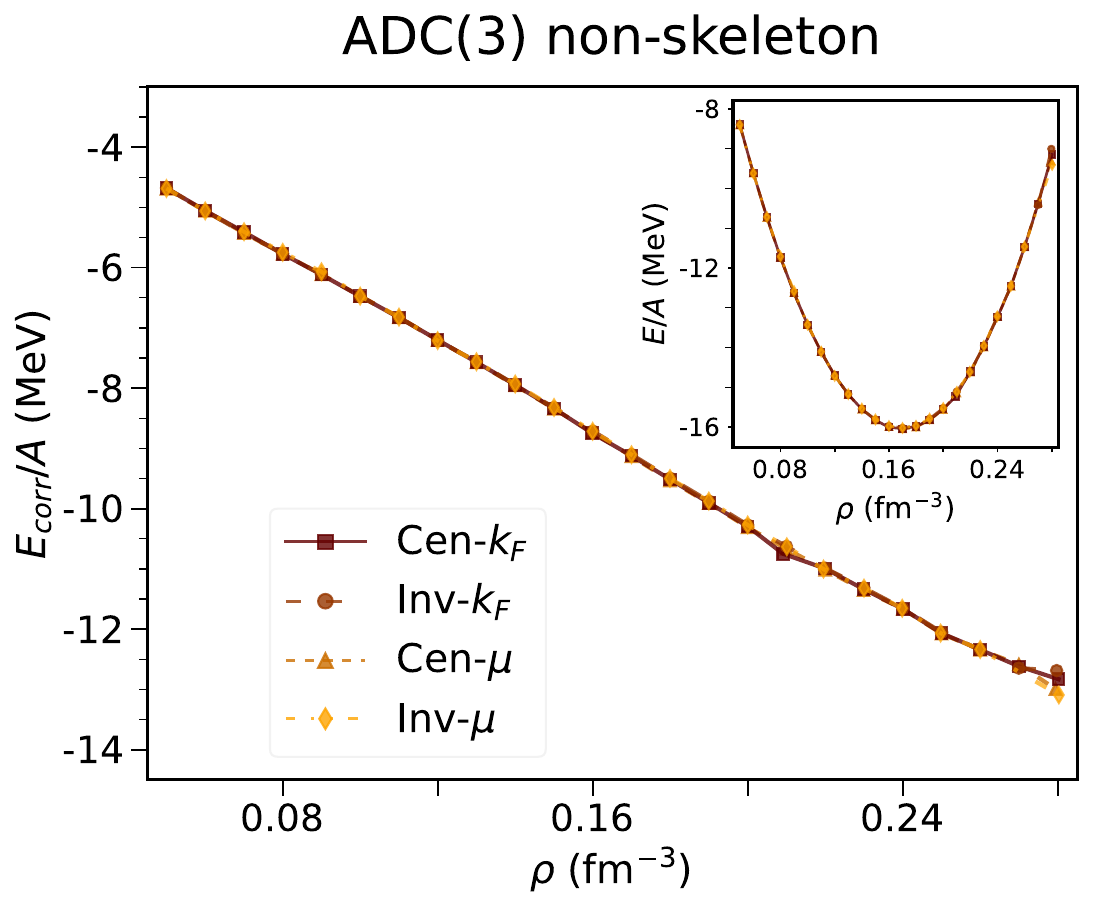}
    \caption{Comparison among the OpRS prescriptions discussed in Sec.~\ref{sec: oprs}
    The main panels show the correlation energy per particle for SNM with the NNLO$_{ \rm{sat} }\,(450)$ interaction. The dynamical self-energy is computed with the standard ADC(3) scheme (top panel) and with the scheme that includes the additional non-skeleton diagrams (bottom panel).
    The corresponding EOS are shown in the insets.
    Each OpRS prescription is labeled by a different marker (see legend), and lines are a guide to the eye.
    }
    \label{fig: all oprs}
\end{figure}

Figure~\ref{fig: all oprs} displays results from using the different OpRS prescriptions from Sec.~\ref{sec: oprs}. 
Calculations are performed in SNM in the standard ADC(3) approximation for the self-energy (top) and ADC(3) with the inclusion of non-skeleton diagrams (bottom).
Correlation energies per particle as a function of the density are reported in the main panels, while energies per particle are shown in the insets. 
In ADC(3), OpRS variants differ slightly by a relatively small amount of a few hundred keV/A at most.
Cen-\oprsMaxuv~ deviates the most from the other three OpRS schemes, especially as the density increases.
The numerical convergence turns out to be the most difficult when this prescription is used, and a much larger number of iterations is needed to converge the OpRS cycle.
We have noticed that this is related to the Cen-\oprsMaxuv~scheme often involving a significant rearrangement in the magnitude and ordering of s.p.~energy levels from one OpRS iteration to another. 
In contrast, calculations with the other three recipes for the OpRS converge smoothly in about 10 iterations or less, and the resulting predictions for the energies remain close to each other over the whole range of densities considered.

The bottom panel of Fig.~\ref{fig: all oprs} shows that by including non-skeleton diagrams in the self-energy all four prescription collapse onto the same curve.
Indeed, non-skeleton contributions should help to reduce the dependence of partially-self-consistent calculations on the choice of the reference state~\cite{Raimondi2017}.
In practice, the ADC(3) truncation of non-skeleton terms has the effect of reverting the s.p.~energies appearing in the $C$ and $D$ matrices to their HF values, as discussed in Sec.~\ref{app: expre non skeleton} and Ref.~\cite{Raimondi2017}. Hence, the differences among possible OpRS prescriptions are reduced by constructions.
Note that the OpRS energies continue to affect the coupling matrices $M$ and $N$, albeit in this case, with negligible contributions to the total energies.

The satisfactory agreement between the OpRS prescriptions testifies to the robustness of our method and its mild sensitivity to implementation details.
In principle, the ``Cen'' prescription may be considered superior since it preserves moments of order $p=1$, and hence the Koltun sum rule, exactly~\cite{Barbieri2022Gorkov}. The ``Inv'' choice is also satisfactory and it has been used routinely for finite nuclei in Dyson-SCGF because it ensures a minimum number of OpRS poles, always in the same number as for the HF solutions~\cite{Barbieri2009,Soma2020Chiral}. However, it can lead to instabilities in cases of half-occupation, of a subshell, as in the case of pairing.
For infinite matter, the combination of translational invariance and of the ``\oprsPH'' or ``\oprsMaxuv'' choices for separating `particle' and `holes' effective poles ensures that also ``Cen'' remains equivalent to a true mean field. Separating poles based on their momentum (\oprsPH) is also more natural for a mean field in momentum space and would not be affected by small fluctuations in occupation near the Fermi surface--which can happen when combining Eqs.~\eqref{eq:e_oprs} and~\eqref
{eq:maxUV}.
In our simulations, we find that Cen-\oprsPH is computationally very stable and that its correlation energy (top panel of Fig.~\ref{fig: all oprs}) is closest to the non-skeleton curves (bottom panel). 
The Cen-\oprsPH ~spectrum also has a clear physical interpretation, as it implies using the centroid s.p.~energies of the dressed spectral functions as the optimized s.p.~spectrum. 
All subsequent calculations will be performed with the Cen-\oprsPH~recipe (see also~\cite{Marino2024})
\footnote{In Ref.~\cite{Marino2024}, the same prescription was denoted as ``Cen-PH''.}.

\subsection{Comparison of many-body truncations}
\label{sec: compare truncations}

\begin{figure}[t]
    \centering
    \includegraphics[width=\columnwidth]{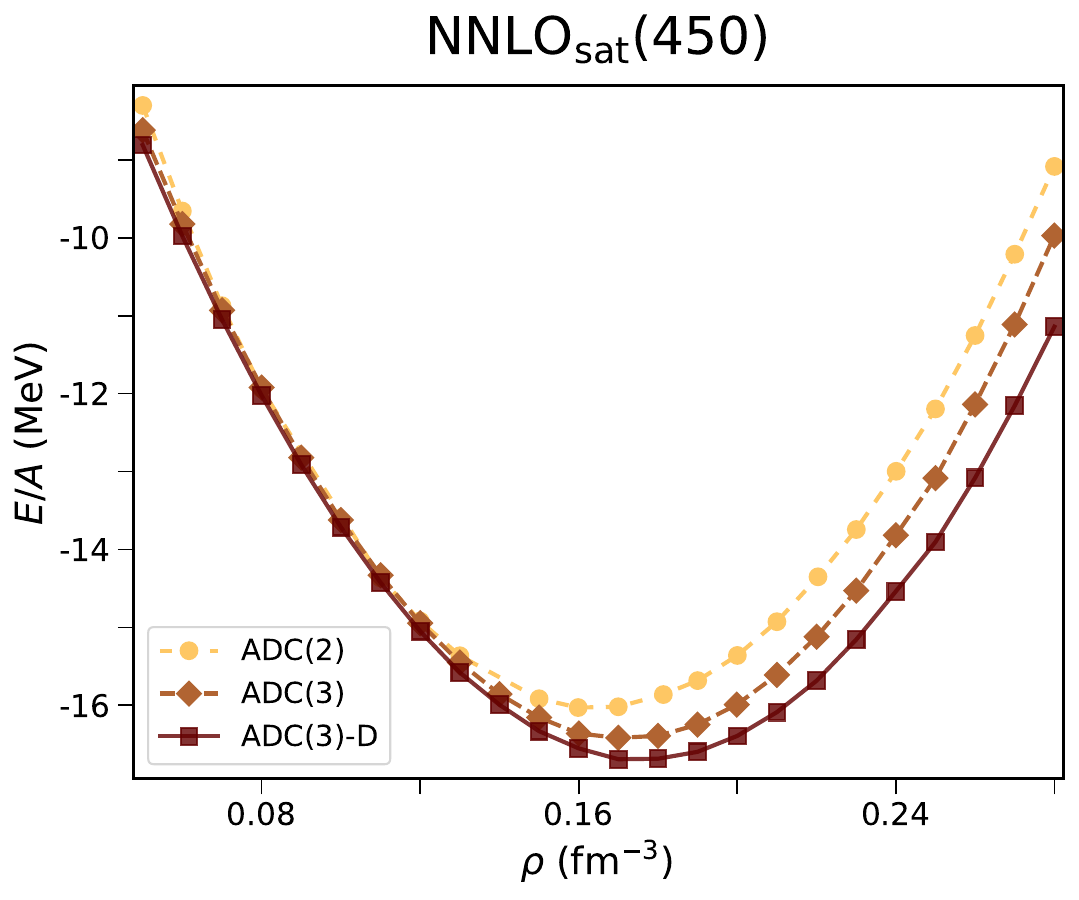}
    \caption{Comparison of ADC truncations for the SNM EOS.
    The NNLO$_{ \rm{sat} }\,(450)$ interaction is employed.
    }
    \label{fig: Cfr_Adc_SNM_NNLOsat}
\end{figure}

\begin{figure*}[ht]
    \centering
    \includegraphics[width=\textwidth]{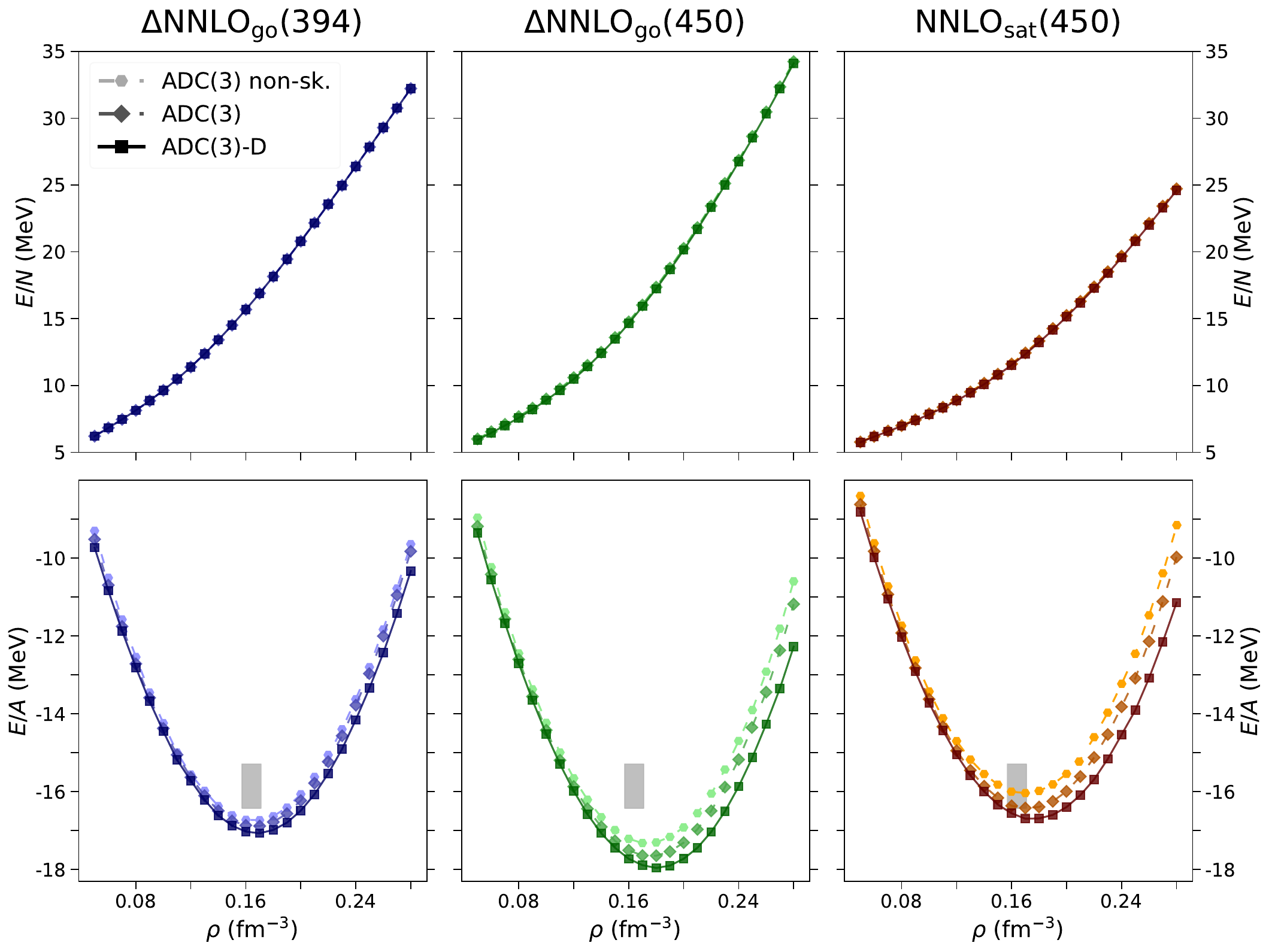}
    \caption{
    EOS in PNM (top row) and SNM (bottom row) for three chiral interactions:
    from left to right,
    $\Delta \rm{ NNLO_{go}(394) }$, $\Delta \rm{ NNLO_{go}(450)}$ and $\rm{ NNLO_{sat} }(450)$ .
    Calculations using the ADC(3) (diamonds), ADC(3) with non-skeleton corrections (circles), and ADC(3)-D (squares) approximations are shown (see legend).
    The ``Cen-\oprsPH'' prescription has been employed in all cases to evaluate the OpRS energies.
    The gray boxes denote the empirical estimates of the SNM saturation point from Ref.~\cite{Drischler2021Review}.
    We use the same vertical scale in each row.
    Lines are a guide to the eye.}
    \label{fig: Eos_ADC}
\end{figure*}

Most many-body methods involve approximations in the way correlations are incorporated.
In our approach, the dynamical self-energy must be truncated to a given expansion order in the ADC($n$) hierarchy.
We gauge different approximation schemes by comparing ADC(2), the first non-trivial ADC level, with the higher-order ADC(3) and ADC(3)-D truncations.
Results are shown in Fig.~\ref{fig: Cfr_Adc_SNM_NNLOsat} for the EOS of SNM, where many-body correlations are stronger than for PNM, and using the relatively hard NNLO$_{ \rm{sat} }\,(450)$ potential.
For densities up to $\rho=0.16$ fm${}^{-\rm 3}$, ADC(2) recovers approximately 90\% of the correlation energy compared to the nearly complete \hbox{ADC(3)-D}. This is in agreement with our previous finding in finite nuclei~\cite{Soma2014Chains,Soma2020}.
At higher densities, ADC(2) misses a larger part of the contribution to the binding energy and is consistently less bound than ADC(3) and ADC(3)-D.
The spread between ADC(3) and ADC(3)-D can be thought of providing an estimate of the errors induced by the self-energy truncation. 
In fact, in light of the comparison with CC of Ref.~\cite{Marino2024}, we expect that the deviations between the two SCGF variants are a conservative estimate of the actual many-body truncation errors, given the very satisfying accordance of ADC(3)-D and CC results including triples.

\section{Results}
\label{sec: results}

Besides total energy, SCGF allows to investigate the dynamics of nucleons in the nuclear medium, by providing complementary information on the spectral functions and momentum distributions.
In Sec.~\ref{sec: eos and rhok}, we report the predictions for the EOS and the occupation numbers in both PNM and SNM using three different interactions at NNLO in the chiral expansion and different ADC variants, based on PBCs.
Sec.~\ref{sec: fs effects} is devoted to testing the more sophisticated sp-TABC and discussing finite-size effects.
Spectral functions and their physical interpretation are discussed in Sec.~\ref{sec: Spectral functions}.

\subsection{Equations of state and momentum distributions}
\label{sec: eos and rhok}

\begin{figure}[ht]
    \centering
    \includegraphics[width=\columnwidth]{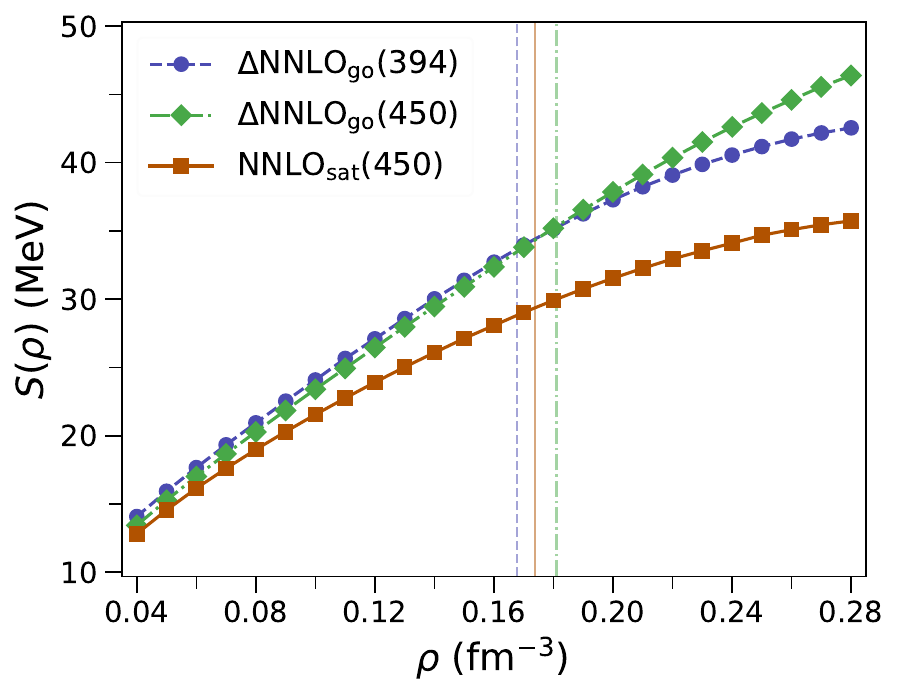}
    \caption{The symmetry energy $S(\rho)$ as a function of the density obtained with the ADC(3)-D scheme for the three interactions.
    Vertical lines denote the SNM saturation density predicted by each potential.}
    \label{fig: symmetry energy}
\end{figure}
The EOS curves for PNM and SNM in all cases are collected in Fig.~\ref{fig: Eos_ADC}.  Tabulated values for all these results are available at~\cite{marino_zenodo}.

For PNM, we notice that the three ADC variants yield substantially indistinguishable predictions of the total energies.
Correlations are indeed relatively weak in neutron matter in this range of nuclear densities, which lies within the domain of validity of $\chi$EFT.
Hence, a good convergence of many-body techniques is expected. 
The ADC self-consistent procedure typically converges in a few OpRS cycles and optimized s.p.~energies are hardly affected by correlations, remaining rather close to the corresponding HF energies.
The deltafull GO interactions from~\cite{DeltaGo2020} are roughly comparable, with the 450 MeV/c model producing slightly higher energies. 
The $\rm{NNLO_{sat}}(450)$ EOS, in contrast, is too soft, and its symmetry energy unrealistic~\cite{Carbone2020}.

Predictions for SNM are displayed in the bottom panels of Fig.~\ref{fig: Eos_ADC}. 
The $\Delta \rm{ NNLO_{go} } (394)$ interaction exhibits a modest dependence on the adopted approximation scheme.
We have already noticed in Sec.~\ref{sec: compare truncations} that additional correlation energy is gained with respect to ADC(3) when using ADC(3)-D.
The impact is more evident for the two interactions with cutoff $\Lambda = 450\,\rm{MeV}/c$, which are less perturbative.
Non-skeleton ADC(3) is systematically less bound than ADC(3).
For example, at $\rho=0.16\,\rm{fm}^{-3}$ with  $\rm{NNLO_{sat}}(450)$, the energies obtained with ADC(3) non-sk, ADC(3) and ADC(3)-D are approximately -16.0, -16.3 and -16.6 MeV, respectively.
A 600 keV discrepancy around saturation density is still a relatively mild deviation, which remains smaller than the interaction dependence.

For SNM, it is important to compare with the empirical saturation point.
\CB{
$\rm{NNLO_{sat}}(450)$ and $\Delta \rm{ NNLO_{go} } (394)$ are compatible with the approximate estimate of the SNM saturation point of $\rho_0 \approx 0.16\,\rm{fm}^{-3}$ and $E_0 \approx -16\,\rm{MeV}$ (see e.g.~Refs.~\cite{ring,ObertelliSagawa}), while our calculations show that $\Delta \rm{ NNLO_{go} } (450)$ somewhat overbinds the saturation energy.
Our results correct earlier estimates~\cite{DeltaGo2020} (see also~\cite{NNLOsatErratum,PbAbInitioAuthorCorrection}).
}

\CB{
Both calculations in model systems~\cite{Barbieri2017} and benchmarks with other methods in nuclear matter~\cite{Marino2024,Marino2025Qnp} suggest that ADC(3)-D is superior to ADC(3).
Instead, the non-skeleton correction is somewhat debatable. Conceptually, it offers a way to address the approximate nature of OpRS by introducing systematic corrections.
In practice, the net effect of non-skeleton ADC(3) is to revert to HF unperturbed energies, see Sec.~\ref{sec: oprs prescriptions}, effectively renouncing to some of the information on higher order correlations that are embedded in the self-consistent propagator. 
}

Fig.~\ref{fig: symmetry energy} displays predictions for the symmetry energy within the quadratic approximation~\cite{Burgio2020,rocamaza2018},
\begin{align}
    S(\rho) = \frac{E_{PNM}}{A}(\rho) - \frac{E_{SNM}}{A}(\rho) .
\end{align}
These results are obtained using the ADC(3)-D truncation, with the vertical lines marking the SNM saturation densities predicted by each potential.
As already noted in Ref.~\cite{Carbone2020}, our calculations highlight that $\rm{ NNLO_{sat} }(450)$ significantly underestimates the density dependence of the symmetry energy~\cite{Oertel2017,Li:2021,Burgio2021}. 
The two GO models predict higher symmetry energies and are close to each other up to $\rho = 0.20\,\rm{fm}^{-3}$~\cite{DeltaGo2020}.
We then expand $S(\rho)$ around the canonical value $\rho_0=0.16\,\rm{fm}^{-3}$,
\begin{align}
    S(\rho) = J + \left( \frac{\rho-\rho_0}{3\rho_0}\right) L,
\end{align}
where the parameters $J$ and $L$ have been introduced following the conventional definition 
and are listed in Tab.~\ref{tab: symmetry energy params}.
This choice makes the comparison with other theoretical calculations easier~\cite{alp2025equationstatefermiliquid}.
However, we stress that the actual saturation point deviates considerably from the empirical $\rho_{0}$ in the case of the $\Delta \rm{ NNLO_{go}(450) }$ force.
Hence, for this potential, the values of the $J$ and $L$ coefficients may be somewhat misleading.
\begin{table}[]
    \begin{tabular}{ p{2.8cm}  p{1.8cm} p{1.8cm} }
    \hline \hline
        \noalign{\vskip 1.mm} 
        Interaction & $J$ (MeV) & $L$ (MeV) \\
        \noalign{\vskip 1.mm} 
        \hline
        \noalign{\vskip 1.5mm} 
        $\Delta \rm{ NNLO_{go}(394) }$ & 32.7 & 62.1 \\
        \noalign{\vskip 1.5mm} 
        $\Delta \rm{ NNLO_{go}(450) }$ & 32.4 & 69.9 \\
        \noalign{\vskip 1.5mm} 
        $\rm{ NNLO_{sat}(450) }$ & 28.1 & 45.6 \\
        \noalign{\vskip 1.mm} 
    \hline \hline        
    \end{tabular}
    \caption{Symmetry energy coefficients $J$ and $L$ for three different interactions, extracted from ADC(3)-D computations. 
    They can be compared with the empirical estimates reported in Refs.~\cite{Oertel2017,Li:2021,Burgio2021}.
    }
    \label{tab: symmetry energy params}
\end{table}

\begin{figure*}[ht]
    \centering
    \includegraphics[width=0.68\columnwidth]{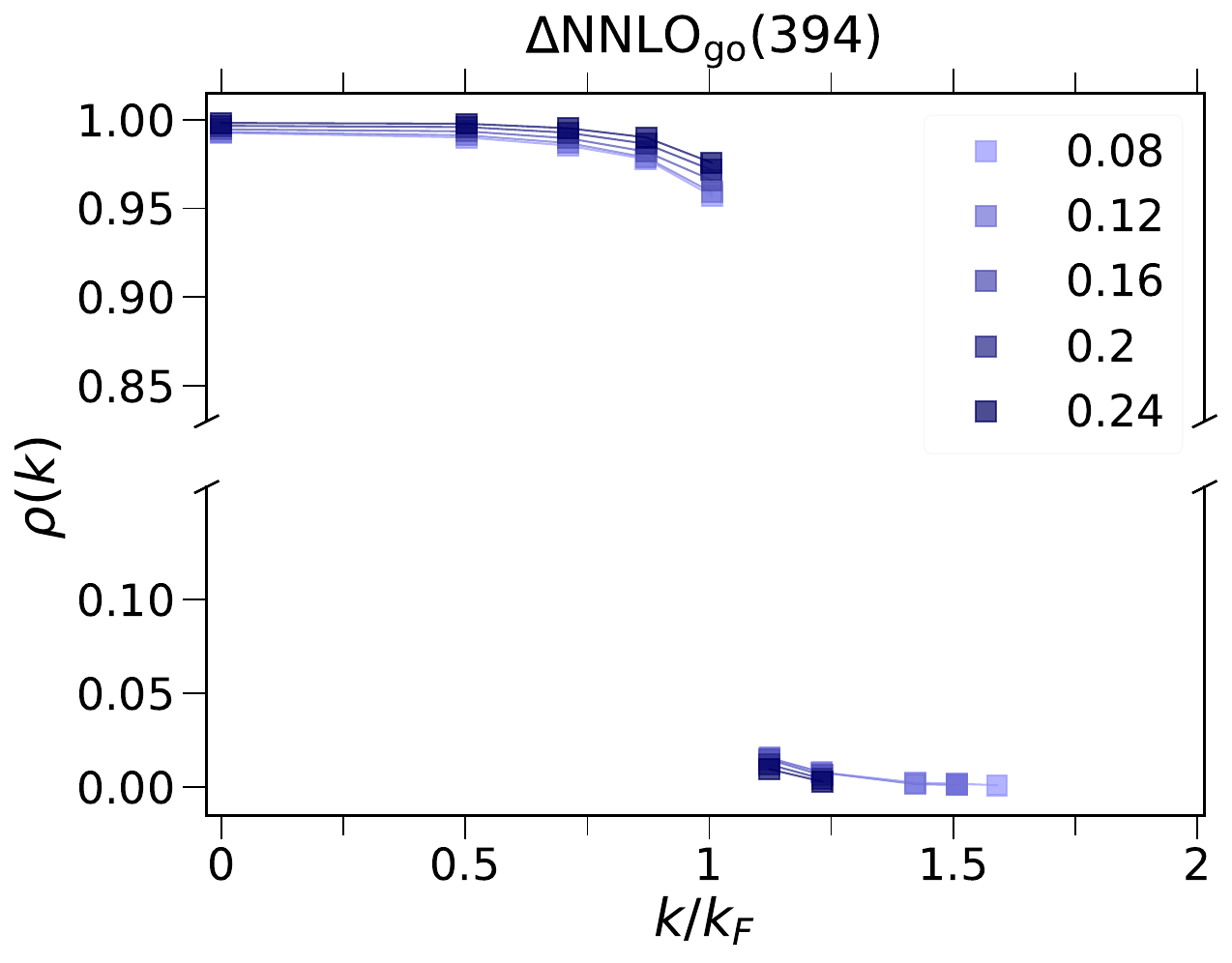}
    \includegraphics[width=0.68\columnwidth]{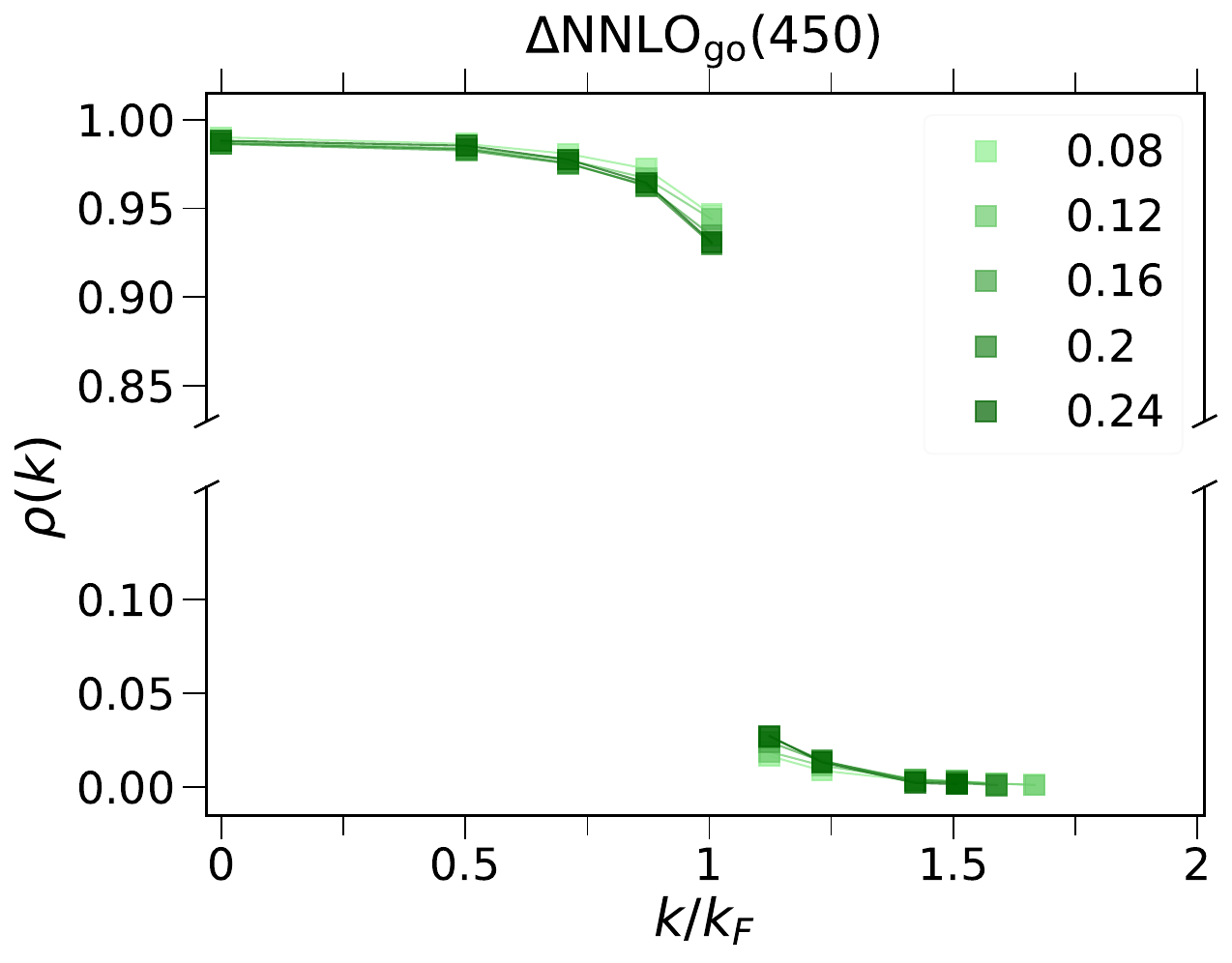}
    \includegraphics[width=0.68\columnwidth]{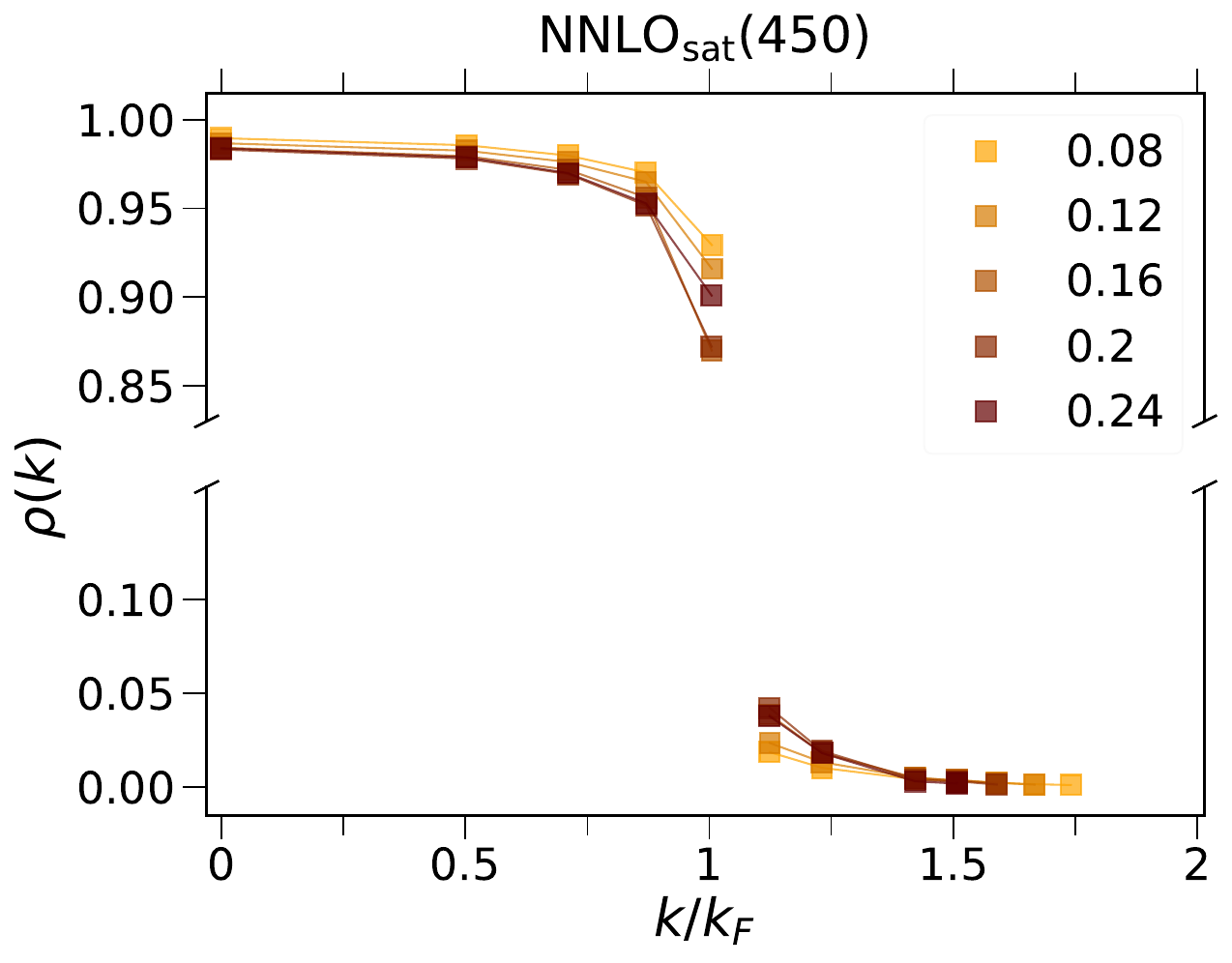}
    \includegraphics[width=0.68\columnwidth]{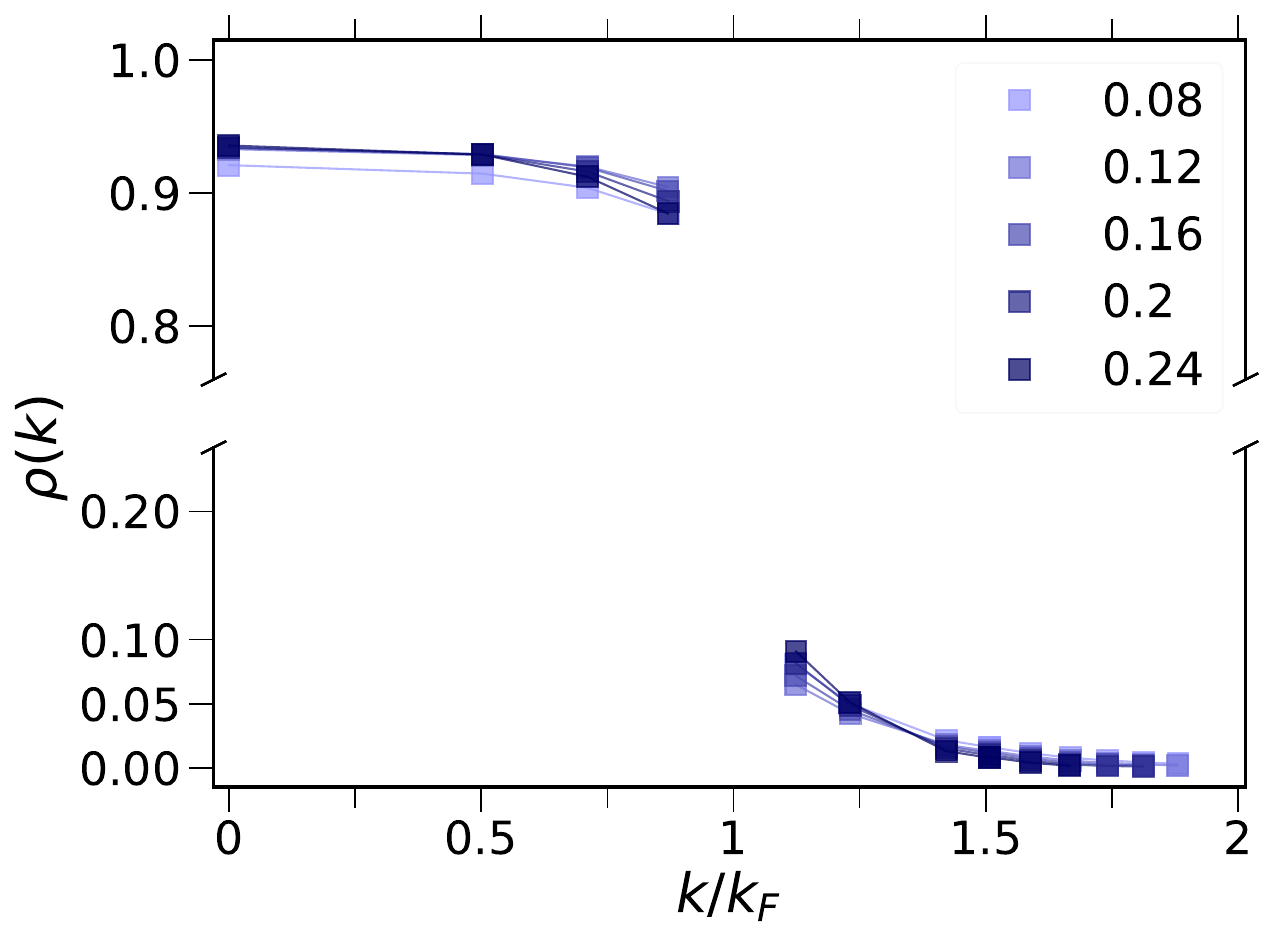}
    \includegraphics[width=0.68\columnwidth]{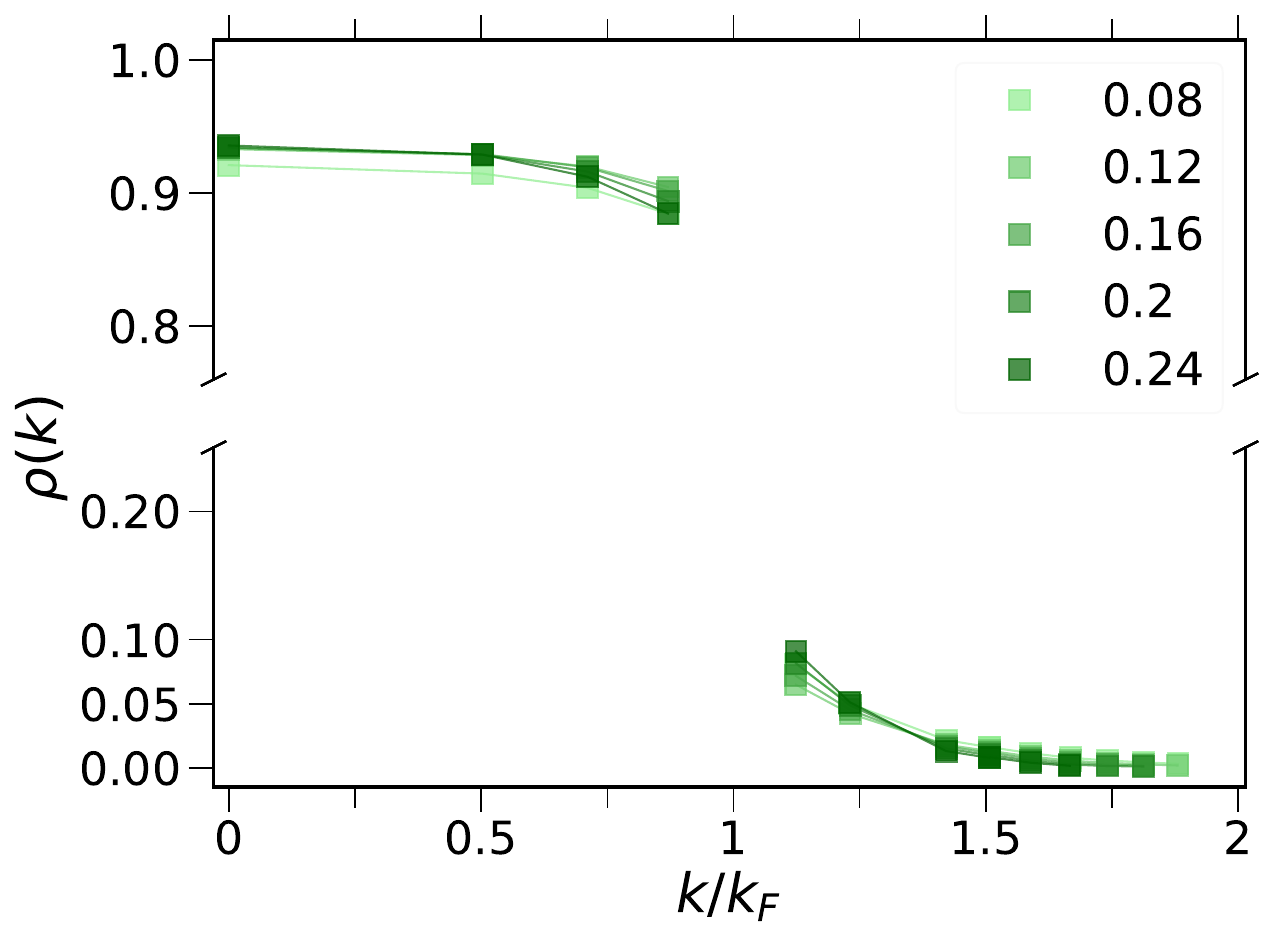}
    \includegraphics[width=0.68\columnwidth]{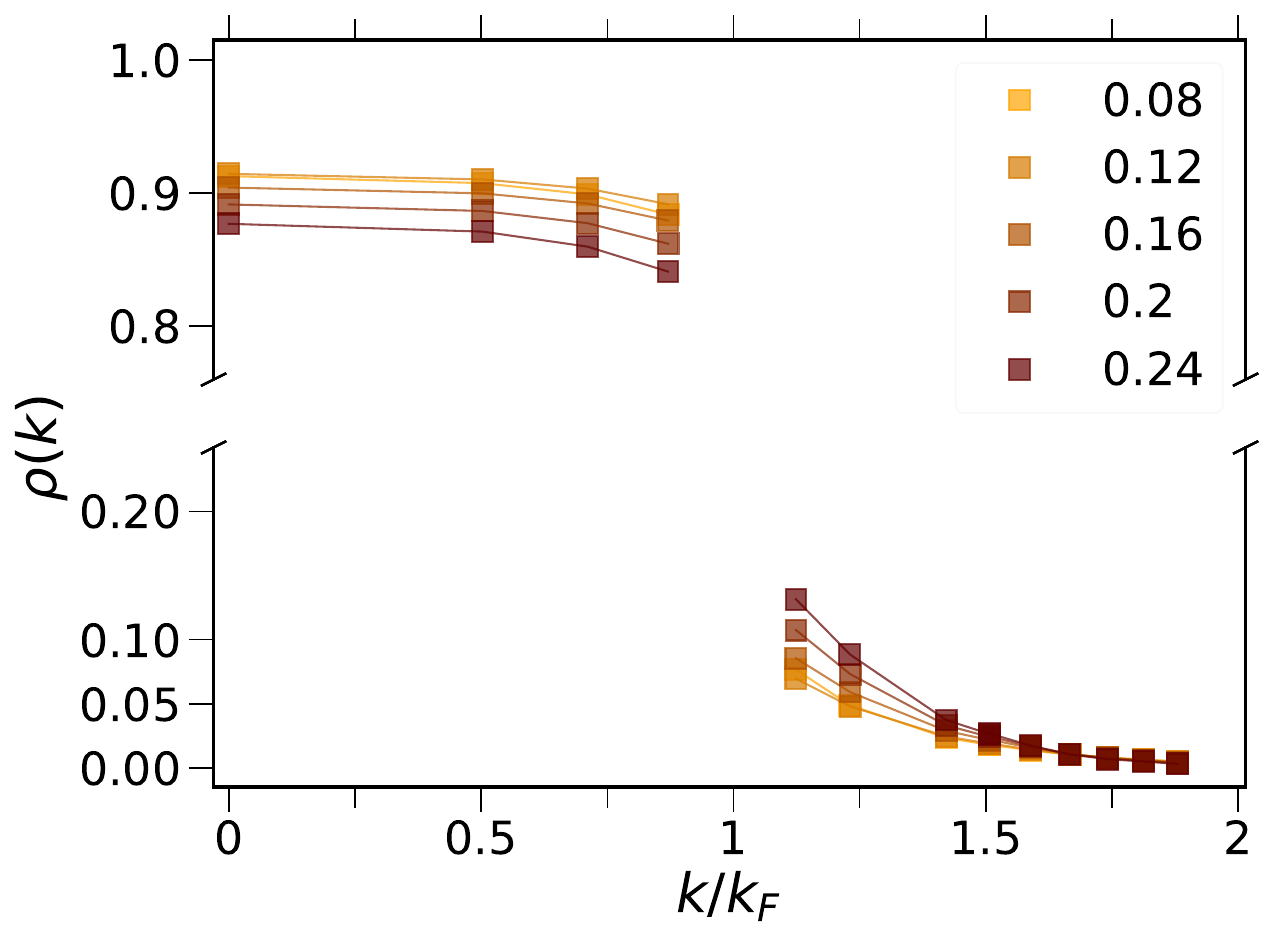}
    \caption{
    Occupation probabilities $\rho(k)$ as a function of the momentum in units of the Fermi momentum $k/k_F$ in PNM (top) and SNM (bottom) for different densities (in fm$^{-3}$), reported in the legend,
    Three different interactions are employed.
    For SNM, neutron distributions are shown.
    All calculations have been performed using PBCs and the ADC(3)-D truncation.
    Different scales are used in the vertical axis for momenta below and above $k_F$, and for top and bottom panels, but these are the same across all three interactions.
    }
    \label{fig: Momentum_Distributions}
\end{figure*}

Next, we present in Fig.~\ref{fig: Momentum_Distributions} our predictions for the momentum distribution, $\rho(k)$. 
In spin-saturated matter, occupations are independent of the spin projection.
Also, due to the approximate isospin invariance, we only show results for neutrons in the SNM case.
Darker lines within each panel correspond to increasing densities between 0.08 and 0.24 in $\rm{fm}^{-3}$.
We highlight that $\rho(k)$ is discontinuous across $k_F$ for a normal Fermi liquid~\cite{Arthuis2023,DickhofVanNeck}, which is known to be case for nuclear matter in the range of densities considered here.
At variance with the HF case, where the occupation probability is a sharp step function, $\rho_{HF}(k) = \theta(k-k_F)$, correlations alter the momentum distribution. The full ADC(3) yields a depletion of $\rho(k)$ below the Fermi surface, in particular $\rho(k=0)<1$.
Correspondingly, a non-zero occupation tail is found at momenta outside the Fermi sphere.
Occupations are non-negligible for momenta $ k \gtrsim k_F$, while they quickly drop at higher momenta.
Qualitatively, the main difference between PNM and SNM is that, in the former, at $k=0$ the occupation is close to 1, while SNM is depleted already at very low momentum, e.g., $\rho(k=0) \approx 0.9$.
From a quantitative point of view, the tail is generally more extended in SNM than PNM and the occupation of hole states, for $k \to k_F^{-}$, is smaller. 
We stress that all the Hamiltonians considered in this work are rather soft and do not induce sizable short-range effects. The occupation of states observed just above $k_F$ is mainly associated with long-range or collective modes that are related to low-energy virtual excitations~\cite{Barbieri2004,Barbieri2009prl}.
When strong short-range repulsion is included in the modeling of the Hamiltonian, additional occupation is found at high-momenta, generally for $k>$2~fm${}^{\rm -1}$ with a tail that is not visible on the linear scale of Fig.~\ref{fig: Momentum_Distributions}~\cite{Barbieri2004}.
As a rule of thumb, correlation effects become stronger with increasing density, which is reflected in larger depletions with a more significant transfer of s.p.~strength from below to above the Fermi momentum.
Finally, we point out that $\rho(k)$ is somewhat sensitive to the choice of the potential, too.
For example, the soft $\Delta \rm{ NNLO_{go} } (394)$ force induces very weak many-body correlations, leading to a  $\rho(k)$ that is less depleted than for the other two interactions and resembles the sharp HF distribution.
To conclude, we remind that all distributions shown in Fig.~\ref{fig: Momentum_Distributions} are available for a few discretized moments due to using PBCs. A better description of the $\rho(k)$ curves is obtained with TABCs, as discuss in the next section.

\subsection{Finite-size effects}
\label{sec: fs effects}

\begin{figure}[ht]
    \centering
    \includegraphics[width=\columnwidth]{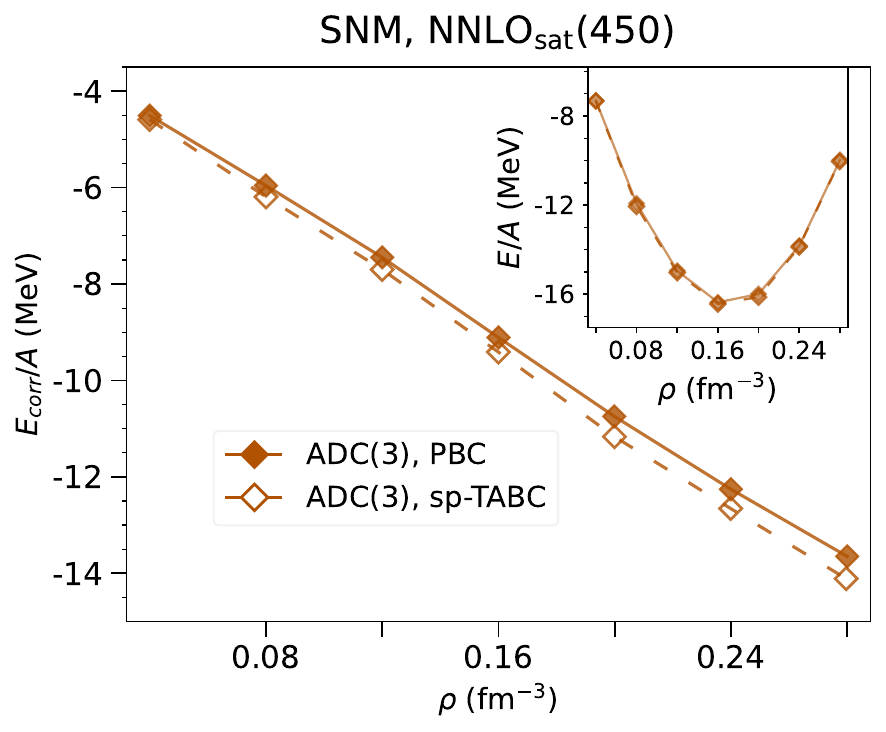}
    \caption{Correlation energies per particle obtained in SNM with $A=132$ nucleons at the ADC(3) truncation level employing the $\rm{ NNLO_{sat} }(450)$ interactions.
    Results with PBC (sp-TABC) are shown as full (empty) symbols.
    In the inset, energies per particle are shown for the same set of calculations.
    Lines are a guide to the eye.
    }
    \label{fig: eos_NNLOsat_Gorkov_SNM_Pbc_vs_Tabc}
\end{figure}

Figure~\ref{fig: eos_NNLOsat_Gorkov_SNM_Pbc_vs_Tabc} demonstrates the effect on the correlations energy when using a sp-TABC basis as opposed to standerd PBCs. 
We focus on the EOS of SNM and report calculations performed at the ADC(3) truncation level.
For sp-TABC, we perform a single calculation in the twisted basis employing the twist angles that allow to best approximate the TL energy at the HF level~\cite{Hagen2014,McilroyChristopher2020Sgfs} (Sec.~\ref{sec: model space}).
In the inset, we compare the total energies for the two choices of boundary conditions.
The main panel displays correlation energies and  highlighting that differences between the two sets of calculations are small: sp-TABCs  yield slightly lower energies than PBCs, but the deviation always remains marginal, of the order of 100-150 keV/A at all the densities considered. 
This comparison allows us to gauge the impact of FS effects on the EOS, and it suggests that these are subleading with respect to uncertainties stemming from the many-body truncation and the interaction dependence. We conclude that standard PBCs -- with $A/g$=33 -- can be safely employed to compute the EOS at typical nuclear densities.

\begin{figure}[ht]
    \centering
    \includegraphics[width=\columnwidth]{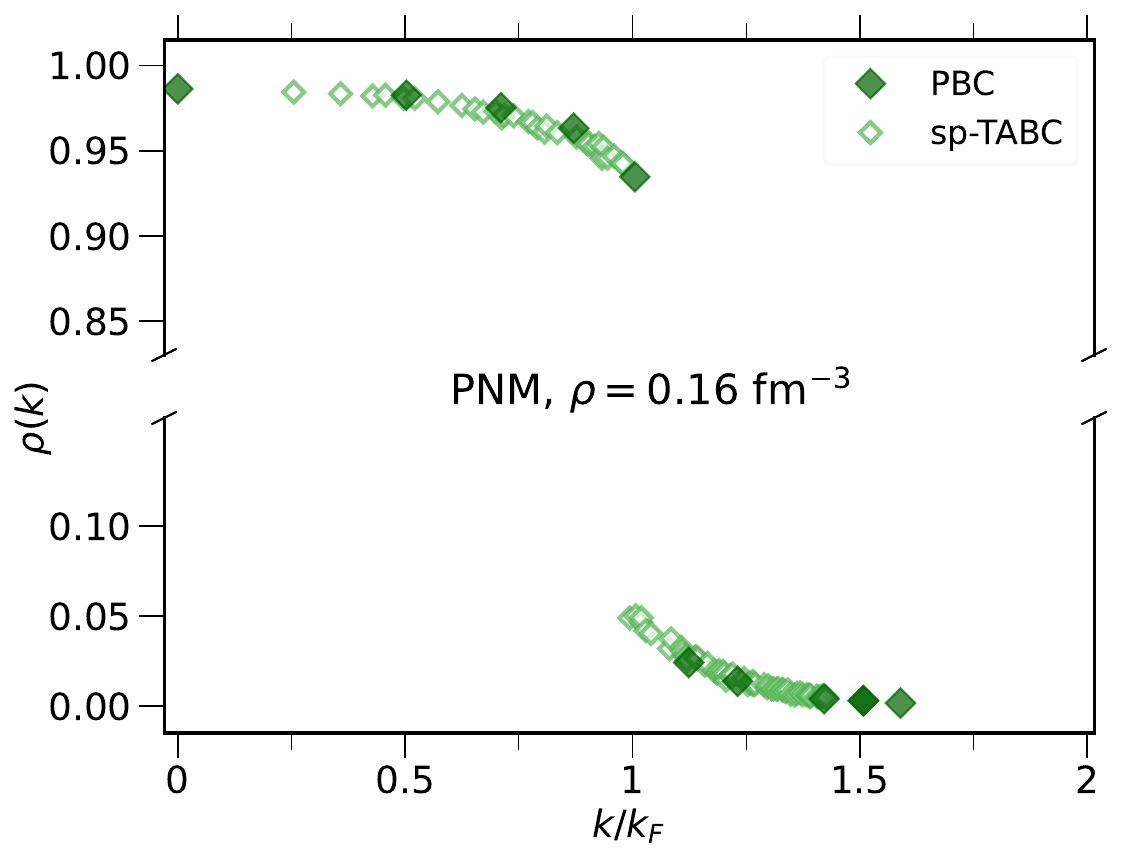}
    \includegraphics[width=\columnwidth]{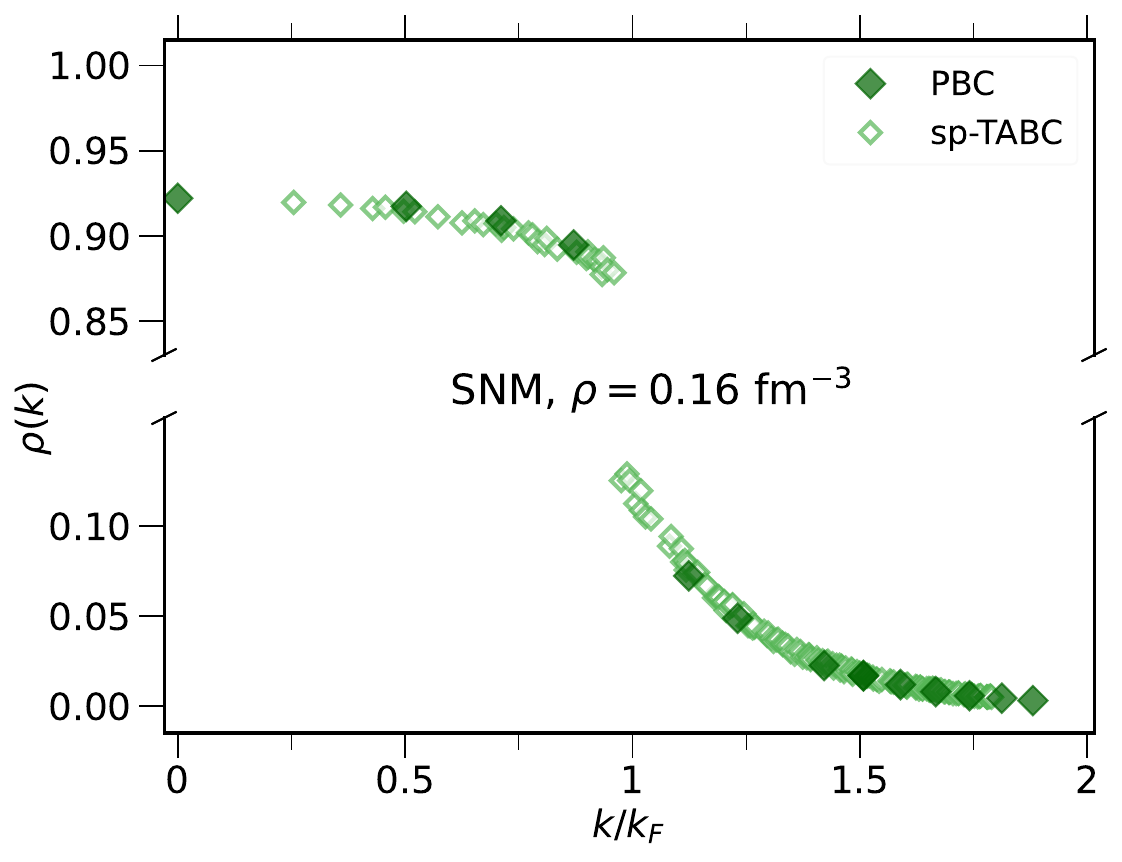}
    \caption{Momentum distributions $\rho(k)$ as a function of $k/k_F$ for PNM (top) and SNM (bottom) at $\rho=0.16\,\rm{fm}^{-3}$. Calculations performed with the ADC(3) approximation using PBC (filled symbols) and TABC (empty symbols)  are compared.
    The $\Delta \rm{ NNLO_{go}(450) }$ interaction is employed.
    Note that different scales are used on the vertical axis below and above the discontinuity, whereas the vertical scale for PNM and SNM is the same.
    }
    \label{fig: mom_distr_pbc_vs_tabc}
\end{figure}

The higher resolution granted by the TABCs is advantageous in describing quantities which are functions of the momentum, like occupation probabilities and spectral functions, but also pairing gaps~\cite{Gezerlis2021Twist}.  
Momentum distributions $\rho(k)$ as a function of $k/k_F$ are compared in Fig.~\ref{fig: mom_distr_pbc_vs_tabc} for PNM and SNM at saturation density and using both PBC  and sp-TABC.
In finite-$A$ simulations, one has a sampling of the occupation probabilities at discrete values of $k$, rather than a continuous function.
With PBCs, the mesh of $k$ points is sparse due to the degeneracy of the s.p.~momenta $\mathbf{k}$.
TABCs break these symmetries and generate a denser mesh.
Consequently, momentum distributions are modeled with greater accuracy. This is apparent close to the Fermi surface, as well as in the high-momentum tail.
In general, predictions of PBCs and TABCs are in good agreement with each other, and the overall trends of the momentum distributions are well-defined both below and above the Fermi surface.

\subsection{Spectral functions}
\label{sec: Spectral functions}

\begin{figure}[ht]
    \centering
    \includegraphics[width=\columnwidth]{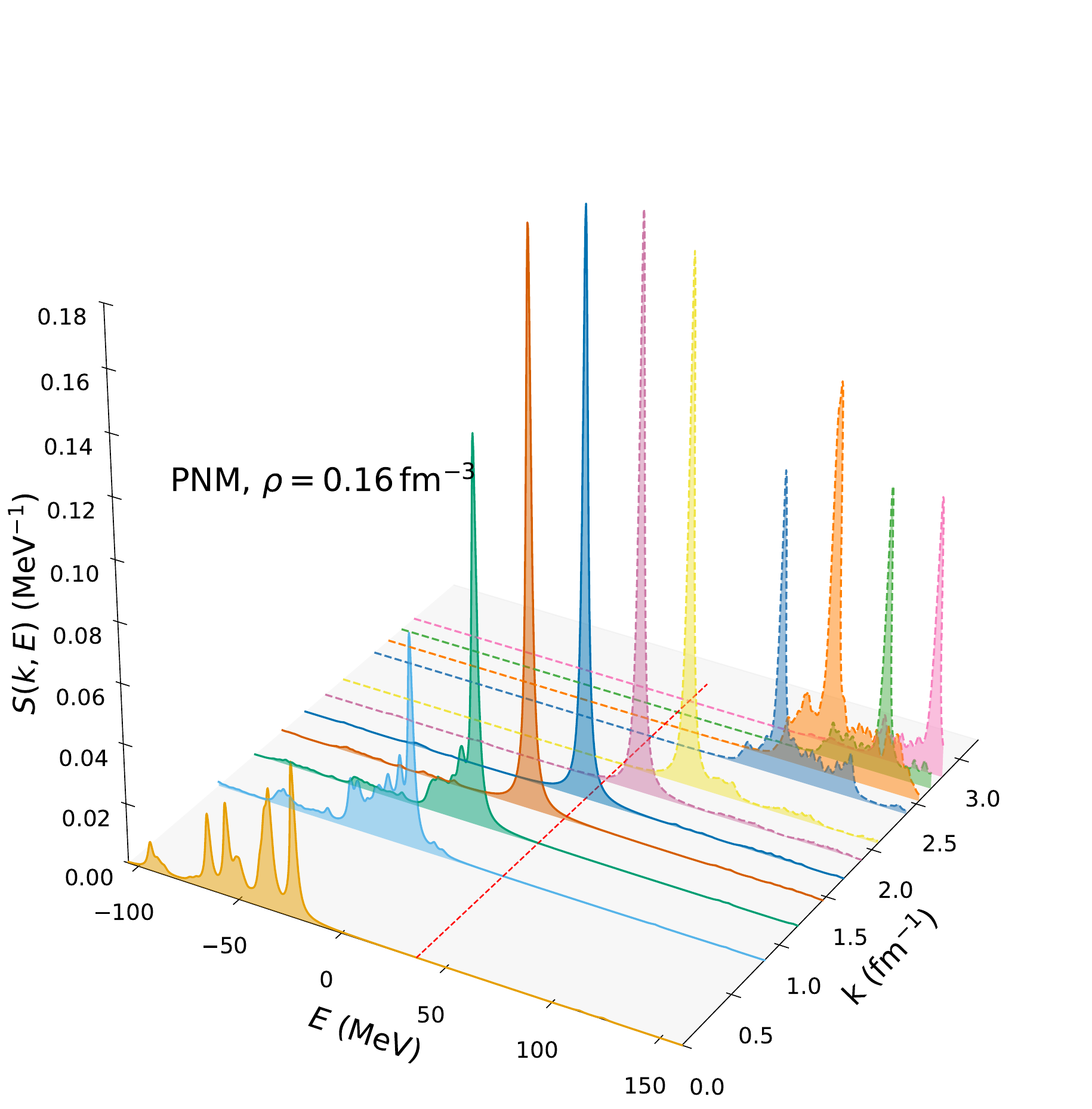}
    \caption{
    Three-dimensional representation of the PNM spectral function at $\rho=0.16\, \rm{fm}^{-3}$.
    Calculations are performed with the  $\Delta \rm{ NNLO_{go}(450)}$ interaction with ADC(3)-D and PBCs. Peaks are convoluted with a Lorentzian with a finite width $\Gamma=1.5$ MeV for display purposes.
    The height $S(k,E)$, in MeV$^{-1}$, represents the value of the spectral function for given momentum $k$ and energy $E$. Each section corresponds to a given inequivalent momentum $k$.
    Quasihole and quasiparticle states are shown with continuous and dashed lines, respectively, and they are separated by the chemical potential, given by a red dashed line at energy $E = \mu$.
    }
    \label{fig: 3d spect func}
\end{figure}

Information about the spectroscopy for addition and removal of a nucleon, as well as for the nucleon mean-free path~\cite{Rios2012}, is contained in the normal component of the spectral function~\cite{Soma2011,Barbieri2017},
\begin{align}
    S_{\sigma_\alpha \tau_\alpha}(\mathbf{k}_\alpha,E) 
    ={}& \mp \frac 1\pi \Im\left\{g^{11}_{\alpha\alpha}(\hbar\omega=E-\mu_\alpha)\right\} \nonumber \\
    ={}& \sum_q \abs{\mathcal{U}_\alpha^q}^2 \delta(E-\mu_\alpha-\hbar\omega_q) \nonumber \\
    &\quad+ \sum_q \abs{\mathcal{V}_\alpha^q}^2 \delta(E-\mu_\alpha+\hbar\omega_q) \,,
    \label{eq:Skw_def}
\end{align}
where the sign $-$~($+$) apply to energies $E>\mu_\alpha$~($E<\mu_\alpha$).  Figure~\ref{fig: 3d spect func} displays  a three 
dimensional representation of the spectral function, as a function of the momentum $k$ and the energy $E=\hbar\omega$, for PNM at $\rho=0.16\, \rm{fm}^{-3}$ with the $\Delta \rm{ NNLO_{go}(450)}$ potential.
Each section of the plot corresponds to the spectral distribution for a given s.p. momentum.
As PBCs are used, momenta are sparse at small $k$ but become progressively denser as $k$ increases.
For display purposes, we smear the energy dependence of the discrete peaks with a Lorentzian function of width $\Gamma=1.5$ MeV (see~\cite{Barbieri2017}).
The chemical potential is shown as a red line at $E = \mu$.
We note that the solution of the Gorkov equations yields negligible anomalous contributions to the self-energy in this case, confirming that PNM can be described as a normal Fermi liquid at densities around saturation~\cite{DickhofVanNeck}.
Interactions with the many-body environment imply that the s.p.~states, which at the HF level are characterized by a well-defined energy, are now spread out into a multitude of energy poles.
We represent the spectral function for states that are occupied (empty) in the mean-field reference with continuous (dashed) lines.
The excitations having hole and particle character are clearly separated by the Fermi energy.
The fragmentation of the s.p.~strength induced by many-body correlations is mostly apparent at very low or very high momenta, e.g., for $k=0$ or $k \ge 2.5.\,\rm{fm}^{-1}$.
In contrast, as the Fermi surface is approached for $k \to k_F$ (either from below or from above), the spectral function is progressively less spread out. States near the Fermi surface feature well-defined narrow resonances that carry most of the s.p.~strength. These dominant peaks are superimposed on an incoherent background of satellite solutions with almost negligible amplitudes.
These narrow, long-lived excitations at the Fermi surface are named quasihole and quasiparticle states in the context of Landau's Fermi liquid theory~\cite{DickhofVanNeck,Rios2012,Schwenk2012}
\footnote{Quasiparticles are predicted to appear for $k \to k_F$ and $E \to \mu$ as a consequence of the vanishing of the imaginary part of the self-energy at the chemical potential, see e.g.~Refs.\cite{Fetter,SomaEpj,Mattuck}.} .

\begin{figure}
    \centering
    \includegraphics[width=\columnwidth]{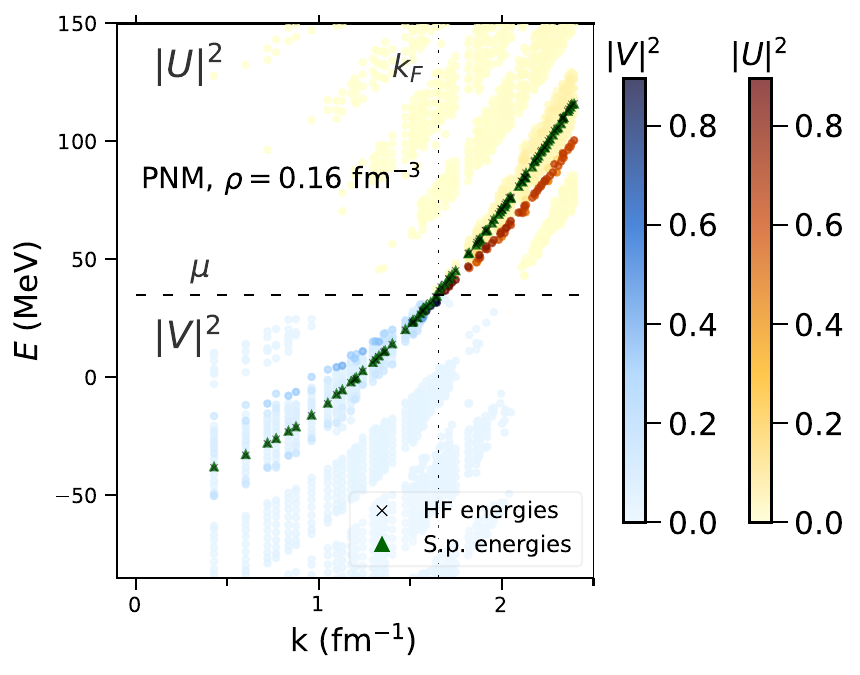}
    \includegraphics[width=\columnwidth]{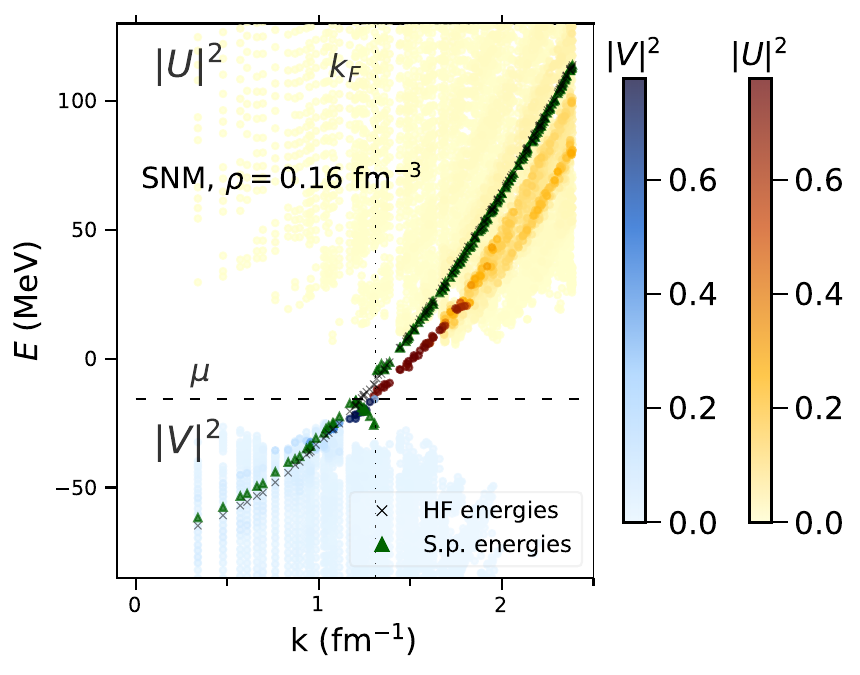}
    \caption{
    Two-dimensional representation of the spectral function for PNM (top) and SNM (bottom) as a function of the momentum $k$ and the energy $E$.
    Calculations are performed at $\rho=0.16\,\rm{fm}^{-3}$ with ADC(3) employing sp-TABCs and the $\Delta \rm{ NNLO_{go}(450) }$ interaction.
    The Fermi momentum $k_F$ and the chemical potential $\mu$ are marked by dotted and dashed lines, respectively.
    The squared amplitudes $\abs{\mathcal{U}^{q}_{\alpha}}^2$ ($\abs{ \mathcal{V}^{q}_{\alpha}}^2$) are shown for poles above (below) the Fermi level. 
    Color scales are shown next to the plot. HF s.p.~energies are represented by crosses.
    Optimized s.p.~energies generated by the Cen-$k_F$ OpRS scheme are shown as triangles and discussed in the text.
    }
    \label{fig: 2d spect funcs}
\end{figure}

We performed further simulations of both PNM and SNM using the finer momentum meshes in sp-TABC. These are displayed in
Fig.~\ref{fig: 2d spect funcs} as two-dimension heat maps and are obtained for $\rho=0.16\,\rm{fm}^{-3}$ with the ADC(3) truncation and the $\Delta \rm{ NNLO_{go}(450)}$ Hamiltonian.
The same calculations also yield the momentum distributions shown in Fig.~\ref{fig: mom_distr_pbc_vs_tabc}.
%
Poles $\epsilon_q = \mu \pm \hbar\omega_{q}$ for each momentum are denoted by dots, and the strength $\abs{\mathcal{V}^{q}_{\alpha}}^2$ ($\abs{ \mathcal{U}^{q}_{\alpha}}^2$) for solutions below (above) the Fermi level $\mu$ is represented by a color scale.
The chemical potential and the Fermi momentum are shown as horizontal and vertical lines, respectively.
Crosses denote the s.p. energies at the HF level.
Triangles refer to the OpRS energies obtained with the Cen-\oprsPH\, recipe.

The PNM spectral functions in Figs.~\ref{fig: 3d spect func} and~\ref{fig: 2d spect funcs} are obtained with different boundary conditions but show consistent predictions, the main difference being the denser mesh of $k$ points in Fig.~\ref{fig: 2d spect funcs} that results from using sp-TABCs.
Quasihole excitations are recognizable for states with $k \lesssim k_F$, and a quasiparticle branch that extends into the larger momentum region is also apparent.
Fragmentation is strongest for momenta well inside the Fermi sphere, as shown by the wide band of secondary peaks below the chemical potential which lie around the HF energies.
OpRS energies are close to the HF ones, with the largest differences being of 1-2 MeV for PNM.

SNM is characterized by stronger correlations, which manifest themselves in an increased fragmentation of the spectral function. The background is more extended and comprises a large number of satellite peaks with low strength $\abs{\mathcal{V}^{q}_{\alpha}}^2$ or $\abs{\mathcal{U}^{q}_{\alpha}}^2$, but it fades for states approaching the Fermi momentum.
At the same time, well-structured excitations at definite energies tend to emerge, as visible by dark circles on top of the incoherent background.
States close to the Fermi momentum feature a single dominant pole with $\abs{\mathcal{V}^{q}_{\alpha}}^2 \simeq 1$ for $k < k_F$ or $\abs{\mathcal{U}^{q}_{\alpha}}^2 \simeq 1$ for $k > k_F$.
For momenta slightly below $k_F$, a few quasihole excitations with energy close to $\mu$ appear. Above the Fermi momentum, a branch of intense quasiparticle peaks extends up to $k \approx 1.5 \,\rm{fm}^{-1}$.
The dominant peaks follow a roughly parabolic trend as a function of the momentum, which can be related to the dominance of the kinetic contributions $\sim k^2$.
Note that the spectral function maxima have a different curvature than the HF or OpRS poles, at least for $k > 1 \,\rm{fm}^{-1}$, and have lower energies than the HF ones.
A detailed investigation of the $k$-dependence of the quasiparticle energies close to the Fermi momentum and the associated effective masses~\cite{Schwenk2012,Rios2020} is left for future studies.

Finally, we have noticed that OpRS energies are close to the HF energies at low momenta and almost indistinguishable from them for $k>k_F$.
However, when approaching the Fermi surface for $k \to k_F^{-}$, the OpRS energies deviate from the monotonic behavior.
This is a consequence of using the Cen-\oprsPH prescription, Eq.~\eqref{eq:ph prescription}, that assigns a hole character and effective s.p. energies below $\mu$ to states with $k<k_F$.  While in HF theory energies are continuous in the TL, the s.p.~spectrum generated by the OpRS can retain the pairing gap across $k_F$.

Overall, comparing the spectral functions in Fig.~\ref{fig: 2d spect funcs} highlights that, at saturation density, PNM is a relatively weakly correlated system.
In contrast, interactions are much stronger in SNM, leading to additional fragmentation of the s.p. strength, a more extended incoherent background in the spectral function, and, correspondingly, a larger depletion of the momentum distribution below the Fermi momentum (see Fig.~\ref{fig: mom_distr_pbc_vs_tabc}).

\section{Conclusions and perspectives}
\label{sec: Conclusions}

In this work, we have developed a novel scheme for Gorkov-SCGF theory and applied it to infinite nuclear matter.
Our framework combines an approximate treatment of pairing correlations, which are handled at first order in a self-consistent way, with dynamical correlations included with the state-of-the-art Dyson-ADC(3) truncation.
Essential is the introduction of an efficient way to approximate dressed Gorkov propagators in terms of particle-number-conserving propagators. 
Moreover, we have investigated two extensions of the standard ADC(3) scheme. First, we have included the contribution of non-skeleton diagrams as a correction for partial self-consistency associated with the choice of an OpRS mean field.
Second, the hybrid ADC(3)-D truncation, which combines ADC(3) and coupled-cluster, has been thoroughly discussed.

By incorporating pairing and dynamical correlations together, we can provide robust predictions in a wide range of densities.
Also, we stress that our approach is capable of handling pairing at the cost of a standard Dyson-ADC(3) or CC computations~\cite{Marino2024}, which is much more favorable than the full Gorkov-ADC or Bogoliubov CC~\cite{Signoracci2015,Tichai2024} approaches.

We have studied PNM and SNM using nuclear potentials, including both 2B and 3B interactions, at NNLO in the chiral EFT expansion.
We have found satisfying predictions for the EOS, in substantial agreement with other many-body techniques~\cite{Marino2024}. 
ADC-SCGF has been firmly established to be a reliable tool to study both PNM and SNM.
At the same time, we stress that one of the strengths of SCGF theory is that quantities like the momentum distributions and the single-particle spectral functions emerge naturally in the formalism, so that the SCGF approach provides a microscopic picture of s.p. properties in the medium. 
In this work, we have shown, in particular, that while PBCs are accurate enough for the EOS, TABCs allow to achieve a better understanding of momentum-resolved quantities, such as the occupation numbers and spectral functions.

Consequently, we aim to apply our ADC-SCGF framework to deepen this understanding and investigate properties like effective masses and quasiparticle lifetimes, which can provide, for instance, a firm microscopic ground to the Fermi liquid description of PNM~\cite{Schwenk2012}.
The Gorkov formalism is also well-suited for giving a consistent description of energies, spectral functions, and pairing gaps in a unified way.
SCGF theory can naturally address extended matter in the superfluid regime, allowing for an accurate \textit{ab initio} description of PNM at very low densities, bearing important consequences for the modeling of the inner crust and the outer core of neutron stars, as well as being connected to the physics of the unitary Fermi gas~\cite{Gandolfi2015,Sedrakian2019}.

Finally, ADC-SCGF simulations can be exploited to ground nuclear EDFs into \textit{ab initio} theory, pushing forward the "Jacob's ladder" program started in Refs.~\cite{Marino2021,MarinoPhdThesis,colo2020}. 
Dynamical and static response functions can be computed, in principle, within the Green's functions formalism, either by adapting the approximate scheme applied to finite nuclei in Refs.~\cite{Barbieri2003ERPA,Raimondi2019}, or by implementing the ADC hierarchy for the polarisation propagator~\cite{Schirmer2018}.
Besides their intrinsic interest (see e.g.~\cite{sobczyk2024spinresponseneutronmatter,Shen2014}), the calculation of response properties provides a possible pathway to firmly constrain the so far purely empirical gradient terms of the EDFs to \textit{ab initio}~\cite{MarinoPhdThesis,Colo:2025ejt,moldabekov2025density}.
Robust predictions of the PNM EOS, pairing gaps, and effective masses can help improving the crucial isovector and pairing sectors of nuclear EDFs, increasing the predictive power of nuclear density functional theory in astrophysically relevant scenarios~\cite{Chamel:2008ca,Burgio2021} and neutron-rich isotopes at the edge of stability.

\section*{Acknowledgements}
The authors thank Rongzhe Hu, Weiguang Jiang, Christopher McIlroy, Arnau Rios, Alessandro Roggero, and Chieh-Jen Yang for useful discussions.
F.M was supported by the Deutsche Forschungsgemeinschaft (DFG, German Research Foundation) – Project-ID 279384907 – SFB 1245, and through the Cluster of Excellence “Precision Physics, Fundamental Interactions, and Structure of Matter” (PRISMA+ EXC 2118/1, Project ID 39083149).
We acknowledge the CINECA awards AbINEF (HP10B3BG09) and RespGF (HP10BQMECT) under the ISCRA initiative, for the availability of high-performance computing resources and support.
This work used the DiRAC Data Intensive service (DIaL3) at the University of Leicester, managed by the University of Leicester Research Computing Service on behalf of the STFC DiRAC HPC Facility (www.dirac.ac.uk). The DiRAC service at Leicester was funded by BEIS, UKRI and STFC capital funding and STFC operations grants. DiRAC is part of the UKRI Digital Research Infrastructure.
Part of the calculations were also performed at the supercomputer Mogon at Johannes Gutenberg Universit\"at Mainz.

\section*{DATA AVAILABILITY}
Part of the data that support the findings of this article are openly available~\cite{marino_zenodo}.

\appendix

\section{Implementation for infinite matter}
\label{app:implementation}

We derived the hybrid Gorkov-Dyson formalism in Sec.~\ref{sec:formalism} without any assumption on the particular system or the choice of the model space, except for assuming a diagonal form of the one-body interaction,~Eq.~\eqref{eq:H0vsTU}. Thus, all relations presented in Secs.~\ref{sec:GkvDys_formalism}, \ref{sec: Self-energy approx} and~\ref{sec:nsk_adc3d} are completely general and can be applied to any many-fermion problem given a (discretized) single particle basis,~$\{\alpha\}$.
For homogeneous matter, translational invariance and the rotational symmetry of the Hamiltonian can be exploited to simplify the implementation. In particular, translational invariance implies that Gorkov propagators and spectral functions are diagonal in the momentum s.p.~basis. Here, we collect in detail the final working equations for our implementation of ADC-SCGF for nuclear matter.

\subsection{Single-particle basis}
\label{app:aq_bases}

Our computations exploit the s.p.~momentum basis defined by the states Eq.~\eqref{eq: sp states}.
Let us first define the magnitude $k_{\alpha} = \abs{ \mathbf{k}_{\alpha} }$ and direction $\hat{\bm{k}}_{\alpha} = \mathbf{k}_{\alpha} / k_{\alpha}$ associated to the wave number $\mathbf{k}_{\alpha}$ of a given state $\alpha$.
In a similar fashion to Refs.~\cite{Soma2011,Raimondi2017,Barbieri2022Gorkov}, we separate the model space quantum numbers as
\begin{align}
\label{eq:split_alpha}
    \alpha = (k_\alpha, \hat{\bm{k}}_{\alpha}, \tau_\alpha, \sigma_\alpha) = (a; \hat{\bm{k}}_{\alpha},\sigma_\alpha), 
\end{align}
where latin indices denote the magnitude and isospin of a given s.p.~state, $a=(k_\alpha, \tau_\alpha)$. Likewise, the quantum numbers $q$ that label the excitation spectrum of states $\ket{\Psi_q}$ with odd particle number, Eq.~\eqref{eq:Om_eig_sts},  can be separated as
\begin{align}
\label{eq:split_q}
    q = (n_q, k_q, \hat{\bm{k}}_q, \tau_q, \sigma_q) = (n_q, \acroExc{}; \hat{\bm{k}}_q,\sigma_q),
\end{align}
where we defined $\acroExc{} =(k_q, \tau_q)$, $n_q$ is a principal quantum number labeling the excitations within the symmetry block defined by $k_q$ and $\tau_q$, $\mathbf{k}_q$ is the total momentum of $\ket{\Psi_q}$, $\sigma_q$ is the spin and $\tau_q$ denotes the isospin of the fermionic species that has odd particle number\footnote{In the most general case of multiple fermion systems, one may encode this information in terms of \emph{one-hot} vectors with its elements marking the particle number parity of each species. Since we only have two-components in our case, proton and neutron, it suffices to label this information in terms of isospin~\cite{Barbieri2022Gorkov}.}.
Particle number annihilation operators $c_\alpha$ create a time-reversed state with inverted momentum and spin. For the ease of notation we also label the corresponding quantum numbers as $\widetilde{\alpha}=(a; -\hat{\bm{k}}_{\alpha},-\sigma_{\alpha})$ and $\widetilde{q} = (n_q, \ell; -\hat{\bm{k}}_q,-\sigma_q)$. 

In this work, we consider infinite and translationally invariant matter in its ground state $\ket{\Psi_0}$, which has vanishing total momentum $\mathbf{K}_0=0$. We also assume spin saturation, $S_0=0$. 
The operators $c^\dagger_{1/2,\sigma_\alpha}$ and $(-)^{1/2+\sigma_\alpha}c_{1/2,-\sigma_\alpha}$ that create and annihilate (with a proper phase) a spin state are known to be the $\sigma_\alpha$ components of spherical tensor operators of rank $1/2$~\cite{Edmonds:1955fi}. Using the Wigner-Eckart theorem and the fact that momentum and spin/isospin projections are additive quantum numbers, the spectroscopic amplitudes~\eqref{eq:UV_def} satisfy 
\begin{subequations} 
\label{eq:UV_redcd}
\begin{align}
     \mathcal{U}_{\alpha}^{q} &{}= \delta_{\alpha\, q } \;\mathcal{U}_{a}^{n_q}
     = \delta_{a \acroExc{} } \, \delta_{\hat{\bm{k}}_{\alpha} \hat{\bm{k}}_{q} } \delta_{\sigma_{\alpha} \sigma_{q}} \mathcal{U}_{a}^{n_q},  \\
     \mathcal{V}_{\alpha}^{q} &{}= \delta_{\alpha\, \widetilde{q} }\; \eta_{\alpha} \mathcal{V}_{a}^{n_q}
     = \delta_{a \acroExc{} }  \, \delta_{\hat{\bm{k}}_{\alpha},-\hat{\bm{k}}_{q} } \delta_{\sigma_{\alpha}, - \sigma_{q}}  \eta_{\alpha} \mathcal{V}_{a}^{n_q},
\end{align}
\end{subequations}
where we have introduced $\delta_{\alpha\,q} = \delta_{\mathbf{k}_{\alpha} \mathbf{k}_{q} }  \delta_{\tau_{\alpha} \tau_{q}} \delta_{\sigma_{\alpha} \sigma_{q}}$ (independent of $n_q$) and the phase factor $\eta_{\alpha} = (-1)^{1/2 - \sigma_{\alpha}}$ to simplify the notation. Rotational invariance implies that the ``reduced'' amplitudes $\mathcal{U}_{a}^{n_q}$ and $\mathcal{V}_{a}^{n_q}$ do not depend on the momentum direction but only on its magnitude.
Likewise $n_q$ depends only on $\acroExc{} =(k_q, \tau_q)$.

The independence of the amplitudes on the momentum direction is strictly correct in the thermodynamic limit.
However, the discrete basis of Eq.~\eqref{eq:momgrid} used in the present work breaks rotational invariance to a limited extent. The independence of Eqs.~\eqref{eq:UV_redcd} from $\hat{\bm{k}}$ is 
preserved for those momentum configurations that can be turned into each other by means of a sequence of 90${}^{\circ}$ rotations around the lattice axes.
For the purpose of numerical implementation, the quantum numbers $k_\alpha$ and $k_q$ that enter Eqs.~\eqref{eq:split_alpha} and~\eqref{eq:split_q} should be interpreted as labelling groups of momentum states that have the same magnitude \emph{and} that are also connected through such rotations of the s.p.~states. Then, the direction quantum numbers $\hat{\bm{k}}_{\alpha}$ and $\hat{\bm{k}}_{q}$ become indices to label the different states within each group~\cite{Barbieri2017,Marino2023}%
\footnote{
In the case of PBCs, states of a given group are connected by a permutation and/or sign inversion of the $n_{i}$ integers ($i=x,y,z$)~\cite{LietzCompNucl,Marino2023}.
For example, given a momentum $\mathbf{k}=\frac{2\pi}{L}\mathbf{n}$, two states with $\mathbf{n}=(n_x,n_y,n_z)$ and $(-n_y,n_x,n_z)$ correspond to a right angle rotation about the $z$-axis and belong to the same group. However, configurations $\mathbf{n}=(2,2,1)$ and $(0,3,0)$ are not equivalent and must be considered as different states even if their magnitude is the same. Typically, introducing twisted angles, as in Eq.~\eqref{eq:momgrid}, is sufficient to lift all degeneracies and each group ``$k$'' is represented by a sole direction ``$\hat{\bm{k}}$'' within the model space.}.
In general, each group of states $a$ has a multiplicity $g_a=\sum_{\hat{\bm{k}}_a}1$ that becomes a volume factor $g_{k_a}\propto 4\pi k_a^2 dk_a$ in the TL.
Note that the momentum configurations in each group are equivalent and it is sufficient to compute the self-energy and propagator for only one of them. This greatly simplifies simulations in standard PBCs. The introduction of sp-TABC effectively lifts this degeneracy and requires separate simulations for each momentum basis state, as each group only contains a pair of states with opposite spin projection.

Because of Eqs.~\eqref{eq:UV_redcd},  solving the Gorkov equation~\eqref{eq:GkvMtxEq} for a given pole $q$ implies finding the time-forward component of the propagator~\eqref{eq:g_SpectRep} with s.p.~state \hbox{$\alpha$} and the backward component for the time reversed state~$\widetilde{\alpha}$.

\subsection{Single-particle quantities}
\label{app:sp_props}

The conservation laws for momentum and particle-number parity and spin symmetry imply that Gorkov propagators are block diagonal in the corresponding quantum numbers~\cite{Fetter,Stein2016}.
Since each combination of $\mathbf{k}_\alpha$,  $\sigma_\alpha$, and $\tau_{\alpha}$ values is associated with a unique s.p.~state $\alpha$, propagators computed in this work depend solely on the symmetry group $a$ (up to a phase factor).

Substituting Eqs.~\eqref{eq:UV_redcd} into \eqref{eq:g_SpectRep}, one finds:
\begin{subequations} 
\label{eq:g_blocks}
\begin{align}
    g^{11}_{\alpha\beta}(\omega)
    ={}& \delta_{\hat{\bm{k}}_{\alpha} \hat{\bm{k}}_{\beta} } \delta_{\sigma_{\alpha} \sigma_{\beta}} \delta_{a b} \, g_a^{11}(\omega), 
     \\
    g^{12}_{\alpha\beta}(\omega) ={}&
    \delta_{\hat{\bm{k}}_{\alpha}, -\hat{\bm{k}}_{\beta} } \delta_{\sigma_{\alpha}, -\sigma_{\beta}} \delta_{a b}  \, \eta_{\beta} \, g_a^{12}(\omega), 
\end{align}
\end{subequations}
where we have introduced
\begin{align}
    g_a^{11}(\omega) ={}& \sum_{n_q}  \delta_{a \acroExc{}}
     \left(
    \frac{ \abs{ \mathcal{U}_{a}^{n_q} }^2  }{ \hbar\omega - \hbar\acroOmega{} + i\eta } +
    \frac{ \abs{ \mathcal{V}_{a}^{n_q} }^2  }{ \hbar\omega + \hbar\acroOmega{} - i\eta }
     \right)\,,
     \\
    g_a^{12}(\omega)  ={}& \sum_{n_q} \delta_{a \acroExc{}}
    \left(
    \frac{ \mathcal{U}_{a}^{n_q} (\mathcal{V}_{a}^{n_q})^* }{ \hbar\omega - \hbar\acroOmega{} + i\eta } 
    - \frac{  \mathcal{U}_{a}^{n_q} (\mathcal{V}_{a}^{n_q})^*  }{ \hbar\omega + \hbar\acroOmega{} - i\eta } 
    \right)\,,
\end{align}
and we have retained the $\delta_{a \acroExc{}}$ factor to stress that the sum over $n_q$ runs over all poles $\acroOmega{}$ consistent with the group indices $a$.

Similarly, the normal and anomalous densities satisfy
$\rho_{\alpha\beta} = \delta_{\alpha\beta} \rho_{a} $ and $\widetilde{\rho}_{\alpha\beta} =   \delta_{\alpha\widetilde{\beta}} \,\eta_{\alpha} \widetilde{\rho}_{a}$, with
\begin{subequations} 
\label{Eq:NormAnomRoh_grps}
\begin{align}
    & \rho_{a} =  \sum_{n_q} \delta_{a \acroExc{}}  \abs{ \mathcal{V}_{a}^{n_q} }^{2}, \\
    & \widetilde{\rho}_{a} = \sum_{n_q}  \delta_{a \acroExc{}} \,  \mathcal{U}_{a}^{n_q} (\mathcal{V}_{a}^{n_q})^* \,.
\end{align}
\end{subequations}

Similarly, the normal spectral function \eqref{eq:Skw_def} can be written in diagonal form as
\begin{align}
    S^{11}_{\alpha\beta}(\omega) ={}&  \delta_{\alpha \beta }\, S^{11}_{a}(\omega)  \\
    ={}& \delta_{\alpha\beta} \sum_{n_q} \delta_{a \acroExc{}}
    \left(
    \abs{ \mathcal{U}_{a}^{n_q} }^{2} \delta(\hbar\omega-\hbar\acroOmega{})  \right. \nonumber \\
    & \qquad \qquad \qquad+ \left. 
    \abs{ \mathcal{V}_{a}^{n_q} }^{2} \delta(\hbar\omega+\hbar\acroOmega{})
    \right) \, . \nonumber
\end{align}

\subsection{Static self-energy}
\label{App:SEinfty_diags}

\begin{figure}
    \centering
    \includegraphics[width=\columnwidth]{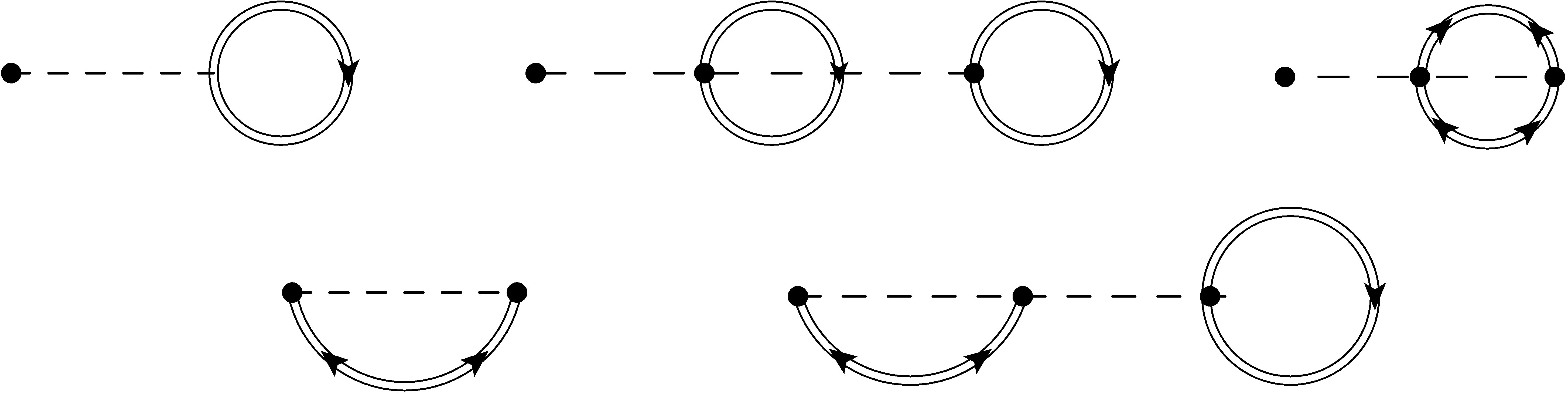}
    \caption{Diagrams contributing to the normal static self-energy $\Lambda$ (top) and its anomalous couterpart $\Delta$ (bottom).
    Dashed lines denote the bare 2B and 3B interactions.
    Dressed normal densities $\rho$ are represented as double lines with a single arrow. 
    Anomalous densities $\widetilde{\rho}$ ($\widetilde{\rho}^{\,*}$) feature two outgoing (incoming) arrows in opposite directions. 
    See e.g.~Ref.~\cite{Soma2011} for the notation.
    }
    \label{fig:diags_sigma_inf_gorkov}
\end{figure}

The static mean field ($\Lambda$) and the pairing field ($\Delta$) that enter Eq.~\eqref{eq:GkvMtxEq} are directly related to the elements of the static self-energy as follows:
\begin{align}
  \Lambda_{\alpha \beta} = \Sigma^{(\infty)\,11}_{\alpha \beta} ={}& - \Sigma^{(\infty)\,22}_{\beta \alpha} = \left(\Lambda_{\beta \alpha}\right)^* \,,\\
  \Delta_{\alpha \beta} = \Sigma^{(\infty)\,12}_{\alpha \beta} ={}& \left(\Sigma^{(\infty)\,21}_{\beta \alpha}\right)^* = - \Delta_{\beta \alpha} \,,
\end{align}
where we also highlight properties from Eq.~\eqref{eq:SigGk_rels}. Note that  $\Lambda$ and  $\Delta$ are Hermitian and antisymmetric matrices, respectively.
We evaluate these fields from the diagrams of Fig.~\ref{fig:diags_sigma_inf_gorkov}, which lead to
\begin{subequations} 
\label{eq:SigGk_static}
\begin{align}
    \label{eq:Sig11_static}
    \Lambda_{\alpha\beta} 
    &= \sum_{\gamma\delta} v_{\alpha\gamma, \beta\delta} \, \rho_{\delta\gamma} \\
    & + \frac{1}{2} \sum_{\gamma\delta\mu\nu}
    w_{\alpha\gamma\mu, \beta\delta\nu} \, \rho_{\delta\gamma} \, \rho_{\nu\mu} + \frac{1}{4} \sum_{\gamma\delta\mu\nu} w_{\alpha\gamma\delta,\beta\mu\nu } \, \widetilde{\rho}_{\mu\nu} \, \widetilde{\rho}_{\gamma\delta}^{\,*},
    \nonumber
\end{align}
and
\begin{align}
    \label{eq:Sig12_static}
    \Delta_{\alpha\beta} = 
    \frac{1}{2} \sum_{\gamma\delta} v_{\alpha\beta, \gamma\delta} \, \widetilde{\rho}_{\gamma\delta} 
    + \frac{1}{2} \sum_{\gamma\delta\mu\nu} w_{\alpha\beta\mu, \gamma\delta\nu} \, \widetilde{\rho}_{\gamma\delta} \, \rho_{\nu\mu}.
\end{align}
\end{subequations} 
Eqs.~\eqref{eq:SigGk_static} generalise the standard HFB potentials (see e.g.~\cite{Signoracci2015,Duguet2015}) by computing the fields in terms of fully correlated (in principle, the exact) densities~\eqref{Eq:NormAnomRoh_grps}, as opposed to being generated by a simpler self-consistent mean field~\cite{Carbone2014,Barbieri2017}. This prescription is often referred to as ``dressed HFB''~\cite{Soma2011}. 
Eqs.~\eqref{eq:SigGk_static} are exact for the contributions coming from 2B forces, while we neglect higher order terms involving the contraction of 3B forces with correlated 2B densities~\cite{Barbieri2017,Carbone2013}. This truncation was shown to have negligible effects on simulations of finite nuclei~\cite{Barbieri2014JPhysG}.

Exploiting the diagonal form of the density matrices $\rho$ and $\widetilde{\rho}$, Eq.~\eqref{Eq:NormAnomRoh_grps}, the expressions for $\Lambda$ and $\Delta$ in spin-saturated infinite matter become
\begin{subequations} 
\label{eq:SigGk_static_grps}
\begin{align}
    \label{eq:Sig11_static_grps}
    \Lambda_{\alpha} 
    ={}& \sum_{g} \rho_g \sum_{\hat{\bm{k}}_{\gamma}\, \sigma_{\gamma}} v_{\alpha\gamma, \alpha\gamma}  
     +  \frac{1}{2}  \sum_{g\, d}  \rho_g  \, \rho_d \sum_{ \substack{\hat{\bm{k}}_{\gamma}\, \sigma_{\gamma}\\ \hat{\bm{k}}_{\delta}\, \sigma_{\delta}} } w_{\alpha\gamma\delta, \alpha \gamma \delta} \nonumber \\
    & + \frac{1}{4} \sum_{g\, d}  (\widetilde{\rho}_g)^*  \, \widetilde{\rho}_d \sum_{\substack{\hat{\bm{k}}_{\gamma}\, \sigma_{\gamma}\\ \hat{\bm{k}}_{\delta}\, \sigma_{\delta}} }
     \eta_\gamma \, \eta_\delta  \, w_{\alpha\gamma\widetilde{\gamma},\alpha\delta\widetilde{\delta} }
\end{align}
and
\begin{align}
    \label{eq:Sig12_static_grps}
    \Delta_{\alpha} ={}&
    \frac{1}{2} \sum_{g} \widetilde{\rho}_g  \sum_{\hat{\bm{k}}_{\gamma}\, \sigma_{\gamma}}  \eta_\gamma \, v_{\alpha\widetilde{\alpha}, \gamma \widetilde{\gamma}}  \nonumber \\
    &+ \frac{1}{2}    \sum_{g\, d}  \widetilde{\rho}_g  \, \rho_d \sum_{\substack{\hat{\bm{k}}_{\gamma}\, \sigma_{\gamma}\\ \hat{\bm{k}}_{\delta}\, \sigma_{\delta}} }     
    \eta_\gamma  \, w_{\alpha \widetilde{\alpha} \delta, \gamma\widetilde{\gamma}\delta}  \,,
\end{align}
\end{subequations} 
where $\widetilde{\gamma}=(g, -\hat{\bm{k}}_{\gamma}, -\sigma_{\gamma})$ and $\widetilde{\delta}=(d, -\hat{\bm{k}}_{\delta}, -\sigma_{\delta} )$.
Note that the static fields retain the same diagonal form of the densities, namely, $\Lambda_{\alpha\beta} = \delta_{\alpha\beta} \Lambda_{\alpha}$ and $\Delta_{\alpha\beta} = \delta_{\widetilde{\alpha}\beta} \Delta_{\alpha}$~\cite{Chamel2008Bcs,Stein2016}.

\section{Dyson-ADC(3) diagrams in infinite nuclear matter}
\label{sec: adc3 diagrams}

\begin{figure}
    \centering  \includegraphics[width=\columnwidth]{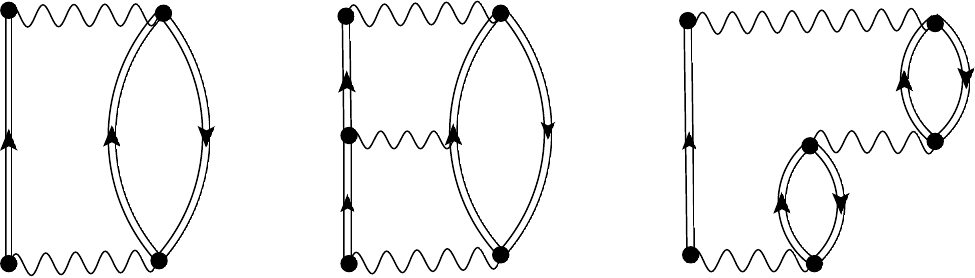}
    \caption[ADC(3) self-energy diagrams]{
    One-particle irreducible, skeleton and interaction-irreducible Feynman diagrams that define the 
    ADC(3) self-energy with up to 2B forces.
    From left to right: second-order contribution, third-order ladder diagram and third-order ring diagram.
    All these propagate 2p1h or 2h1p ISCs. The wiggly lines denote the effective 2B interaction~\eqref{eq: eff interaction 2B} and the double lines with an arrow represent a general dressed propagator, which is substituted by the OpRS, Eq.~\eqref{eq:gOpRS} in this work. }
    \label{fig: adc3 diagrams}
\end{figure}

Expressions~\eqref{eq:HybridSig} and~\eqref{eq:GkvMtxEq} require the dynamical self-energy in the Dyson representation. The complete formalism for 3B Hamiltonians, such as Eq.~\eqref{eq:H0_H1}, truncated at the ADC(3) level has been derived in detail in Ref.~\cite{Raimondi2017}. In the following, we summarize the explicit forms of the coupling vertices $M$ and $N$, as well as of the interaction matrices $C$ and $D$, employed in the present work. These quantities originate from the \emph{interaction-irreducible} diagrams shown in Figs.~\ref{fig: adc3 diagrams} and~\ref{fig:non-skeleton}, and are here specialized to a mean-field-like OpRS propagator of the form given in Eq.~\eqref{eq:gOpRS}.

All diagrams in Figs.~\ref{fig: adc3 diagrams} and~\ref{fig:non-skeleton} incorporate 3B interactions in terms of effective 2N matrix elements ($\widetilde{v}$). We compute the latter via the NO2B approximation~\cite{Carbone2013,Hebeler3nf,Marino2024,Hagen2014},
\begin{align}
    \label{eq: eff interaction 2B}
    \widetilde{v}_{\alpha \beta, \gamma \delta} =
    {v}_{\alpha \beta, \gamma \delta} + \sum_{h\in F}
    {w}_{\alpha \beta h,  \gamma \delta h} \,,
\end{align}
where the 3B force is averaged only over the s.p. states ($h$) that are occupied in the mean-field reference state.
Note that Eq.~\eqref{eq: eff interaction 2B} could be improved by averaging over the correlated density matrix $\rho_{\mu\nu}$, Eq.~\eqref{eq: norm density def}~\cite{Carbone2013,Barbieri2014JPhysG}. 
The static self-energies are instead always computed directly from the bare 2B and 3B interactions (App.~\ref{App:SEinfty_diags}).

With the above prescriptions, our ADC($n \le 3$) calculations entail couplings between s.p.~states and intermediate state configurations of the 2p1h and 2h1p type. The former are indexed by the collective label $r = (n_1 < n_2, k_3)$, where $n$ and $k$ denote, respectively, the particle and hole poles of Eq.~\eqref{eq:Dys_SpectRep} in the general case, and of Eq.~\eqref{eq:gOpRS} for the OpRS reference adopted in this work. The 2h1p ISCs are identified by the triplets $s = (k_1 < k_2, n_3)$. Enforcing an ordered pair for the first two sub-indices, which correspond to poles of the same type, eliminates the need for additional normalization factors in the relations for the matrices $M$, $N$, $C$, and $D$ given below~\cite{Barbieri2017}.

For a general reference state with fragmented s.p. strength, the contributions to ADC($n \le 3$) can be expressed in terms of 2B interactions among the quasiparticle and quasihole indices $n$ and $k$. These are found by contracting the 2B matrix elements Eq.~\eqref{eq: eff interaction 2B} with the corresponding spectroscopic amplitudes~\cite{Raimondi2017}. For example,
\begin{subequations} 
\label{eq:dressedV}
\begin{align}
    \widetilde{v}_{n_1 n_2,  k_3 k_4} ={}&
    \sum_{\alpha \beta, \gamma \delta}
    (\mathcal{U}^{n_1}_\alpha \mathcal{U}^{n_2}_\beta)^* \, \widetilde{v}_{\alpha\beta,\gamma\delta} \,(\mathcal{V}^{k_3}_\gamma \mathcal{V}^{k_4}_\delta)^*\,, \\
    \widetilde{v}_{n_1 n_2,  n_3 n_4} ={}&
    \sum_{\alpha \beta, \gamma \delta}
    (\mathcal{U}^{n_1}_\alpha \mathcal{U}^{n_2}_\beta)^*\, \widetilde{v}_{\alpha\beta,\gamma\delta} \;\mathcal{U}^{n_3}_\gamma \mathcal{U}^{n_4}_\delta \,, \\
    \widetilde{v}_{n_1 k_4,  k_2 n_3} ={}&
    \sum_{\alpha \beta, \gamma \delta}
    (\mathcal{U}^{n_1}_\alpha)^* \, \mathcal{V}^{k_4}_\beta \, \widetilde{v}_{\alpha\beta,\gamma\delta} \, (\mathcal{V}^{k_2}_\gamma)^* \, \mathcal{U}^{n_3}_\delta \,,
\end{align}
\end{subequations} 
and similarly for other cases. Then, the  $t^{(0)}$ 2p-2h amplitudes, Eq.~\eqref{eq: amplitude mbpt}, can be written as
\begin{align}
    (t^{(0)})^{n_1\,n_2}_{k_3\,k_4} = 
    \frac{
    \widetilde{v}_{n_1 n_2, k_3 k_4} }
    { \varepsilon^-_{k_3}+\varepsilon^-_{k_4}-\varepsilon^+_{n_1}-\varepsilon^+_{n_2}} \,.
\end{align} 
We specialize the ADC equations to an OpRS propagator of the form Eq.~\eqref{eq:gOpRS}, with spectroscopic amplitudes given by Kronecker deltas, Eq.~\eqref{eq:OpRS_UV}. Hence, Eqs.~\eqref{eq:dressedV} reduce to the matrix elements~\eqref{eq: eff interaction 2B} with respect to the model space states. Also,
the excitation energies $\epsilon_{n}^{+}$, $\epsilon_{k}^{-}$ are given by the respective OpRS s.p.~energies.

The ADC(3) scheme with 2B effective interactions is defined by the diagrams in Fig. \ref{fig: adc3 diagrams}. 
The second-order diagram generates only coupling matrices that are of order 1 in the interaction, $M^{(1)}$ and $N^{(1)}$. 
The ladder and ring diagrams contribute to both second-order coupling matrices and to the interaction matrices. The different contributions are labelled as $M^{(2,pp)}$, $N^{(2,hh)}$, $C^{(pp)}$ and $D^{(hh)}$, for the ladder term, and $M^{(2,ph)}$, $N^{(2,ph)}$, $C^{(ph)}$ and $D^{(ph)}$, for the ring term.

The ADC(2) is fully defined by the first order vertices $M^{(1)}$ and $N^{(1)}$,
\begin{align}
    \label{eq: M fw adc2}
    & M^{(1)}_{r \, \alpha} = \widetilde{v}_{n_1 n_2, \alpha k_3}, \\
    \label{eq: N bk adc2}
    & N^{(1)}_{\alpha\,s} = \widetilde{v}_{\alpha n_3, k_1 k_2}
\end{align}
and the energies of the unperturbed ISCs, $E^{>}$ and $E^{<}$,
\begin{subequations}
\label{eq:Efw_bk_mats}
\begin{align}
    \label{eq: Efw mat}
    & E^{>}_{r r^\prime} = ( \epsilon_{n_1}^{+} + \epsilon_{n_2}^{+} - \epsilon_{k_3}^{-} ) \,\delta_{rr^{\prime}}, \\
    \label{eq: Ebk mat}
    & E^{<}_{s s^\prime} = ( \epsilon_{k_1}^{-} + \epsilon_{k_2}^{-} - \epsilon_{n_3}^{+} ) \,\delta_{ss^{\prime}} \, .
\end{align}
\end{subequations}

The ADC(3) receives additional contributions from the ladder and ring diadrams. The forward-in-time coupling vertices and interactions are given by
\begin{align}
    \label{eq: contr M ADC3}
    & M = M^{(1)} + M^{(2,pp)} + M^{(2,ph)}, \\
    \label{eq: contr C ADC3}
    & C = C^{(pp)} + C^{(ph)},
\end{align}
where
\begin{align}
    \label{eq: M fw ladder}
    & M^{(2,pp)}_{r \alpha} =
    \frac{1}{2} \sum_{k_4 k_5} (t^{(0)})^{n_1 n_2}_{k_4 k_5} \widetilde{v}_{k_4 k_5, \alpha k_3},  \\
    \label{eq: M fw ring}
    & M^{(2,ph)}_{r \alpha} =
    \mathcal{A}_{12}
    \left[
    \sum_{n_5 k_6}
    (t^{(0)})^{n_2 n_6}_{k_3 k_5} \,
    \widetilde{v}_{n_1 k_5, \alpha n_6}
    \right],
\end{align}
and
\begin{align}
    \label{eq: C fw ladder} 
    & C^{(pp)}_{r r^\prime} = 
    \widetilde{v}_{n_1 n_2, n_1^{\prime} n_2^{\prime} } \,
    \delta_{k_3 k_3^\prime } , \\
    \label{eq: C fw ring} 
    & C^{(ph)}_{r r^\prime} =
    \mathcal{A}_{12} \mathcal{A}_{1^\prime 2^\prime}
    \left[
    \widetilde{v}_{n_1 k_3^{\prime}, k_3 n_1^{\prime} } \,
    \delta_{n_2 n_2^\prime}
    \right].
\end{align}
In the equations above, $\mathcal{A}_{ij}$ is the antisymmetrization operator,
\begin{align}
    \mathcal{A}_{ij} f(i,j,k) = f(i,j,k) - f(j,i,k).
\end{align}
For the backward-in-time self-energy,
\begin{align}
    \label{eq: contr N ADC3}
    & N = N^{(1)} + N^{(2,hh)} + N^{(2,ph)}, \\
    \label{eq: contr D ADC3}
    & D = D^{(hh)} + D^{(ph)},
\end{align}
where
\begin{align}
    \label{eq: N bk ladder}
    & N^{(2,hh)}_{\alpha,s} =
    \frac{1}{2}
    \sum_{n_4 n_5}
    (t^{(0)})^{n_4 n_5}_{k_1 k_2}
    \widetilde{v}_{\alpha n_3, n_4 n_5},   \\
    & N^{(2,ph)}_{\alpha,s} =
    \mathcal{A}_{12}
    \left[
    \sum_{n_4 k_5}
     (t^{(0)})^{n_4 n_3}_{k_5 k_2 }
   \widetilde{v}_{\alpha k_5, k_1 n_4}
    \right]
\end{align}
and 
\begin{align}
    \label{eq: D bk ladder} 
    & D^{(hh)}_{s s^\prime} = 
    - \, \widetilde{v}_{k_1 k_2, k_1^{\prime} k_2^{\prime} } \,
    \delta_{n_3 n_3^\prime} ,  \\
    \label{eq: D bk ring} 
    & D^{(ph)}_{s s^\prime} = 
    - \mathcal{A}_{12} \mathcal{A}_{1^\prime 2^\prime}
    \left[
    \widetilde{v}_{k_1 n_3^{\prime}, n_3 k_1^{\prime} } \,
    \delta_{k_2 k_2^\prime}
    \right].
\end{align}

\subsection{Expression of the non-skeleton contributions}
\label{app: expre non skeleton}

The non-skeleton contributions to the dynamical ADC(3) self-energy are shown in Fig.~\ref{fig:non-skeleton} and detailed in Ref.~\cite{Raimondi2017}.
In the special case of infinite matter with an OpRS propagator~\eqref{eq:gOpRS}, corrections to the coupling matrices vanish, while contributions to the forward and backward interaction matrices are given by
\begin{align}
    & C^{\widetilde{U}} = C^{ \widetilde{U} p } + C^{ \widetilde{U} h }, \\
    & C_{rr^\prime}^{ \widetilde{U} p } = 
    \mathcal{A}_{12} \mathcal{A}_{1^\prime 2^\prime}
    \left[
    \widetilde{u}^{(1)}_{n_1} 
    \delta_{n_1 n_1^\prime } \delta_{n_2 n_2^\prime }
    \right]
    \delta_{k_3 k_3^\prime}, \\
    & C_{rr^\prime}^{ \widetilde{U} h } =
    - \mathcal{A}_{12} 
    \left[
    \delta_{n_1 n_1^\prime } \delta_{n_2 n_2^\prime }
    \right]
    \widetilde{u}^{(1)}_{k_3} 
    \delta_{k_3 k_3^\prime}.
\end{align}
and
\begin{align}
    & D^{\widetilde{U}} = D^{ \widetilde{U} p } + D^{ \widetilde{U} h }, \\
    & D_{ss^\prime}^{ \widetilde{U} h } =
    \mathcal{A}_{12} \mathcal{A}_{1^\prime 2^\prime}
    \left[
    \widetilde{u}^{(1)}_{k_1}  \delta_{k_1 k_1^\prime } \delta_{k_2 k_2^\prime }
    \right]
    \delta_{n_3 n_3^\prime} , \\
    & D_{ss^\prime}^{ \widetilde{U} p } =
    - \mathcal{A}_{12} 
    \left[
    \delta_{k_1 k_1^\prime } \delta_{k_2 k_2^\prime }
    \right]
    \widetilde{u}^{(1)}_{n_3} 
    \delta_{n_3 n_3^\prime},
\end{align}
respectively, with $\widetilde{u}^{(1)}$ given by Eq.~\eqref{eq:u1tilde_practical}.

Formally, including non-skeleton correction consisting in the substitutions $C + E^{>} \to C + E^{>} + C^{\widetilde{U}}$ and $D + E^{<} \to D + E^{<} + D^{\widetilde{U}}$.
In practice, this can be implemented straightforwardly by employing the HF energies in the diagonal matrices $E^{>(<)}$, Eqs.~\eqref{eq:Efw_bk_mats}, instead of the OpRS energies used in standard ADC(3).


\section{Gorkov quantities in dual space}
\label{app:g_vs_G}

Gorkov propagators can also be derived in dual space, where the single particle basis set $\{\alpha\}$ is complemented by a dual basis
$\{\widetilde\alpha\}$ that is in one-to-one correspondence with the former one~\cite{Soma2011,Soma2014Numerical,Barbieri2022Gorkov}. Typically---but not necessarily---states $\alpha$ and $\widetilde\alpha$ are related by the inversion on some quantum number associated with time-reversal symmetry.  Moreover, one introduces an anti-unitary phase $\eta_\alpha$ such that $\eta_\alpha \eta_{\widetilde\alpha} = -1$.

The Gorkov propagators~\eqref{eq:gkv_def} can then be re-defined in dual space as
\begin{align}
\mathbf{G}_{\alpha \beta}(t,t') =& -\frac i\hbar 
\mel{\Psi_0} {T [ 
    \begin{pmatrix}
        c_\alpha(t) c^\dagger_\beta(t') & c_\alpha(t) \bar{c}_\beta(t') \\
        \bar{c}^\dagger_\alpha(t) c^\dagger_\beta(t') & \bar{c}^\dagger_\alpha(t) \bar{c}_\beta(t')
    \end{pmatrix}
]} {\Psi_0} ,
\label{eq:Gkv_def_dual}
\end{align}
where the dual creation and annihilation operators are:
\begin{align}
    \label{eq:g_vs_Gbar}
  \bar{c}_\alpha(t) =\,& \eta_\alpha \, {c}_{ \widetilde{\alpha} }(t), \\
    \bar{c}^\dagger_\alpha(t) =\,& \eta_{\alpha} \, {c}^{\dagger}_{ \widetilde{\alpha} }(t)\, .
\end{align}

With the above definitions the relation between propagators \eqref{eq:gkv_def} and \eqref{eq:Gkv_def_dual} become
\begin{subequations} \label{eq:g_G_equivalence}
\begin{align}
        g^{11}_{\alpha \beta}(\omega) ={}& \quad \quad \, G^{11}_{\alpha \beta}(\omega) \\
        g^{12}_{\alpha \beta}(\omega) ={}& ~~\; - \, G^{12}_{\alpha \widetilde\beta}(\omega) \, \eta_\beta \\
        g^{21}_{\alpha \beta}(\omega) ={}& - \eta_\alpha \, G^{21}_{\widetilde\alpha \beta}(\omega) \\
        g^{22}_{\alpha \beta}(\omega) ={}& \quad \, \eta_\alpha \, G^{22}_{\widetilde\alpha \widetilde\beta}(\omega)\, \eta_\beta \, ,
\end{align}
\end{subequations}
which apply to both time and frequency representations.

The Lehmann representation of propagators $\mathbf{G}_{\alpha \beta}(\omega)$ leads to introduce four spectroscopic amplitudes
\begin{subequations} \label{eq:UV_def_dual}
\begin{align}
    & {}^{(D)}\mathcal{U}_{\alpha}^{q} = \mel{\Psi_0}{c_\alpha}{\Psi_q}, \\
    & {}^{(D)}\mathcal{V}_{\alpha}^{q} = \mel{\Psi_0}{\bar{c}^\dagger_\alpha}{\Psi_q}, \\
    & {}^{(D)}\bar{\mathcal{U}}_{\alpha}^{q} = \mel{\Psi_0}{\bar{c}_\alpha}{\Psi_q} = \quad \eta_\alpha \, {}^{(D)}\mathcal{U}^q_{\widetilde\alpha}, \\
    & {}^{(D)}\bar{\mathcal{V}}_{\alpha}^{q} = \mel{\Psi_0}{c^\dagger_\alpha}{\Psi_q} = -\, \eta_\alpha \, {}^{(D)}\mathcal{V}^q_{\widetilde\alpha},
\end{align}
\end{subequations}
where the subscript $(D)$ marks quantities defined in dual space. These are simply related to Eqs.~\eqref{eq:UV_def} by 
\begin{align}
    \mathcal{U}_{\alpha}^{q} = {}^{(D)}\mathcal{U}_{\alpha}^{q} 
    \qquad \hbox{and} \qquad
    \mathcal{V}_{\alpha}^{q} = {}^{(D)}\bar{\mathcal{V}}_{\alpha}^{q} \,.
   \label{eq:UV_equivalence}
\end{align}

Finally, comparing Eqs.~\eqref{eq:g_G_equivalence} with the Gorkov equation~\eqref{eq:GorkovEq} allows to extract the correspondence between self-energy components in normal and dual spaces:
\begin{subequations} \label{eq:Sigma_equivalence}
\begin{align}
        \Sigma^{\star\, 11}_{\alpha\beta}(\omega) ={}& \quad \, {}^{(D)}\Sigma^{\star\, 11}_{\alpha\beta}(\omega) \,, \\
        \Sigma^{\star\, 12}_{\alpha\beta}(\omega) ={}& \quad \, {}^{(D)}\Sigma^{\star\, 12}_{\alpha \widetilde{\beta}}(\omega) \, \eta_{\widetilde\beta} \,, \\
        \Sigma^{\star\, 21}_{\alpha\beta}(\omega) ={}& \eta_{\widetilde{\alpha}} \, {}^{(D)}\Sigma^{\star\, 21}_{\widetilde{\alpha} \beta}(\omega) \,, \\
        \Sigma^{\star\, 22}_{\alpha\beta}(\omega) ={}& \eta_{\widetilde\alpha} \, {}^{(D)}\Sigma^{\star\, 22}_{\widetilde{\alpha} \widetilde{\beta}}(\omega)\, \eta_{\widetilde\beta} \, .
\end{align}
\end{subequations}
The latter relations apply separately to the static and dynamic components of the self energy, $\Sigma^{(\infty)}$ and $\widetilde\Sigma(\omega)$.
Combining Eqs.~\eqref{eq:SigGk_rels} and~\eqref{eq:Sigma_equivalence} imposes the following Lehmann representation of the self-energy in dual space~\cite{Barbieri2022Gorkov}:
\begin{align}
  {}^{(D)}\widetilde{\mathbf{\Sigma}}_{\alpha\beta}&(\omega) =
  \nonumber \\
  ={}&\sum_{\nu \,\nu'}
  \begin{pmatrix}
        \mathcal{C}_{\alpha\,\nu} \\
        \mathcal{D}_{\alpha\,\nu}^T
    \end{pmatrix} 
    \left[
  \frac{1}{\hbar\omega\IdentityMat - \mathcal{E} + i\eta}
  \right]_{\nu \nu'}
     \begin{pmatrix}
        \mathcal{C}_{\nu'\beta}^\dagger &
        \mathcal{D}_{\nu'\beta}^*
    \end{pmatrix} \nonumber \\
    +{}& \sum_{\nu \,\nu'}
  \begin{pmatrix}
        \bar{\mathcal{D}}_{\alpha\,\nu}\dagger  \\
        \bar{\mathcal{C}}_{\alpha\,\nu}^*
    \end{pmatrix} 
    \left[
  \frac{1}{\hbar\omega\IdentityMat + \mathcal{E}^T - i\eta}
  \right]_{\nu \nu'}
     \begin{pmatrix}
        \bar{\mathcal{D}}_{\nu'\beta} &
        \bar{\mathcal{C}}_{\nu'\beta}^{\,T}
    \end{pmatrix} \,,
  \label{eq:DynSE_Gk_dual} 
\end{align}
with%
\footnote{Eqs.(30) of Ref.~\cite{Barbieri2022Gorkov} are also correct provided that the phase is both real and anti-unitaty (that is, $\eta_\alpha=\pm 1$). This is the case in most applications. Relations~\eqref{eq:CDbar_vs_CD}, given here, are more general since they only assume the anti-unitarity.}
\begin{subequations} \label{eq:CDbar_vs_CD}
\begin{align}
  \bar{\mathcal{C}}_{\alpha,\nu} ={}& - \eta_{\widetilde{\alpha}} \; \mathcal{C}_{\widetilde{\alpha},\nu}  \,, \\     
  \bar{\mathcal{D}}_{\nu,\alpha} ={}& \quad \,\eta_{\widetilde{\alpha}} \, \mathcal{D}_{\nu,\widetilde{\alpha}}     \,.   
\end{align}
\end{subequations}

Comparing Eqs.~\eqref{eq:Sigma_equivalence} and~\eqref{eq:DynSE_Gk_dual} to \eqref{eq:DynSE_Gk} allows to identify the corresponding coupling matrices of the standard formulation as
\begin{subequations} \label{eq:MN_vs_CDbar}
\begin{align}
  \mathcal{M}_{\nu,\alpha} ={}&  \mathcal{C}^\dagger_{\nu,\alpha} \,,   \\     
  \mathcal{N}_{\alpha,\nu} ={}& \bar{\mathcal{D}}^\dagger_{\alpha,\nu} \,.
\end{align}
\end{subequations}
Working equations up to ADC($3$) level were derived in Ref.~\cite{Barbieri2022Gorkov} in dual space space. Their counterpart in simple model space can be easily obtained by expressing coupling matrices $\mathcal{C}$ and $\bar{\mathcal{D}}$ in terms of the ${}^{(D)}\mathcal{U}$ and ${}^{(D)}\bar{\mathcal{V}}$. One then finds the self-energy Eq.~\eqref{eq:DynSE_Gk} by using Eqs.~\eqref{eq:MN_vs_CDbar} and taking the correspondence Eq.~\eqref{eq:UV_equivalence} into account.

\bibliography{bibliography.bib}

\end{document}